\documentclass[%
reprint,
superscriptaddress,
amsmath,amssymb,
aps,
prx,
floatfix,
]{revtex4-1}

\usepackage{amstext}
\usepackage{graphicx}
\usepackage{color}
\usepackage{soul}
\usepackage{MnSymbol}
\usepackage{subfiles}
\usepackage{bm}
\usepackage{hyperref}
\usepackage{layouts}
\usepackage{enumitem}
\usepackage{natbib}

\newcommand{\dd}{\mathrm{d}}
\newcommand{\e}{\mathrm{e}}
\newcommand{\sech}{\operatorname{sech}}
\newcommand{\sign}{\operatorname{sign}}
\newcommand{\E}[1]{\left\langle #1 \right\rangle}
\DeclareMathOperator*{\saddle}{saddle}

\begin{document}

\title{Unlearnable Games and ``Satisficing'' Decisions: \\
A Simple Model for a Complex World}

\author{Jérôme Garnier-Brun}
\email{jerome.garnier-brun@polytechnique.edu}
\affiliation{Chair of Econophysics and Complex Systems, \'Ecole polytechnique, 91128 Palaiseau Cedex, France}
\affiliation{LadHyX, CNRS, École polytechnique, Institut Polytechnique de Paris, 91120 Palaiseau, France}

\author{Michael Benzaquen}
\affiliation{Chair of Econophysics and Complex Systems, \'Ecole polytechnique, 91128 Palaiseau Cedex, France}
\affiliation{LadHyX, CNRS, École polytechnique, Institut Polytechnique de Paris, 91120 Palaiseau, France}
\affiliation{Capital Fund Management, 23 Rue de l’Universit\'e, 75007 Paris, France}
\author{Jean-Philippe Bouchaud}
\affiliation{Chair of Econophysics and Complex Systems, \'Ecole polytechnique, 91128 Palaiseau Cedex, France}
\affiliation{Capital Fund Management, 23 Rue de l’Universit\'e, 75007 Paris, France}
\affiliation{Académie des Sciences, 23 Quai de Conti, 75006 Paris, France}

\date{\today}

\begin{abstract}
As a schematic model of the complexity economic agents are confronted with, we introduce the ``SK-game'', a discrete time binary choice model inspired from mean-field spin-glasses. We show that even in a completely static environment, agents are unable to learn collectively-optimal strategies. This is either because the learning process gets trapped in a sub-optimal fixed point, or because learning never converges and leads to a never ending evolution of agents intentions. Contrarily to the hope that learning might save the standard ``rational expectation'' framework in economics, we argue that complex situations are generically {\it unlearnable} and agents must do with {\it satisficing} solutions, as argued long ago by Herbert Simon \cite{simon1955behavioral}. Only a centralized, omniscient agent endowed with enormous computing power could qualify to determine the optimal strategy of all agents. 
Using a mix of analytical arguments and numerical simulations, we find that (i)  
long memory of past rewards is beneficial to learning whereas over-reaction to recent past is detrimental and leads to cycles or chaos; (ii) increased competition destabilizes fixed points and leads first to chaos and, in the high competition limit, to quasi-cycles; (iii) some amount of randomness in the learning process, perhaps paradoxically, allows the system to reach better collective decisions; (iv) non-stationary, ``aging'' behaviour spontaneously emerge in a large swath of parameter space of our complex but static world. On the positive side, we find that the learning process allows cooperative systems to coordinate around {\it satisficing} solutions with rather high (but markedly sub-optimal) average reward. However, hyper-sensitivity to the game parameters makes it impossible to predict {\it ex ante} who will be better or worse off in our stylized economy. 
\end{abstract}

\maketitle

{\it La complexité de l’ensemble fait que tout ce qui peut arriver est vraiment, malgré l’expérience acquise, impossible à prévoir, encore plus à imaginer. Il est inutile de tenter de le décrire, car on peut concevoir n’importe quelle solution.\footnote{The complexity of the whole means that anything that can happen is really, despite the experience gained, impossible to predict, let alone imagine. It is pointless
attempt to describe it, because any solution is conceivable.}

Boris Vian, in ``L’Automne à Pékin''.}

\tableofcontents

\section{Introduction}
\label{sec:introduction}
\subsection{Rationality, Complexity \& Learning}

Classical economics is based on the idea that rational agents make optimal decisions, i.e. optimize their expected utility over future states of the world, weighted by their objective probabilities. Such an idealisation of human behaviour has been criticized by many (see e.g. \cite{simon1955behavioral, grandmont1998expectations, kirman2010complex, hommes2021behavioral, dosi2019more, king2020radical}). In particular, assuming that all agents are rational, allowing one to use game theoretic arguments to build such optimal strategies -- often the result of complicated mathematical calculations -- is implausible, to say the least. 

A way to possibly save the rational expectation paradigm is to posit that agents are able to learn best responses from past experience. Yes, agents are only ``boundedly'' rational, but they learn and in the long run, they act  ``as if'' they were rational \cite{marcet1989convergence}. This is clearly expressed by Evans and Honkapohja in their review paper on the subject \cite{evans2013learning}. They note that {\it [i]n standard macroeconomic models rational expectations can emerge in the long run, provided the agents’ environment remains stationary for a sufficiently long period}.

While seemingly reasonable, this proposition is by no means guaranteed to be legitimate. Indeed, the hypothesis that the environment should be stationary over ``sufficiently long periods'' can be restated in terms of the speed of convergence of the learning process, that should be short enough compared to the correlation time scale of the environment. However, in many circumstances and in particular in complex games, the convergence of the learning process to a collectively optimal state can be exceedingly long, or may in fact {\it never} take place. For example, reasonable learning rules can trap the system in some sub-optimal regions of the (high dimensional) solution space, see e.g. \cite{pemantle2007survey,lamberton2004can,moran2020force, colon2022radical}. In other words, the learning process itself can be non-ergodic, even if the environment is described by an ergodic, stationary process. Another possibility is that agents' strategies, even probabilistic, evolve chaotically forever, as was found by T. Galla \& D. Farmer \cite{galla2013complex} in the context of competitive multi-choice two-player games, or by one of us (JPB) with R. Farmer  in a simple binary choice, multiplayer game \cite{bouchaud2023self}. In such cases, the probabilities governing the different possible choices are not fixed but must themselves be described by probabilities.

This is in fact a generic feature of ``complex systems''. As proposed by G. Parisi \cite{PARISI1999557,parisi2007physics}, the description of such systems requires the introduction of \textit{probabilities of probabilities}, as their statistical behaviour themselves (and not only individual trajectories) are highly sensitive to the small changes in parameters, initial conditions, or time. The inability to describe such systems with knowable probabilities was coined ``radical complexity'' in \cite{bouchaud2021radical}. 

The sensitivity of optimal solutions to the parameters of the problem, or to the algorithm used to find them, has a very real consequence: one can no longer assume that all agents, even fully rational, will make the same decision, since any small perturbation may lead to a completely different solution, although similar in performance. In other words, the ``common knowledge'' assumption is not warranted.\footnote{``Common knowledge'' means that ``You know what I know and I know that you know what I know''.} This has already been underlined for example in the context of portfolio optimisation in \cite{galluccio1998rational,garnier2021new}, or in the context of networked economies \cite{sharma2021good,colon2022radical}, but is expected to be of much more general scope, as anticipated by Keynes long ago and emphasized by many heterodox economists in the more recent past \cite{simon1955behavioral, kirman2010complex, dosi2019more, king2020radical}. 

Here we want to dwell on this issue in the context of a multi-player binary game -- the ``SK-game'' --, understood as an idealisation of the economic world where agents strongly interact in such a way that their payoffs depend non trivially on the action of others. In our setting, some relationships are mutually beneficial, while others are competitive. Agents have to learn how to coordinate to optimize their expected gains, which they do in a standard reinforcement way by observing the payoff of their past actions and adapting their strategies accordingly.   

We find that our stylized model gives rise to a very wide range of dynamical behaviour: sub-optimal fixed points (a.k.a. Nash equilibria), but also limit cycles when learning is too fast, and ``chaos'', meaning that individual mixed strategies never converge. Chaos can be deterministic, when learning is noiseless, or stochastic when random noise is introduced in the learning process. In such chaotic situations, the environment of each agent changes over time, not because of exogenous shocks but because of the endogenous learning dynamics of other agents. As also argued in \cite{colon2022radical}, a purely static network of interactions can generate an apparent never-ending evolution of the world in which agents live, although it may appear stationary for very long periods of time. In such a case, one speaks of ``aging'', which we address in a dedicated section below.

Even when learning is efficient and fixed points are reached, the average pay-off of agents -- although better than random -- is noticeably lower than the maximum possible value, that could be obtained if a perfectly informed social planner with colossal computing power was dictating their strategies to each agent. 

Hence, our model provides an explicit example of Herbert Simon's {\it satisficing} principle \cite{simon1955behavioral}: with reasonable, boundedly rational methods, agents facing complex problems can only reach some satisfying and sufficing but sub-optimal solutions. In fact, in the absence of the omniscient and omnipotent social planner alluded to above, agents can hardly do better -- the world in which they live in is {\it de facto unlearnable}. This is the main message of our paper. 

A further interesting twist of the model is that the fixed point reached by the learning process depends sensitively on the initial condition and/or the detailed structure of the interaction network. As a consequence, which agents will do better than average and which will do worse  is totally unpredictable, even when the economic interactions between agents are fixed and known. Correspondingly, small changes in the interaction network (as a result of -- say -- small exogenous shocks, or as some agents die and other are born \cite{bouchaud2023self}) can lead to very large changes in individual strategies that agents would have to re-learn from scratch. 

This feature actually corresponds to another possible definition of a ``complex'' system \cite{parisi2007physics}: small perturbations can lead to disproportionally large reconfigurations of the system -- think of sand (or snow) avalanches as an archetype of this type of behaviour: a single grain of sand might leave the whole slope intact or, occasionally, trigger a landslide \cite{bak2013nature} (see also, e.g., \cite{anderson2018economy, bak1993aggregate, bookstaber2017end, kirman2010complex, shiino1990replica, dosi2019more, moran2019may, sharma2021good, dosi2023foundations} for early and more recent discussions of complexity ideas in the context of economics).

\subsection{Related Ideas}

Our work follows the footsteps of several important contributions on the subject of learning in economics, beyond the papers already quoted above, such as \cite{simon1955behavioral, evans2013learning}, and in particular the complex two-player game of Galla \& Farmer \cite{galla2013complex}. Another relevant work is the study of Grandmont \cite{grandmont1998expectations} on the stability of economic equilibrium under different learning rules. Although his general conclusions are somewhat similar to ours, there are important differences, due to the fact that complexity, in our model, arises from the strong interaction between agents. The main differences are the following: 
\begin{itemize}
    \item While our agents can in some cases (i.e. when interactions are reciprocal) converge to a locally stable equilibrium, this equilibrium is rather inefficient compared to the best theoretical one, which is unattainable without the help of a central planner with super-powers. Furthermore, instead of a single equilibrium in the Grandmont case, the equilibrium reached by our learning agents is one among an exponential number of possible equilibria, that are each sensitively dependent on the parameters of the model. In other words, which equilibrium is reached by the agents in totally unpredictable.
    \item As soon as some small amount of noise is present, Grandmont style instabilities set in and the system starts exploring the set of possible equilibria, albeit in a slow, non-ergodic way. In other words, the stability of the visited equilibria increases with the age of the system.
    \item In the case where interactions are sufficiently non-reciprocal, there are no fixed points to the learning dynamics except the trivial one where agents play randomly at each turn (like ``rock-paper-scissors''). However, this is {\it not} what agents agree to do. They keep holding strong beliefs at each time step, such beliefs evolving chaotically  in time -- but now in an ergodic way.\footnote{For other examples of models where learning leads to chaotic dynamics, see e.g. \cite{brock1997rational, brock1998heterogeneous}.} 
\end{itemize} 

Another somewhat related idea can be found in the recent paper of Hirano \& Stiglitz \cite{hirano2022land}, where there can be a plethora of equilibrium trajectories, neither converging to a steady state or even to a limit cycle, what they call  ``wobbly” macro-dynamics. We may finally mention the work of Dosi \textit{et al.} \cite{dosi2020rational}, which showcases that sophisticated learning rules can fail at improving the individual and collective outcomes when interacting agents are heterogeneous. Although the context is quite different, such a finding appears to be consistent with the saturation of the average reward at sub-optimal satisficing levels that we observe when the memory loss rate of our learning agents vanishes.

\subsection{Outline of the paper}

The layout of the paper is as follows. In Sec.~\ref{sec:model}, we introduce our spin-glass inspired game. We then review the model's main features and present what we believe are the most important conceptual results in the socio-economic context in Sec.~\ref{sec:summary}. In Sec.~\ref{sec:statics}, we enter the technical analysis of the ``SK-game'' by discussing the existence and abundance of fixed point solutions. A similar analysis is conducted for short limit cycles when there are no fluctuations in the system in Sec.~\ref{sec:Cycles}. Having determined when these solutions may or may not exist, we take interest in the $N\to \infty$ dynamics in Sec.~\ref{sec:DMFT} by writing the Dynamical Mean-Field Theory of the problem. Combining the static and dynamic pictures, we focus on the fully rational (or ``zero temperature'') limit and in particular on the role of learning on the collective dynamics in Sec.~\ref{sec:deterministic}. Section~\ref{sec:fluctuations} finally considers the effect of fluctuations stemming from the bounded rationality of agents in their decision making process. In Sec.~\ref{sec:conclusion}, we summarize our findings and discuss future perspectives as well as the relevance of our model for other complex systems such as biological neural networks.

\section{A simple model for a complex world}
\label{sec:model}

\subsection{Set-up of the Model}

As a minimal, stylized model for decision making in a complex environment of interacting agents, we restrict ourselves to binary decisions, as in many papers on models with social interactions, see e.g. \cite{brock2001discrete, challet2004minority, gordon2009discrete, bouchaud2013crises, bouchaud2023self}. At every timestep $t$, each agent $i$ plays $S_i(t) = \pm 1$, with $i = 1,\dots,N$, which can be thought of, for example, as the decision of an investor to buy or to sell the stock market, or the decision of a firm to increase or to decrease production, etc. The incentive to play $S_i(t)=+1$ is $Q_i(t)$ and is the agent's estimate of the payoff associated to $S_i(t)=+1$ compared to that of $S_i(t)=-1$. The actual decision of agent $i$ is probabilistic and drawn using the classic ``logit'' rule \cite{anderson1992discrete}, i.e. sampled from a Boltzmann distribution over the choices,
\begin{equation}
    \mathbb{P}\left[ S_i(t) = \pm 1\right] = \frac{\e^{\pm \beta Q_i(t)}}{\e^{\beta Q_i(t)} + \e^{- \beta Q_i(t)}} = \frac12 \Big[ 1 \pm \tanh \left(\beta Q_i(t)\right) \big],
    \label{eq:logit}
\end{equation}
or, equivalently, the expected choice $m_i$ (or ``intention'') of agent $i$ at time $t$ is given by
\begin{equation}
    m_i(t) := \langle S_i(t) \rangle = \tanh \left( \beta Q_i(t) \right).
    \label{eq:m_i_Q}
\end{equation}
Parameter $\beta$, assumed to be independent of $i$ henceforth, is analogous to the inverse temperature in statistical physics and represents the agent's \textit{rationality} or \textit{intensity of choice}.\footnote{For a complete review of the motivations behind this rule, see \cite{anderson1992discrete, bouchaud2013crises} and references therein.}  The limit $\beta \to \infty$  corresponds to perfectly rational agents, in the sense that they systematically pick the choice corresponding to their preference (given by the sign of $Q_i(t)$), while setting $\beta = 0$ gives erratic agents that randomly pick either decision with probability $1/2$ regardless of the value of $Q_i(t)$. This will in fact turn out the be the case, in our model, in a whole region of parameter space: when $\beta$ is smaller than some critical value $\beta_c$, all intentions $m_i$ do converge to zero, leading to a random string of decisions. 

The evolution of the preference $Q_i(t)$ is where the learning takes place. We resort in so-called ``$Q$-learning'' \cite{watkins1992q}, i.e. reinforcement learning with a memory loss parameter $\alpha$. Given the (yet unspecified) reward $\pm R_i(t)$  associated to making the choice $\pm 1$ at time $t$, the evolution of incentives (and, in turn, beliefs) is given by
\begin{equation}
    Q_i(t+1) = (1-\alpha) Q_i(t) + \alpha R_i(t).
    \label{eq:learning_general}
\end{equation}
This map amounts to calculating an Exponentially Weighted Moving Average (EWMA) on the history of rewards $R_i(t)$. Taking $\alpha = 0$, the agent's preferences are fixed at their initial values, and we thus restrict ourselves to $\alpha > 0$. When $\alpha \to 0$, $Q_i(t)$ is approximately given by the average reward over the last $\alpha^{-1}$ time steps. Note here that this \textit{averaging} of past rewards is not exactly the same as the \textit{accumulation} rule (where the reward would not be multiplied by $\alpha$ in Eq. \eqref{eq:learning_general}) appearing in some forms of ``Experience Weighted Attraction'' that are popular in the socio-economic context \cite{camerer1999experience}. In fact one can always rescale the prefactor $\alpha$ into $\beta$, so our choice just means that $Q$ does not artificially explode as $\alpha \to 0$, which would impose that $\beta$ effectively diverges in that limit.

Now, the missing ingredient is the specification of the rewards, that encodes heterogeneity and non-reciprocity of interactions. Inspired by the theory of spin-glasses, in particular by the Sherrington-Kirkpatrick (SK) model \cite{sherrington1975solvable,mezard1987spin}, we set
\begin{equation}
    R_i(t) = \sum_{j=1}^N J_{ij} S_j(t).
    \label{eq:reward}
\end{equation}
Here, the matrix elements $J_{ij}$ specify the mutually beneficial or competitive nature of the interactions between $i$ and $j$. (Note that $J_{ij}$ measures the impact of the decision of $j$ on the reward of $i$.) 

In the context of firm networks, a client-supplier relation would correspond to $J_{ij} > 0$, whereas two firms $i,j$ competing for the same clients would correspond to $J_{ij} < 0$. In the so-called ``Dean problem'', $J_{ij} > 0$ means that agents $i$ and $j$ get along well whereas $J_{ij} < 0$ means that they are in conflict \cite{panchenko2015introduction}. The sign of $S_i$ determines in which of the two available rooms agent $i$ should sit, in order to minimize the number of possible conflicts. A predator-prey situation is when $J_{ij} \times J_{ji} < 0$, meaning that if $i$ makes a gain, $j$
makes a loss and vice versa.

Note that reward $R_i(t)$ depends on the actual (realized) decision of other players, and not their expected decisions or intentions. In other words, agents resort in \textit{online} learning, which differs from \textit{offline} learning where other players' decisions $S_i(t)$ are averaged over large batch sizes during which their inclinations would be assumed constant and replaced by their expectation $m_i(t)$. 

Based on the learning dynamics, agents thus make a decision based on an imperfectly learned approximation of what other players are likely to do. Indeed, using Eq.~\eqref{eq:reward} to express $Q_i(t)$ as the time-weighted sum (EWMA) of past rewards, plugging Eq.~\eqref{eq:learning_general},  and finally replacing in Eq.~\eqref{eq:m_i_Q}, one finds that the evolution of agent $i$'s intention writes
\begin{equation}
    m_i(t+1) = \tanh\left( \beta \sum_{j=1}^N J_{ij} \tilde{m}^\alpha_j(t) \right)
    \label{eq:NMFE_general},
\end{equation}
where $\tilde{m}^\alpha_i(t)$ can be interpreted as the estimate of agent $j$'s expected decision at time $t+1$ based on its past actions up to time $t$,
\begin{equation}
    \tilde{m}^\alpha_i(t) = \alpha \sum_{t' \leq t} (1-\alpha)^{t-t'} S_i(t').
\end{equation}
Expressed in this form, it is clear that there is characteristic timescale $\tau_\alpha \sim 1/\alpha$ over which past choices contribute  to the moving average $\tilde{m}^\alpha_i(t)$. Note that \textit{offline} learning would correspond to a different evolution equation, namely
\begin{equation}
    m_i(t+1) = \tanh\left( \beta \sum_{j=1}^N J_{ij} {m}_j(t) \right),
    \label{eq:NMFE_offline}
\end{equation}
although the two coincide in the $\alpha \to 0$ limit.

At this stage, it may be useful to compare and contrast the present model with previous work. On the one hand, the learning procedure closely resembles the original proposition by Sato and Crutchfield \cite{sato2005stability}, and its treatment by Galla \cite{galla2009intrinsic,galla2011cycles,galla2013complex} and others \cite{vilone2011chaos,kianercy2012dynamics,burridge2015limit,pangallo2019best}, however these authors considered games comprising  only two players with many strategies. The subsequent case explored by Galla and Farmer considering a larger number of players \cite{sanders2018prevalence} therefore lies closer to our setting, but still consider many strategies, while the similar binary decision models proposed by Semeshenko \textit{et al.} \cite{semeshenko2006choice,semeshenko2008collective} are restricted to perfectly rational agents and homogeneous interactions $J_{ij} = J_0 > 0 \; \forall i,j$. Note that all these works also consider accumulated rewards and offline learning, in contrast with our averaged rewards and online learning (for a comparison between the two and other learning dynamics, see \cite{pangallo2019best}). On the other hand, replicator models with random non-symmetric interactions between a large number of species \cite{opper1992phase,galla2006random} share many features with the system at hand, but the prescribed dynamics are inherently linked to evolutionary principles such as extinction that are not present in our model. Finally, other Ising-inspired games such as that introduced in \cite{leonidov2021ising} are conceptually similar, in particular in their extension with myopic strategy revision (meaning updating based on future expectations and not directly passed realizations as done here) \cite{leonidov2022strategic}. So far, however, these models have been studied without heterogeneities and therefore do not present the ``radical complexity'' related to the presence of a very large number of possible solutions discussed hereafter. 

\subsection{The Interaction Matrix}

In order to rely on known results about the SK model, we will assume in the following that the all agents randomly interact with one another, meaning that all elements of the matrix $\mathbf{J}$ are non-zero. Sparse matrices, corresponding to low-connectivity interaction matrices, would probably be more realistic in an economic context. However, we expect that many of the conclusions reached below will qualitatively hold in such cases as well.

We choose interactions $J_{ij}$ between $i$ and $j$ to be random Gaussian variables of order $N^{-1/2}$, with $J_{ij}$ in general different from $J_{ji}$, accounting for possible non-reciprocity of interactions. More precisely, we introduce the parameter $\varepsilon$ and write the interaction matrix as
\begin{equation}
    J_{ij} = \left(1 - \frac{\varepsilon}{2}\right) J_{ij}^S + \frac{\varepsilon}{2} J_{ij}^A,
    \label{eq:matrix_full}
\end{equation}
with $\mathbf{J}^S$ a symmetric matrix and $\mathbf{J}^A$ an anti-symmetric matrix.  The entries of both these matrices are independent and sampled from a Gaussian distribution of mean 0 and variance $\sigma^2/N$ (the case of a non-zero average value of $\mathbf{J}^S$ will be briefly discussed below). This defines what we will call the ``SK-game'' henceforth.   

In the following we set $\sigma =1$ without loss of generality. The resulting variance of $J_{ij}$ is thus given by
\begin{equation}
    \upsilon(\varepsilon):=N \mathbb{V}[J_{ij}] = 1-\varepsilon + \frac12 \varepsilon^2. 
\end{equation}
The specific cases $\varepsilon = \{0,1,2\}$ hence correspond to fully symmetric ($J_{ij}=J_{ji}$), a-symmetric (i.e. $J_{ij}$ and $J_{ji}$ independent) and anti-symmetric ($J_{ij}=-J_{ji}$) interactions respectively. We can thus also characterize the correlation between $J_{ij}$ and $J_{ji}$ through parameter $\eta$,
\begin{equation}
    \eta = \frac{\overline{J_{ij}J_{ji}}}{\overline{J_{ij}^2}} = \frac{1 - \varepsilon}{\upsilon(\varepsilon)}, \label{eq:eta}
\end{equation}
where overlines indicate an average over the disorder.

It may actually be insightful to allow for a non-zero average value to the interaction parameters, and define the reward $R_i(t)$ as
\begin{equation}
    R_i(t) = \sum_{j=1}^N J_{ij} S_j(t) + J_0 M(t); \quad M(t):=\frac1N \sum_{j=1}^N S_j(t)
    \label{eq:reward2}.
\end{equation}
Note that if only the $J_0$ term is present, the game becomes simple, in the sense that either $J_0 > 0$ and $\beta J_0 > 1$ and all agents converge to the same strategy $m_i \equiv m$ with $m$ a non zero solution of $m = \tanh(\beta J_0 m)$, or they converge to the random ``rock-paper-scissor'' strategy $m_i \equiv 0$.

Finally, one may think that agents have some idiosyncratic preferences, or different costs associated to the two possible decisions $S_i = \pm 1$. This would amount to adding to the reward $R_i(t)$ a time independent term $H_i$, where $H_i$ favors $S_i=+1$ if positive and $S_i=-1$ if negative. In the present paper we will restrict to $H_i \equiv 0$, $\forall i$, but one expects from the literature on spin-glasses that the main results discussed below would still hold for small enough $H_i$s. Beyond some threshold value, on the other hand, agents end up aligning to their {\it a priori} preference, i.e. $m_i H_i > 0$.

\section{Basic intuition and main results}
\label{sec:summary}

\begin{figure}
    \centering
    \includegraphics[width=\linewidth]{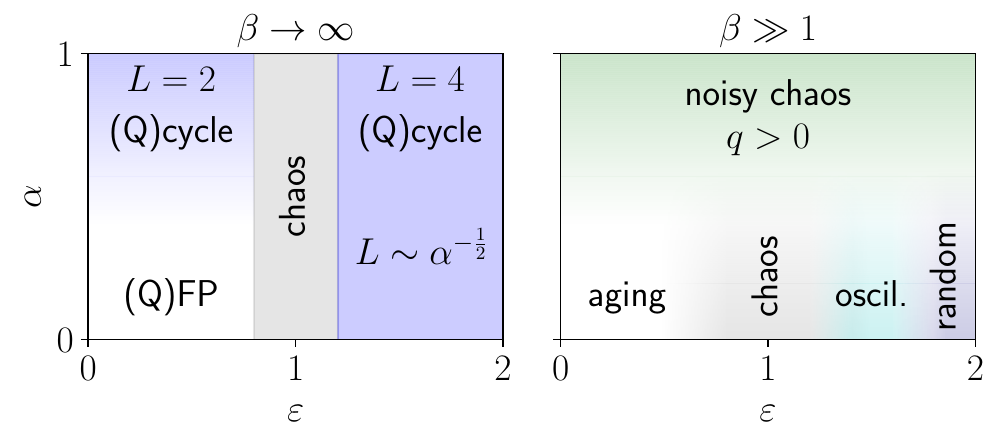}
    \caption{Qualitative phase diagram in the $(\varepsilon,\alpha)$ plane for the SK-game in the noiseless (left) and weak noise (right) regimes. FP  and (Q) refer to ``fixed point'' and ``quasi'' respectively ($m_i(t) = 0 \; \forall t$). ``Random'' refers to the case where agents pick $\pm 1$ with equal probability, whereas ``chaos'' means that at a given instant of time players have well defined intentions $m_i$ but these evolve chaotically with time.}
    \label{fig:qualitative_phase_diag}
\end{figure}

In this section, we review the most salient properties of the ``SK-game'' introduced above, and why these results may be of interest in the context of socio-economic modeling. This section is intended to be give a broad, non-technical overview, whereas the following sections will delve in more details, using the language and methods of statistical physics. Readers more interested in conceptual results and less in technical details are invited to read this section and skip the subsequent ones, and jump directly to the final discussion Sec.~\ref{sec:conclusion}. 

In the SK-game, the payoff of each agent is a random function of the decisions of all other agents. Hence, learning the optimal strategy (in terms of the probability for agent $i$ to play $+1$ or $-1$) is bound to be extremely difficult. Defining the average reward per agent as 
\begin{equation}
\mathcal{R}_N := \frac{1}{N} \sum_i S_i R_i = \frac{1}{N} \sum_{i,j=1}^N S_i J_{ij} S_j,
\end{equation}
and noting that $\sum_{i,j=1}^N S_i J_{ij} S_j= \left(1-\frac{\varepsilon}{2}\right)\sum_{i,j=1}^N S_i J^S_{ij} S_j$, the largest possible average reward $\mathcal{R}_\infty$ for $N \to \infty$
can be exactly computed using the celebrated Parisi solution of the classical SK model and reads \cite{mezard1987spin}:
\begin{equation}
    \lim_{N \to \infty} \mathcal{R}_N := \mathcal{R}_\infty = 0.7631...\times (2 - \varepsilon). 
    \label{eq:E_GS}
\end{equation}
However, in practice this optimal value can only be reached using algorithms that need a time exponential in $N$.\footnote{In fact, it was recently shown that one can devise a smart algorithm with run-time growing like $K(\epsilon) N^2$ to find configurations $\{S_i\}$ that reach a value of at least $1-\epsilon$ times the optimum \eqref{eq:E_GS}, with $\epsilon > 0$ independent of $N$ \cite{montanari2021optimization}.} Hence, it is expected that simple learning algorithms will inevitably fail to find the true optimal solution. Nevertheless, we also know from the spin-glass folklore \cite{mezard1987spin} (see below for more precise statements) that many configurations of $\{S_i\}$'s correspond to quasi-optima, or, in the language of H. Simon, satisficing solutions \cite{simon1955behavioral} (see also \cite{garnier2021new}). It is in a sense the proliferation of such sub-optimal solutions that prevent simple algorithms to find the {\it optimum optimorum}. Furthermore, if learning indeed converges (which is not the case when $\varepsilon$ is too large, i.e. when interactions are not reciprocal enough), the obtained fixed point heavily depends on the initial condition and/or the specific interaction matrix $\mathbf{J}$. 

\subsection{Phase Diagram in the Noiseless Limit}

Let us first consider the case where agents always choose the action that would have had the best average reward in the past $Q_i(t)$ (assuming that other agents still played what they played). This corresponds to the noiseless learning limit $\beta \to \infty$. In this case, the iteration map Eq. \eqref{eq:NMFE_general} becomes 
\begin{equation}
    S_i(t+1) = \text{sign} \left( \sum_{t' \leq t} (1-\alpha)^{t-t'} \sum_{j=1}^N J_{ij} S_j(t') \right),
    \label{eq:NMFE_T=0}
\end{equation}
and the model is fully specified by two parameters: $\alpha$ (controlling the memory time scale of the agents) and $\varepsilon$ (controlling the reciprocity of interactions). For $N$ not too large, the evolution of Eq. \eqref{eq:NMFE_T=0} leads to either fixed points, or oscillations, or else chaos. The schematic phase diagram in the plane $(\alpha,\varepsilon)$ is shown in Fig. \ref{fig:qualitative_phase_diag}. 

\begin{figure}
    \centering
    \includegraphics[width=\linewidth]{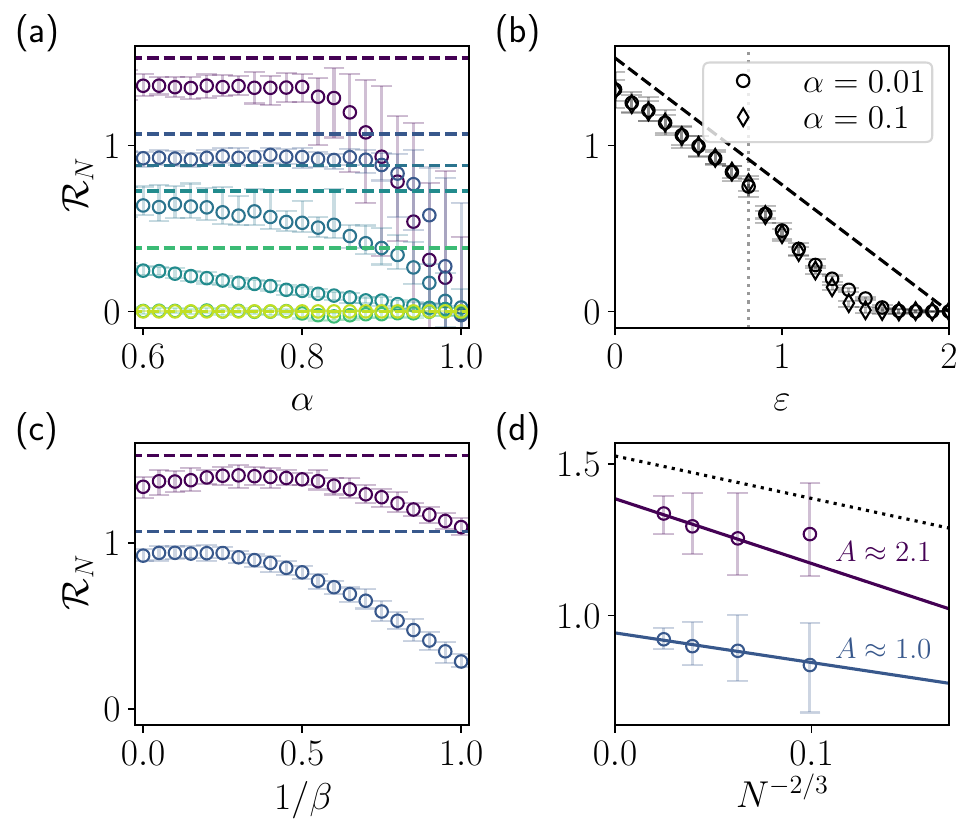}
    \caption{Evolution of the average reward with (a) the memory loss rate $\alpha$, for $\varepsilon = \{0,0.6,0.85,1.05,1.5,2\}$ from dark purple to light green, $\beta \to \infty$; (b) the asymmetry $\varepsilon$ for $\alpha = 0.01$ and $\alpha = 0.1$, $\beta \to \infty$, the vertical dotted line indicates the value $\varepsilon_c$ above which the system becomes chaotic \cite{hwang2019number}; (c) the noise level $1/\beta$ for $\alpha = 0.01$, $\varepsilon = \{0,0.6\}$ (dark purple and blue respectively), note the non-monotonic behaviour; (d) system size $N$ for $\alpha = 0.01$, $\beta \to \infty$, $\varepsilon = \{0,0.6\}$ (dark purple and blue respectively), continuous lines showing fits  $\mathcal{R}_N = \mathcal{R}_\infty - {A}{N^{-2/3}}$, excluding $N=32$. The dotted line indicates the best fit for the SK ground state, for which $A \approx 1.5$ \cite{boettcher2010simulations}. In (a), (b) and (c) $N=256$ and the dashed line represents the Parisi solution for $\mathcal{R}_{\infty}$.}
    \label{fig:avg_reward}
\end{figure}

One clearly sees a region for $(\alpha,\varepsilon)$ small where learning reaches a fixed point, in which the average reward is close, but significantly below the theoretical optimum $\mathcal{R}_\infty$ given by Eq. \eqref{eq:E_GS}, see Fig. \ref{fig:avg_reward}.\footnote{Note that there are finite $N$ corrections that must be taken into account for such a comparison, which read \cite{boettcher2010simulations}
\begin{equation}
   \mathcal{R}_N \approx \mathcal{R}_\infty - \frac{A}{N^{2/3}}, \quad A \approx 0.75 \times (2 - \varepsilon).
\end{equation} } Note that learning definitely helps: for $\varepsilon=0$, most fixed points are characterized by a typical reward $\mathcal{R} \approx 1.01$ \cite{bray1980metastable}, significantly worse than the value $\approx 1.40$ reached by our learning agents, extrapolated to $N\to \infty$. 

As $\varepsilon$ and $\alpha$ are varied one observes the following features:
\begin{itemize}
\item When $\varepsilon$ is not too large (interactions sufficiently reciprocal) and $\alpha$ increases (shorter and shorter memory) learning progressively ceases to converge and oscillations start appearing: impatient learning generates cycles. This leads to a sharp decrease of the average reward (see Fig. \ref{fig:avg_reward}(a)), as agents over-react to new information and are no longer able to coordinate on a mutually beneficial equilibrium. A similar effect was observed in dynamical models of supply chains, where over-reaction leads to oscillating prices and production -- the so-called bullwhip effect \cite{Sterman1989} (see also \cite{sharma2021good, dessertaine2022out}).

\item Conversely, when $\alpha$ is small (long memory) and $\varepsilon$ increases, the probability to reach a fixed point progressively decreases, and when a fixed point is reached, the average reward is reduced -- see Fig. \ref{fig:avg_reward}(b). Beyond some threshold value, the dynamics becomes completely chaotic, leading to further loss of reward. Note that ``chaos'' here means that  although agents have well defined intentions $m_i(t) \neq 0$ at any given instant of time, these intentions evolve chaotically forever.\footnote{Note that the role of $\varepsilon$ is somewhat similar to that of parameter $\Gamma$ in \cite{galla2013complex}: increasing competition leads to chaos.} 

\item Surprisingly, oscillations {\it reappear} when $\varepsilon$ becomes larger than unity, i.e. when interactions are mostly anti-symmetric, ``predator-prey'' like.  Perhaps reminiscent of the famous Lotka-Volterra model, agents' decisions and payoffs become periodic, with a period that scales anomalously as $\alpha^{-1/2}$, i.e. much shorter than the natural memory time scale $\alpha^{-1}$. Although not the Nash equilibrium $m_i \equiv 0$, this oscillating state allows the average reward to be positive, even when at each instant of time, some agents have negative rewards.

\item Only in the extreme competition limit $\varepsilon=2$ (corresponding to a zero sum game, Eq.~\eqref{eq:E_GS}) and for small $\alpha$, are agents able to learn that the unique Nash equilibrium is to play random strategies $m_i \equiv 0$ (see Fig. \ref{fig:qualitative_phase_diag}, bottom right region).\footnote{Note that this is strictly speaking true only when $\beta$ is large, but not infinite. For infinite $\beta$ and finite $N$ one always gets oscillations of length $L=4$.}

\item Finally, in the extreme (and unrealistic) case $\alpha=1$, where agents choose their strategy only based on the last reward, the system evolves, as $\varepsilon$ increases from zero, from high frequency oscillations with period $L=2$ to ``weak chaos'', to ``strong chaos'' when $\varepsilon \approx 1$, and finally back to oscillations of period $L=4$ when $\varepsilon \to 2$. 
\end{itemize}

In order to characterize more precisely such temporal behaviours, it is useful to introduce 
the two-point auto-correlation function of the \textit{expected} decisions or {\it intentions}:
\begin{equation}
    C(t,t+\tau) = \frac{1}{N} \sum_i \overline{\langle m_i(t) m_i(t+\tau) \rangle}, \label{eq:Cdef}
\end{equation}
where the angular brackets now refer to an average over initial conditions.\footnote{Of course, in the $\beta \to \infty$ limit discussed in this subsection, one can replace $m_i(t)$ by the actual decision $S_i(t)$ in the definition of $C(t,t+\tau)$. More generally, the spin-spin correlation function is given by $(1-q(t)) \delta(\tau) + C(t,t+\tau)$ with $q(t)= \overline{\langle m_i^2(t) \rangle}$.} In cases where the dynamics are assumed to be time-translation invariant, we will write $C(\tau)$ which corresponds to the above quantity averaged over time after the system has reached a steady-state.

\begin{figure}
    \centering
    \includegraphics[width=\linewidth]{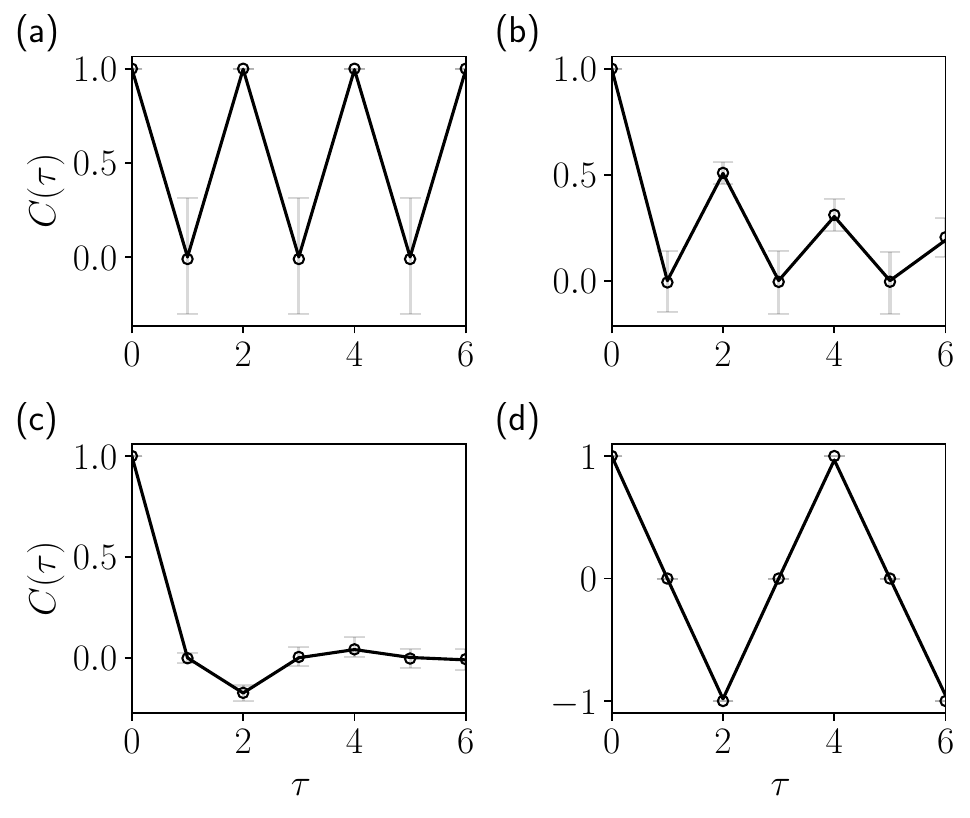}
    \caption{Evolution of the steady-state two-point correlation function for $\alpha = 1$, $\beta \to \infty$, markers indicating simulations of the game at $N = 256$ averaged over 128 samples of disorder and initial conditions, error-bars showing 95\% confidence intervals, and continuous lines representing the solution of the DMFT equation (see Eq. \eqref{eq:DMFT_discrete} below). (a) $\varepsilon = 0.1$  ($\varepsilon < \varepsilon_c \approx 0.8$), cycles of length $L = 2$. (b) $\varepsilon = 0.85$ ($\varepsilon_c < \varepsilon < 1$), ``weakly'' chaotic behavior. (c) $\varepsilon = 1.05$ ($1 < \varepsilon < 2 - \varepsilon_c$), ``strongly'' chaotic behavior. (d) $\varepsilon = 1.5$ ($\varepsilon > 2-\varepsilon_c$), cycles of length $L = 4$.}
    \label{fig:correls_0T}
\end{figure}

The autocorrelation function corresponding to the different cases described above are plotted in Fig. 
\ref{fig:correls_0T}. Note that the signature of oscillations of period $L$ is that $C(nL) \equiv 1$ for all integer $n$. However, note that when $\varepsilon < \varepsilon_c$, not all spins flip at each time step. The fact that $C(2n+1)=0$ means that half of the spins in fact remain fixed in time, while the other half oscillate in sync between $+S_i$ and $-S_i$.\footnote{Note the rather large error bar on $C(2n+1)$, meaning that there are actually substantial fluctuations of the number of idle spins around the value $N/2$ when $N$ is finite.} In the chaotic phases, $C(\tau)$ tends to zero for large $\tau$, with either underdamped or overdamped oscillations. Hence in these cases, the configuration $\{S_i\}$ evolves indefinitely with time, and hardly ever revisit the same states. 

\subsection{Finite $N$, Large $t$ vs. Finite $t$, Large $N$}

The previous results hold in the long time limit for finite $N$, i.e. when we take formally the limit $t \to \infty$ before $N \to \infty$. There is however another regime for which the $N \to \infty$ limit is taken first, which might be more appropriate in the context of firm networks (say) when the number of firms is very large. 

Interestingly, specific analytical tools are available to treat this regime -- the so-called Dynamic Mean-Field Theory (DMFT) that we will use below. 

In this infinite $N$ regime, new types of behaviour appear that one can call quasi-fixed points or quasi-cycles. In the case of quasi-fixed points, learning does not strictly speaking converge, but actions $S_i(t)$ fluctuate around time-independent averages. In other words, the two-point correlation $C(\tau)$ is not equal to one for all $\tau$ (which would be the case for a fixed point) but reaches a positive plateau value for large $\tau$: $C(\tau \to \infty) = C_\infty > 0$. Only when $\varepsilon=0$ does one find $C_\infty=1$. The same holds for quasi-cycles of length $L$ if one considers the correlation function computed for $\tau = nL$, with $n$ an integer: $C(nL \to \infty) = C_\infty > 0$, with $C_\infty=1$ when $\varepsilon=2$.

The schematic phase phase diagram drawn in Fig. \ref{fig:qualitative_phase_diag} then continues to hold in the large $N$ finite $t$ limit, provided one interprets ``fixed points/cycles'' as ``quasi-fixed points/cycles'' in the sense defined above. More details on the subtle role of $\alpha$, $N$ and $t$ will be provided in the next, technical sections. 

\subsection{Noisy Learning \& ``Aging''}
\label{sec:summary_aging}

In the presence of noise, the ``convictions'' $|m_i|$ of agents naturally decrease, and in fact become zero (i.e. decisions are totally random) beyond a critical noise level that depends on the asymmetry parameter $\varepsilon$: more asymmetry leads to more fragile convictions (see Fig. \ref{fig:q_phase_diag} below for a more precise description). 

When the noise is weak but non-zero, strict fixed points do not exist anymore, but are replaced (for $\varepsilon$ small enough) by quasi-fixed points -- in the sense that the intentions $m_i$ fluctuate around some plateau value for very long times, before evolving to another configuration completely uncorrelated with the previous one. This process goes on forever, albeit at a rate that slows down with time: plateaus become more and more permanent. This is called ``aging'' in the context of glassy systems. 

In a socio-economic context, it means that a form of quasi-equilibrium is temporarily reached by the learning process, but such a quasi-equilibrium will be completely disrupted after some time, even in the absence of any exogenous shocks. This is very similar to the quasi-nonergodic scenario recently proposed in Ref. \cite{bouchaud2023self}, although in our case the evolution time is not constant but increases with the ``age'' of the system, i.e. the amount of time the game has been running. 

Perhaps counter-intuitively, however, the role of noise is on average beneficial when $1/\beta$ is not too large. Indeed, as shown in Fig. \ref{fig:avg_reward}, the average reward first increases as a small amount of noise is introduced, before reaching a maximum beyond which ``irrationality'' becomes detrimental. The intuition is that without noise, the system gets trapped by fixed points with large basins of attraction, but lower average rewards. A small amount of noise allows agents to reassess their intentions and collectively reach more favorable quasi-fixed points, much as with simulated annealing or stochastic gradient descent.   

When learning leads to a chaotic evolution, i.e. when $J_{ij}$ and $J_{ji}$ are close to uncorrelated ($\varepsilon \sim 1$), noise in the learning process does not radically change the evolution of the system: deterministic chaos just becomes noisy chaos. However, there is still a distinction between a low-noise phase where at each instant of time, agents have non-zero expected decisions $m_i$ (that will evolve over time) from a high-noise phase where agents always make random choices between $\pm 1$ with probability $1/2$ (see Fig. \ref{fig:qualitative_phase_diag}).

Finally, in the case where learning leads to cycles, any amount of noise irremediably disrupts the synchronisation process, and cycles are replaced by pseudo-cycles, with either underdamped or overdamped characteristics. In the limit $\varepsilon \to 2$, fluctuations drive the system to a paramagnetic state where $q = C(0) = 0$ (see Fig.~\ref{fig:q_phase_diag} below), meaning the agents remain undecided. 

\subsection{Individual Rewards}

\begin{figure}
    \centering
    \includegraphics[width=\linewidth]{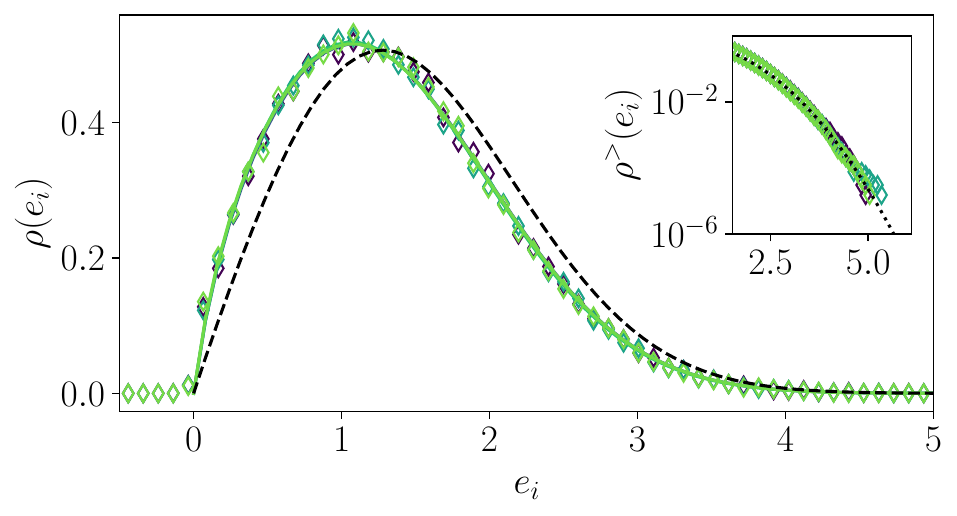}
    \caption{Distribution of individual rewards at $\beta \to \infty$ for $N = 512$ and 128 initial conditions and realizations of the disorder in the fully symmetric case $\varepsilon = 0$ for $\alpha = \{0.5,0.1,0.01\}$ (dark to light coloring). The dashed line is the Sommers-Dupont analytical solution to the SK model \cite{sommers1984distribution}. Inset: associated survival function in a lin-log scale and focusing on the right tail, dotted line corresponding to a Gaussian fit.}
    \label{fig:distrib_ei}
\end{figure}

\begin{figure}
    \centering
    \includegraphics[width=\linewidth]{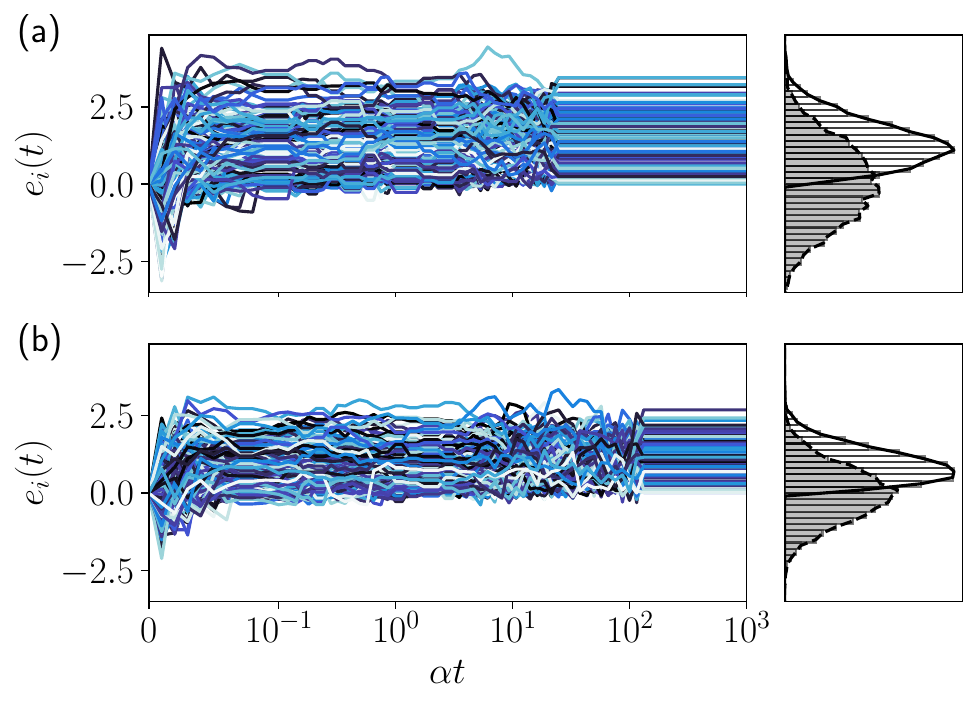}
    \caption{Evolution of individual rewards in time for $N = 256$, $\alpha = 0.01$, $\beta \to \infty$, (a) $\varepsilon = 0$ and (b) $\varepsilon = 0.6$. Right: histogram of the individual rewards after a single timestep (shaded) and at the final time (unshaded).}
    \label{fig:individual_rewards_traj}
\end{figure}

As we have noted above, the average reward is close, but significantly below the theoretical optimum $\mathcal{R}_\infty$ given by Eq.~\eqref{eq:E_GS}. However, some agents are better off than others, in the sense that the individual reward $e_i$ at the fixed point (when fixed points exist), is different from agent to agent. Noting that 
\begin{equation}
    e_i := S_i^\star R_i^\star  = \left\vert \sum_j J_{ij} S_j^\star \right\vert,
\end{equation}
where the second equality holds because at the fixed point one must have $S_i^\star = \sign(R_i^\star)$, it is clear that in the fully reciprocal case $\varepsilon=0$, all rewards $e_i$ are positive. The {\it distribution} $\rho(e)$ of these rewards over agents is expected to be self-averaging for large $N$, i.e. independent of the specific realisation of the $J_{ij}$ and of the initial condition. Such distribution is shown in Fig. \ref{fig:distrib_ei}. One notices that $\rho(e)$ vanishes linearly when $e \to 0$
\[
\rho(e) \underset{e \to 0}{\approx} \kappa e, \qquad \kappa \approx 1.6,
\]
as for the standard SK model, although the value of $\kappa$ is distinctly different from the one obtained for the true optimal states of the SK model, for which $\kappa_{\mathrm{SK}} \approx 0.6$ \cite{sommers1984distribution,pankov2006low}. Such a discrepancy is expected, since the fixed points are obtained as the long time limit of the learning process -- in particular, since $\kappa > \kappa_{\mathrm{SK}}$, the number of poorly rewarded agents is too high compared to what it would be in the truly optimal state. Note that once $\alpha$ is sufficiently small for the system to reach a fixed point, its precise value does not seem to have an impact on the distribution of rewards and $\kappa$.

Another important remark is that the distribution of rewards $\rho(e)$ does not develop a ``gap'' for small $e$, i.e. a region where $\rho(e)$ is exactly zero. In other words, although all agents have positive rewards, some of them are very small. This is associated with the so-called ``marginal stability'' of the equilibrium state \cite{muller2015marginal}, to wit, its fragility with respect to small perturbations, as discussed in more details in the next subsection. 

For very large $e$, the distribution $\rho(e)$ decreases like a Gaussian (Fig.~\ref{fig:distrib_ei} inset), corresponding to a Central Limit Theorem behaviour in that regime, as for the SK model.\footnote{One would expect a different behaviour in the possibly relevant case of a fat-tailed distribution of the $J_{ij}$, see \cite{cizeau1993mean, janzen2010thermodynamics, neri2010phase}. We leave this question for further investigations.} Fig.~\ref{fig:individual_rewards_traj} shows how the rewards of individual agents evolve from an initially random configuration, before settling to constant (but heterogeneous) values at the fixed point.

\begin{figure*}
    \centering
    \includegraphics[width=\textwidth]{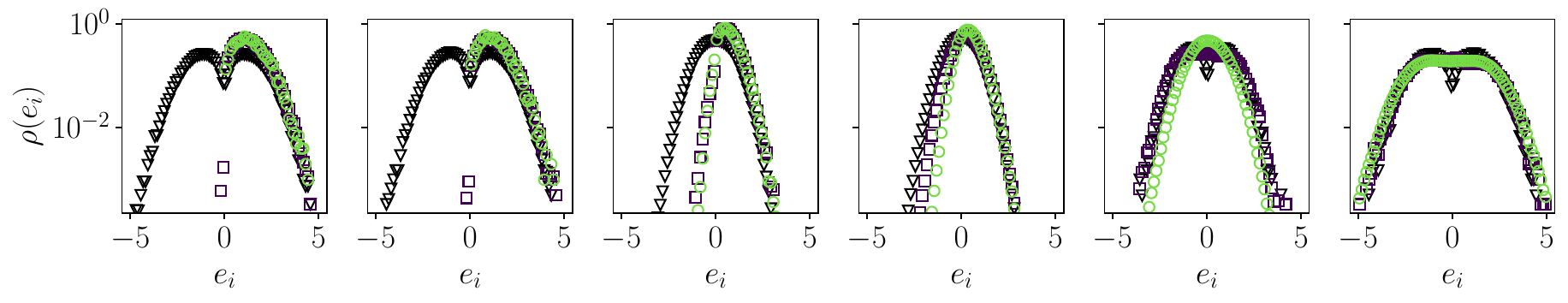}
    \caption{Distribution of individual rewards for $N = 256$ and $\alpha = \{1,0.5,0.01\}$ represented by black triangles, purple squares and green circles respectively, $\beta \to \infty$, measured over 32 initial conditions and realizations of the disorder. From left to right: $\varepsilon = \{0,0.1,0.85,1.05,1.5,2\}$.}
    \label{fig:individual_rewards_epsilon}
\end{figure*}

As competitive effects get stronger (i.e. as $\varepsilon$ increases) and the system  ceases to reach a fixed point, the distribution $\rho(e)$ develops a tail for negative values of $e$, meaning that some agents make negative gains, see Fig. \ref{fig:individual_rewards_epsilon}. In the extreme ``predator-prey" limit $\varepsilon=2$, the distribution $\rho(e)$ becomes perfectly symmetric around $e=0$, as expected -- see  Fig. \ref{fig:individual_rewards_epsilon}, rightmost plot. However, note that there is no persistence in time of the winners: the individual reward autocorrelation function eventually decays to zero, possibly with oscillations in the competitive region.

\subsection{Unpredictable Equilibria}

\begin{figure}
    \centering
    \includegraphics[width=\linewidth]{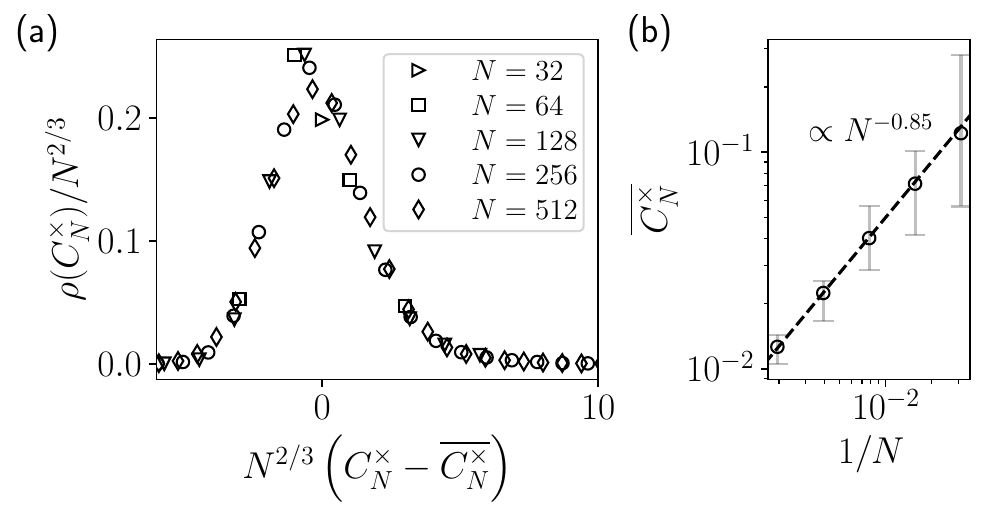}
    \caption{Overlap between solutions for different initial conditions and identical draws of interactions, for $\alpha = 0.01$, $\beta \to \infty$, $\varepsilon = 0$. (a) Distribution of overlaps shifted by the mean and rescaled with the system size $N$ to the power $2/3$. (b) Average overlap as a function of system size in log-log coordinates, with the best regression line $N^{-0.85}$. Error-bars show the 95\% confidence interval over 16 different draws of the disorder.}
    \label{fig:reward_overlaps}
\end{figure}

Now, the interesting point about our model is that the final rewards are highly dependent on the initial conditions and/or the realisation of the $J_{ij}$'s. In other words, successful agents in one realisation of the game become the losers for another realisation obtained with different initial conditions. A way to quantify this is to measure the cross-sectional correlation of final rewards for two different realisations, i.e.
\begin{equation}
    C_N^\times := \frac1N \sum_{i} (e_i^{a} - \langle e^a \rangle) (e_i^{b} - \langle e^b \rangle), \quad a \neq b
\end{equation}
where $a,b$ corresponds to two different initial conditions and $\langle e \rangle$ corresponds to the cross-sectional average reward. As shown in Fig.~\ref{fig:reward_overlaps}, $C_N^\times$ goes to zero at large $N$, indicating that the final outcome of the game, in terms of the winners and the losers, cannot be predicted. The dependence on $N$ appears to be non-trivial, with different exponents governing the decay of the mean overlap $\overline{C_N^\times}$ (decaying as $N^{-0.85}$) and its standard deviation (decaying as $N^{-2/3}$).

A similar effect would be observed if instead  of changing the initial condition one would randomly change the interaction matrix $\mathbf{J}$ by a tiny amount $\epsilon$. The statement here is that for any small $\epsilon$, $C_N^\times$ goes to zero for sufficiently large $N$. This is called ``disorder chaos'' in the context of spin-glasses \cite{krzkakala2005disorder}; by analogy with known results for the SK model, we conjecture that $C_N^\times$ is a decreasing function of $N \epsilon^\zeta$, where $\zeta$ is believed to be equal to 3 in the SK case \cite{aspelmeier2008bond}. This means that when $N \gg \epsilon^{-\zeta}$ the rewards between two systems with nearly the same interaction structure, starting with the same initial conditions, will be close to independent. 

Such a sensitive dependence of the whole equilibrium state of the system (in our case the full knowledge of the intentions $m_i$ of all agents) prevents any kind of ``common knowledge'' assumption about what other agents will decide to do in a specific environment. No reasonable learning process can lead to a predictable outcome; even the presence of a benevolent social planner assigning their optimal strategy to all agents would not be able to do so without a perfect knowledge of all interactions between agents and without exponentially powerful (in $N$) computing abilities. Such a ``radically complex'' situation leads to ``radical uncertainty'' in the sense that the behaviour of agents, even rational, cannot be predicted. Learning agents can only achieve satisficing solutions, that are furthermore hypersensitive to details. As we have seen in Sec.~\ref{sec:summary_aging}, any amount of noise in the learning process will make the whole system ``jump'' from one satisficing solution to another in the course of time.  

\subsection{Increasing Cooperativity}
\label{sec:cooperativity}
A way to help agents coordinate is to use rewards given by Eq. \eqref{eq:reward2} with $J_0 > 0$, representing a non-zero average cooperative contribution to rewards. This term obviously helps agents finding mutually beneficial strategies. (Note that with our normalisation, the $J_0$ term is in fact $N^{-1/2}$ times smaller than the random interaction terms $J_{ij}$.)

\begin{figure}
    \centering
    \includegraphics[width=\linewidth]{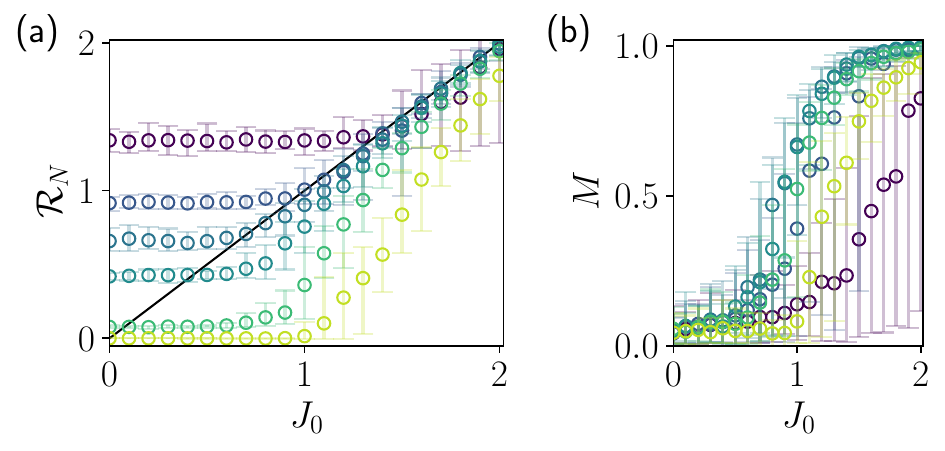}
    \caption{(a) Evolution of the average reward with the incentive to cooperate $J_0$ for $N = 256$, $\alpha = 0.01$, $\beta \to \infty$ and $\varepsilon = \{0,0.6,0.85,1.05,1.5,2\}$ with colors ranging from purple to light green with increasing $\varepsilon$. (b) Average intention in the long time limit $M = \lim_{t\to \infty} M(t)$ for the same parameters.}
    \label{fig:avg_reward_J0}
\end{figure}

The impact of such a term is well understood in the context of the SK model for $\varepsilon=0$ \cite{toulouse1980mean}, which nicely translates into the current dynamical framework. For $\beta = \infty$, one finds that whenever $J_0 \leq 1$, the average intention $M(t)$ remains zero for large $N$ and one expects that the learning process is not affected by such a ``nudge''. When $J_0 > 1$, on the other hand, the situation changes as all agents start coordinate on one of the two possible choices. As shown in Fig.~\ref{fig:avg_reward_J0}, the average intention becomes non-zero, although a finite fraction of agents still play opposite to the majority because of their own idiosyncratic rewards. 

For $J_0 \gg 1$, radical complexity disappears and learning quickly converges to the obvious optimal strategy where all agents make the same move $S_i = +1$ or $S_i = -1$, $\forall i$. In this case, $\mathcal{R}_N = J_0$ as $M(t)$ eventually reaches unity, see Fig.~\ref{fig:avg_reward_J0}. For $\varepsilon > 0$, the same occurs albeit for different values of $J_0$.

In the case $J_0 < 0$ with $|J_0| \gg 1$, the only solution of Eq.~\eqref{eq:NMFE_offline} (valid for $\alpha \to 0$) is $m_i= 0$ for all $i$, i.e. agents cannot coordinate and play random strategies. 

\subsection{Habit Formation}
\label{sec:self-reinforcement}

Up to this point, all results have assumed that there is no self-interaction, $J_{ii} = 0$. Nonetheless, it is interesting to consider the possibility of having an $O(1)$ positive diagonal term in the interaction matrix. In the socio-economic context, such a contribution is relevant as it represents self-reinforcement of past choices, which is also called ``habit formation'' where agents stick to past choices, a popular idea in behavioural science, see e.g. \cite{pemantle2007survey, moran2020force} and refs. therein. 

The introduction of a diagonal contribution has important consequences for the problem. Assuming the self-interaction is identical for all agents, $J_{ii} = J_{\rm d} > 0$, it will rather intuitively favor the emergence of fixed points since agents are tempted to stick to past choices. It is for instance known that in the case of fully random interactions $\varepsilon = 1$, fixed points will start to appear when $J_{\rm d}$ is sufficiently large \cite{ZecchinaLesHouches}. Interestingly, these fixed points can be very difficult to reach dynamically with standard Hopfield dynamics ($\alpha = 1$).

Adding such diagonal contribution to our learning dynamics, we have observed that the fraction of trajectories converging to seemingly dynamically inaccessible configurations significantly increases, specially when $\alpha \ll 1$. While further work is required to precisely assess the effectiveness of learning with self-reinforcement (particularly as finite size effects appear to play a significant role), such a result is consistent with the overall influence of learning reported here.

\subsection{Core Messages}

In line with the conclusions of Galla \& Farmer \cite{galla2013complex}, our multi-agent binary decision model provides an explicit counter-example to the idea  that learning could save the rational expectation framework (cf. Sec.~\ref{sec:introduction} and Ref. \cite{evans2013learning}). 

Learning, in general, does not converge to any fixed point, even when the environment (in our case the interaction matrix $\mathbf{J}$) is completely static: non-stationarity is self-induced by the complexity of the game that agents are trying to learn, as also recently argued in \cite{colon2022radical}.

When learning does indeed converge (which requires {\it a minima} a high level of reciprocity between agents) the collective state reached by the system is far from the optimal state, which only a benevolent, omniscient social planner with formidable powers can achieve. In other words, even more sophisticated learning rules would not really improve the outcome: the SK-game is {\it unlearnable} and -- as argued by H. Simon \cite{simon1955behavioral} -- agents must resort to suboptimal, {\it satisficing} solutions.  

Furthermore, any small random perturbation (noise in the learning process, or slow evolution in the environment) eventually destabilises any fixed point reached by the learning process, and completely reshuffles the collective state of the system: in the long run, agents initially favoring the $+1$ decision end up favoring $-1$, and better-off agents end up being the underdogs, and vice-versa (much as in the simpler model of Ref. \cite{bouchaud2023self}).

Finally, even in the most favourable case of a fully reciprocal game with slow learning, the average reward is in fact {\it improved} when some level of noise (or irrationality) is introduced in the learning rule, before degrading again for large noise.  

\section{Fixed Point Analysis and Complexity}

\label{sec:statics}
We have seen that our model displays a wide variety of complex collective dynamics. Only in some cases does learning converge to non-trivial fixed points where strategies are probabilistic but with time independent probability $p_\pm$ to play $\pm 1$, such that  $p_\pm = (1 \pm m_i^\star)/2$ for agent $i$. Such a steady-state would be analogous to an \textit{economic equilibrium} (although it is essential to dissociate this notion from that of a thermodynamic equilibrium, which may only exist in the case of fully reciprocal interactions, $\varepsilon=0$).

We will mostly focus, in the following, on the long memory case $\alpha \ll 1$ which is most relevant for thinking about learning in a (semi-)realistic context. In this case, one can show that the exponential moving average on the realized values $S_i(t)$ converges to one on the expected values $m_i(t)$. Indeed, as detailed in Appendix~\ref{appendix:NMFE},
\begin{equation}
    \E{\left(\alpha \sum_{t'\leq t} (1-\alpha)^{t-t'} (m_i(t') - S_i(t')) \right)^2 } \leq \frac{\alpha}{2 - \alpha} \xrightarrow[\alpha \to 0]{} 0.
    \label{eq:meanfield}
\end{equation}
This means that up to fluctuations of order $\sqrt{\alpha}$, we can describe the dynamics of the system through a deterministic iteration on $m_i(t)$, in fact corresponding to {\it offline learning}. (We will see below that the neglected fluctuations are of order $\sqrt{\alpha}/\beta$.)

Further making the ansatz that the mean-field dynamics will eventually reach a fixed point $m_i(t) = m_i^\star \, \forall i$ given sufficient time, Eq.~\eqref{eq:NMFE_general} then yields
\begin{equation}
    m_i^\star = \tanh \left( \beta \sum_j J_{ij} m_j^\star \right).
    \label{eq:NMFE_FP}
\end{equation}
This equation is known in the spin-glass literature as the Naive Mean-Field Equations (NMFE) when $\varepsilon = 0$, and defines a so-called static Quantal Response Equilibrium, similar to its fully mean-field equivalent ($J_{ij} = J/N$) studied in \cite{leonidov2021ising}.

To the reader familiar with the physics of disordered systems, this equation is immediately reminiscent of the celebrated Thouless-Anderson-Palmer (TAP) equation \cite{thouless1977solution} describing the mean magnetization in the Sherrington-Kirkpatrick (SK) spin-glass \cite{sherrington1975solvable}. Physically, the NMFE equation is satisfied when minimizing the free energy of a system of $N$ sites comprising $M \to \infty$ binary spins, with sites interacting through an SK-like Hamiltonian \cite{bray1986naive}. Despite being seemingly simpler than its previously mentioned TAP counterpart, which includes an additional Onsager ``reaction term'', the NMFE shares many of its properties. Relevant to our problem, both the NMFE and the TAP equations have a paramagnetic phase ($m_i^\star = 0 \, \forall i$) for $\beta < \beta_c$, while above this critical value there is a spin-glass phase where $q^\star = N^{-1}\sum_i (m_i^\star)^2 > 0$ and solutions are exponentially abundant in $N$ \cite{bray1986naive,takayama1990spin,nishimura1990metastable}. The NMFE has a critical temperature $1/\beta_c = 2$ as opposed to $1/\beta_c = 1$ in the TAP case, while the two equations become strictly equivalent in the $\beta \to \infty$ limit.

Using known properties from the spin-glass literature, we can therefore already establish that \textit{if} the system reaches a fixed point when interactions are fully reciprocal ($\varepsilon=0$) and memory is long ranged, it will be either a trivial fixed point where agents continue making random decisions for ever ($m_i^\star = 0 \, \forall i$), or, when learning is not too noisy ($\beta > \beta_c$) the number of fixed points is $\sim \exp[\Sigma(\beta) N]$, where $\Sigma(\beta)$ is called the ``complexity''. In this second case, the fixed point actually reached by learning depends sensitively on the initial conditions and the interaction matrix $\mathbf{J}$. 

How is this standard picture altered when interactions are no longer reciprocal? In such cases, the system cannot be described using the equilibrium statistical mechanics machinery.

\subsection{Critical Noise Level}

In order to extend the notion of critical noise $\beta_c$ to $\varepsilon > 0$, one can naively look at the linear stability of the paramagnetic solution $m_i^\star = 0 \, \forall i$ to Eq.~\eqref{eq:NMFE_FP}. Just as in the TAP case \cite{mezard1987spin}, expanding the hyperbolic tangent to the second order and projecting the vector of $m_i$ on an eigenvector of $\mathbf{J}$, the stability condition can be expressed with the largest eigenvalue of the interaction matrix. Adapting known results from random matrix theory to our specific problem formulation, the spectrum of $\mathbf{J}$ can be expressed as an interpolation between a Wigner semi-circle on the real axis ($\varepsilon = 0$), the Ginibre ensemble ($\varepsilon = 1$) and a Wigner semi-circle on the imaginary axis ($\varepsilon = 2$) \cite{sommers1988spectrum}. The resulting critical ``temperature'' is then given by
\begin{equation}
    T_c(\varepsilon) = \frac{1}{\beta_c(\varepsilon)} = \frac{1}{2} \frac{(2 - \varepsilon )^2}{\sqrt{\upsilon(\varepsilon)}},
    \label{eq:critical_temp}
\end{equation}
recovering the known result $1/\beta_c = 2$ for the case $\varepsilon = 0$. (We recall that we have set the interaction variance $\sigma^2$ to unity throughout the paper. If needed $\sigma$ can be reinstalled by the rescaling $\beta \to \beta \sigma$.) 

\subsection{The Elusive Complexity}

To determine if there are still an exponential number of fixed point to reach below the candidate critical noise level, i.e. if there is a spin-glass phase, for $\beta > \beta_c(\varepsilon)$ when $\varepsilon > 0$, we should study the complexity, defined for a single realization of the disorder as
\begin{equation}
    \Sigma(\beta,\varepsilon) = \lim_{N\to \infty} \frac{1}{N} \log \mathcal{N}_J(N,\beta,\varepsilon), \label{eq:complexity_def}
\end{equation}
where $\mathcal{N}_J$ is the number of fixed points in the system for a given interaction matrix. There are then two ways to compute an average of this quantity over the disorder: the ``quenched'' complexity, where the mean of the logarithm of the number of solutions is considered, and its ``annealed'' counterpart, where the logarithm is taken on the mean number of solutions. The former is usually considered to be more representative, as unlikely samples leading to an abnormally large number of solutions can be observed to dominate the latter (see e.g. \cite{garnier2021new,ros2022generalized} for recent examples), but requires a more involved calculation with the use of the so-called ``replica trick'' \cite{mezard1987spin}. In the TAP case, quenched and annealed complexities coincide for solutions above a certain free-energy threshold \cite{bray1980metastable} (where most solutions lie but importantly not the ground state).  

As a matter of fact, even in the annealed case, the computation of the TAP complexity has proved to be a formidable task, and has sparked a large amount of controversy, as the original solution computed by Bray and Moore (BM) \cite{bray1980metastable} has been put into question before being (partially) salvaged by the metastability of TAP states in the thermodynamic limit \cite{aspelmeier2004complexity}. For a relatively up to date summary of the situation, we refer the reader to G. Parisi's contribution in \cite{bovier2006mathematical}.

While the BM approach can be adapted to the NMFE \cite{takayama1990spin,waugh1990fixed}, several aspects of the calculation remain unclear, particularly as the absence of a sub-dominant reaction term means that the argument of the metastability of states in the $N \to \infty$ limit is no longer valid \textit{a priori}, although numerical results support the marginally stable nature of NMFE fixed points in the thermodynamic limit \cite{takayama1990spin}. We leave its extension to $\varepsilon > 0$ to a later dedicated work.

Nevertheless, the previously introduced critical $\beta_c$ and the existing computation of the number of fixed points as a function of $\varepsilon$ in the $\beta \to \infty$ limit \cite{gutfreund1988nature,hwang2019number} can be used to conjecture the boundaries of the region in $(\beta,\varepsilon$) space where the complexity $\Sigma$ is non-vanishing. Indeed, in the zero temperature case, it has been shown \cite{hwang2019number} that the annealed complexity can be expressed as a function of the asymmetry parameter $\eta$ defined in Eq. \eqref{eq:eta} as
\begin{equation}
    \Sigma(\eta) = -\frac{1}{2} \eta x^2 + \log 2 + \log \Phi(\eta x),
    \label{eq:complexity_0T}
\end{equation}
with $\Phi$ the Gaussian cumulative density  and $x$ is the solution to
\begin{equation}
    x \Phi(\eta x) = \Phi'(\eta x), \quad \Phi(x) := \frac{1}{2} \operatorname{erfc}\left(-\frac{x}{\sqrt{2}} \right)
    \label{eq:saddle_0T}
\end{equation}
The main insight provided by this result is that the complexity vanishes at $\eta = 0$, corresponding to $\varepsilon = 1$, where the paramagnetic fixed point is supposed to be unstable as $\beta_c(\varepsilon = 1) = \sqrt{2}$. As the complexity is a decreasing function of temperature, this therefore means that $\varepsilon = 1$ is an upper limit for the existence of fixed points when $\beta$ is finite. This conjecture is also consistent with the breakdown of fixed point solutions to the dynamical mean field theory below the critical noise level that will be discussed in Sec.~\ref{subsec:DMFT_FP}, as well as the saddle point equations obtained when adapting the BM calculation to $\varepsilon > 0$. 

\begin{figure}
    \centering
    \includegraphics[width=\linewidth]{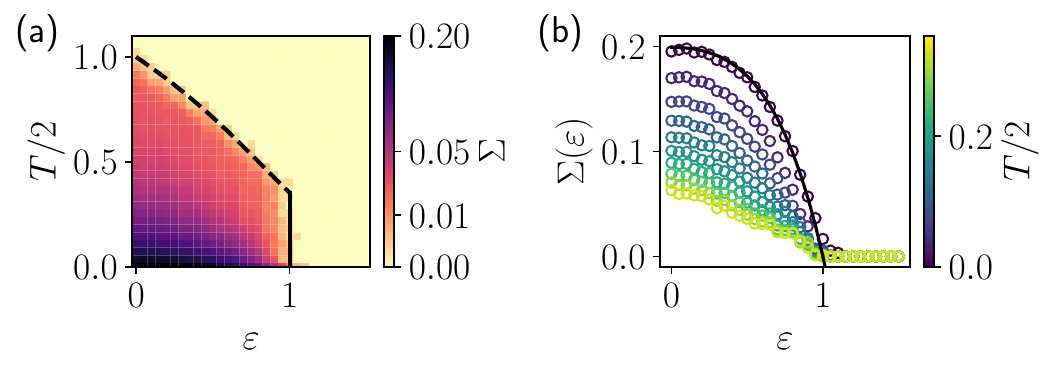}
    \caption{(a) Annealed complexity of the Naive Mean-Field Equation in $(T,\varepsilon)$ space, where we recall $T = \beta^{-1}$, measured numerically for $N = 40$. The dashed line represents the critical temperature $T_c(\varepsilon)$ for which the paramagnetic fixed point ceases to be stable, while the continuous line indicates $\varepsilon = 1$ inferred from the $\beta \to \infty$ result. (b) Annealed complexity as a function of $\varepsilon$ for varying temperatures $T < 1/\sqrt{2}$, i.e. in the bottom region of (a) where the complexity vanishes at $\varepsilon = 1$. The continuous line represents the $\beta \to \infty$ analytical solution, recovering the result of Tanaka and Edwards \cite{tanaka1980analytic} $\Sigma \approx 0.1992$ for $\varepsilon = 0$. For $\varepsilon > 1$ and $T=0$, the only possible fixed point is $m_i^\star = 0$, $\forall i$.}
    \label{fig:NMFE_FP}
\end{figure}

Combining these two somewhat heuristic delimitation for the existence of a large number of non-trivial fixed points, we obtain the critical lines shown in Fig.~\ref{fig:NMFE_FP}(a). Overlaying these borders with the annealed complexity measured numerically, we find a very good agreement. In particular, the vanishing of the complexity at $\varepsilon = 1$ in  appears to be consistent for $T < T_c(\varepsilon = 1) = 1/\sqrt{2}$, as shown in Fig.~\ref{fig:NMFE_FP}(b). The agreement with the $\beta \to \infty$ analytical result, represented by the continuous line, also appears to validate our counting method at low temperatures. Note that one can in fact show that for $\varepsilon > 1$ and $N = \infty$, the only fixed point (or Nash equilibrium) is the ``rock-paper-scissors'' equilibrium $m_i^\star = 0$, $\forall i$. 

\section{Counting Limit Cycles}
\label{sec:Cycles}

In the previous section, we have established the region of parameter space where exponentially numerous fixed points exist, which might possibly be reached by learning in the slow limit $\alpha \ll 1$.  However, limit cycles of various lengths turn out to also be exponentially numerous when $\varepsilon < 1$, so we need to discuss them as well before understanding the long term fate of the learning process within our stylized complex world. 

\subsection{Cycles without Memory ($\alpha=1$)} \label{sec:cycles1}

In the memory-less limit, the dynamics becomes that of the extensively studied Hopfield model \cite{hopfield1982neural,gutfreund1988nature,bastolla1998relaxation} where the binary variable represents the activation of a neuron evolving as
\begin{equation}
    S_i(t+1) = \sign \bigg(\sum_j J_{ij} S_j(t) \bigg),
    \label{eq:TAP_0T}
\end{equation}
with \textit{parallel} updates. Counting limit cycles of length $L$ is even more difficult than counting fixed points (which formally correspond to $L=1$). Some progress have been reported by Hwang \textit{et al.} \cite{hwang2019number} in the memory-less case $\alpha=1$. The notion of fixed point complexity $\Sigma$ (defined in Eq. \eqref{eq:complexity_def}) can be extended to limit cycle complexity $\Sigma_L$ for limit cycles of length $L$, with $\Sigma_{L=1} \equiv \Sigma$.  The results of Hwang et al. \cite{hwang2019number} can be summarized as follows:
\begin{itemize}
    \item When $\varepsilon < 1$, limit cycles with $L=2$ have the largest complexity, which is exactly twice of the fixed point complexity: $\Sigma_2 = 2 \Sigma_1$ (as was in fact previously shown by Gutfreund \textit{et al.} \cite{gutfreund1988nature}).
    \item The complexities $\Sigma_L(\varepsilon)$ all go to zero when $\varepsilon=1$.
    \item When $1 < \varepsilon \leq 2$, limit cycles with $L=4$ dominate, with $\Sigma_4(\varepsilon) \geq \Sigma_2(2-\varepsilon)$.
    \item Close to $\varepsilon=1$, the cut-off length $L_c$, beyond which limit cycles become exponentially rare, grow exponentially with $N$: $L_c \sim e^{aN}$, where $a$ weakly depends on $\varepsilon$.  
\end{itemize}
From this analysis, one may surmise that:
\begin{enumerate}[label=\alph*. ]
    \item When a limit cycle is reached by the dynamics, it is overwhelmingly likely to be of length $L=2$ for $\varepsilon < 1$ and of length $L=4$ for $1 < \varepsilon \leq 2$.
    \item Even if exponentially less numerous, exponentially long cycles will dominate when $e^{aN} > e^{N \Sigma_2}$, which occurs when $\varepsilon_c < \varepsilon < 2 - \varepsilon_c$, with $\varepsilon_c \approx 0.8$. 
\end{enumerate}
These predictions are well obeyed by our numerical data, see Figs. \ref{fig:correls_0T}, \ref{fig:chaos_0T}. Note however the strong finite $N$ effects that show up in the latter figure, which we  discuss in the next sections. 

\subsection{Cycles with Memory ($\alpha < 1$)} \label{sec:cycles2}

When $\alpha < 1$ and $\beta=\infty$, the update of $S_i(t)$ is given by Eq. \eqref{eq:NMFE_T=0}, have the same fixed points independently of $\alpha$, but of course different limit cycles, which may in fact cease to exist when $\alpha$ is small. In this section, we attempt to enumerate the number of cycles of length $L$ in the spirit of the calculation of Hwang \textit{et al.} \cite{hwang2019number} for $\alpha < 1$. As detailed in Appendix~\ref{appendix:LC_complexity}, we write the number of these cycles as a sum over all possible trajectories of a product of $\delta$ functions ensuring the $\alpha < 1$ dynamics of $Q_i$ are satisfied between two consecutive time-steps, while a product of Heaviside step functions enforces $S_i(t) = \sign(Q_i(t))$. Introducing the integral representation of the $\delta$ function, averaging over the disorder and taking appropriate changes of variable to decouple the $N$ dimensions, the (annealed) complexity of cycles of length $L$ writes
\begin{widetext}
\begin{equation}
    \Sigma_L(\alpha,\eta) = \saddle_{\hat{R},\hat{K},\hat{V}}\bigg\{ \sum_{s < t} i\hat{R}(t,s) i\hat{K}(t,s) - \frac{\eta}{2} \sum_{t,s} \hat{V}(t,s) \hat{V}(s,t) + \log  \mathcal{I}_L\bigg\},
\label{eq:enum_cycles}
\end{equation}
\end{widetext}
where $\hat{R}(t,s)$ and $\hat{K}(t,s)$ are symmetric matrices while $\hat{V}(t,s)$ is not \textit{a priori}, and $\mathcal{I}_L$ is explicitly given in the Appendi, and where we can notably identify $-i\hat{R}(t,s) = C(\lvert t-s \rvert)$. As a sanity check, one can verify that the $L=1$ case, corresponding to the fixed point complexity, is indeed independent of $\alpha$ and is given by the same expression as Eq.~\eqref{eq:complexity_0T}, see Appendix \ref{appendix:LC_complexity_FP}. In a similar vein, one can recover $\Sigma_2(\alpha = 1) = 2\Sigma_1(\eta)$.

\begin{figure}
    \centering
    \includegraphics[width=\linewidth]{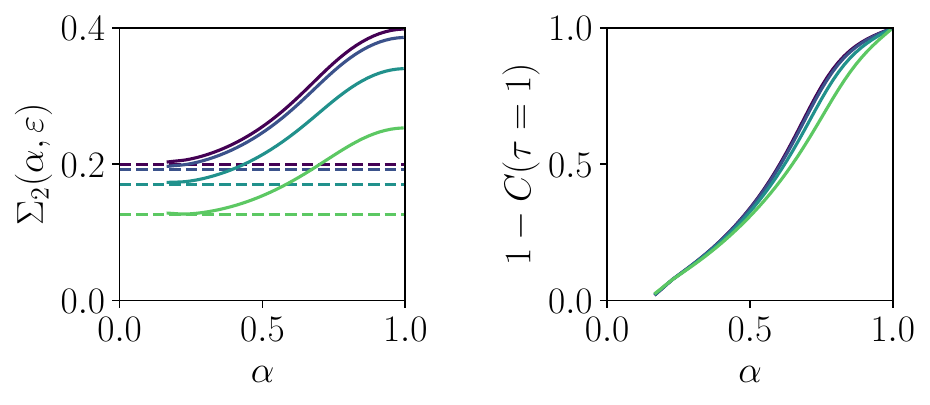}
    \caption{(a) $L = 2$ cycle complexity from the numerical resolution of the saddle point equations for $\varepsilon = \{0,0.2,0.4,0.6\}$ from dark purple to light green, dashed lines indicating the fixed point complexity associated to each parameter. (b) Order parameter at the $L=2$ saddle, showing the nontrivial coalescence of the cycle and fixed point saddle point as $\alpha$ is decreased. The numerical resolution appears to breakdown when we get close to $\alpha = 0$.}
    \label{fig:L2_cycles}
\end{figure}

Numerically solving the saddle point equations for decreasing values of $\alpha$, it appears that $\Sigma_2(\alpha,\eta) \to \Sigma(\eta)$ when $\alpha \to 0^+$, see Fig.~\ref{fig:L2_cycles}. While numerical difficulties prevent us from exploring very small values of $\alpha$, it seems clear that the saddle point corresponding to the $L = 2$ cycles eventually coalesces with the fixed point saddle (which is known to be a sub-dominant saddle point when $\alpha = 1$, see \cite{hwang2019number}  and the discussion above). In any case, and perhaps surprisingly, there does not appear to be a critical value of $\alpha$ below which fixed points become more abundant than cycles. We therefore expect a progressive crossover and not a sharp transition.

\section{Dynamical Mean-Field Theory}
\label{sec:DMFT}

We have thus seen that both fixed points and limit cycles are exponentially numerous. However, the question remains as to what happens \textit{dynamically}, as the existence of a large number of fixed points or limit cycles by no means guarantees that these will be reached at long times. 

\begin{figure}
    \centering
    \includegraphics[width=\linewidth]{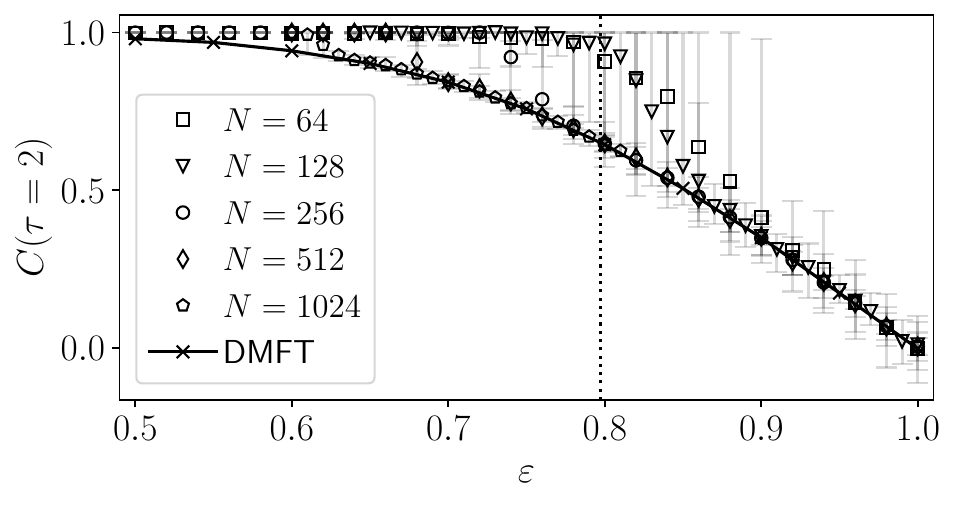}
    \caption{Steady-state two-point correlation between configurations shifted by $\tau = 2$ time-steps in the $\alpha = 1$, $\beta \to \infty$ limit from finite $N$ numerical simulations averaged over 128 samples of disorder and initial conditions, error bars showing 95\% confidence intervals. The $N=\{64,128\}$ simulations are run for $t = 10^8$ time-steps to illustrate taking the $t\to\infty$ limit before $N\to \infty$, whereas $N=\{256,512,1024\}$ have been simulated for $t = 5\times 10^6$ time-steps to recover the $N\to \infty$ before $t\to \infty$ regime. The continuous line represents the $N\to \infty$ DMFT solution integrated numerically, while the vertical dotted line corresponds to the critical value $\varepsilon_c$ found by Hwang \textit{et al.} \cite{hwang2019number}.} 
    \label{fig:chaos_0T}
\end{figure}

In fact, the number of agents $N$ is expected to play a major role in determining the long term fate of the system. In particular, there are strong indications that the time $\tau_r$ needed to reach a fixed point or a limit cycle grows itself exponentially with $N$, at least when $\alpha = 1$ \cite{bastolla1998relaxation}. More precisely, 
\begin{equation}
    \tau_r \sim N^s e^{N B(\varepsilon)},
\end{equation}
where $s$ is an exponent (possibly dependent on $\varepsilon$) and $B(\varepsilon)$ an effective barrier such that $B(\varepsilon=0)=0$. Hence one expects that as $N$ grows, fixed points/limit cycles will in fact never be reached, even if they are numerous. This is in fact what happens numerically, see Fig. \ref{fig:chaos_0T}.

What is then to be expected in the limit $N \to \infty$? In order to study the complicated learning dynamics that takes place, we will resort to Dynamical Mean-Field Theory (DMFT). In a nutshell, DMFT allows deterministic or stochastic dynamics in discrete or continuous time of a large number $N$ of interacting degrees of freedom to be rewritten as a one-dimensional stochastic process with self-consistent conditions. While difficult to solve both analytically and numerically due to their self-consistent nature, DMFT equations have proved very effective at describing a very wide range of complex systems -- see \cite{cugliandolo2023recent} for a recent review. Note however that such an approach is only valid when $N \to \infty$; as will be clear later, strong finite size effects can appear and change the conclusions obtained using DMFT. 

In our case, we write the DMFT for the evolution of the incentives $Q_i(t)$, which directly yield $m_i(t) = \tanh(\beta Q_i(t))$. In order to do so, we rewrite our \textit{online} learning process, which depends on the realized $S_i(t)$, as an expression solely in terms of $m_i(t)$ with additional fluctuations, 
\begin{equation}
    \sum_j J_{ij} S_j(t) = \sum_j J_{ij} m_j(t) + \eta_i(t), \quad \eta_i(t) = \sum_j J_{ij}\xi_i(t),
\end{equation}
with $\xi_i(t) = S_i(t) - m_i(t)$ and hence $\langle \xi_i(t) \rangle = 0$ and $\langle \xi_i(t) \xi_i(s) \rangle = (1 - (m_i(t))^2)\delta_{t,s}$. Now, assuming the central limit theorem holds, the random variables $\eta_i$ become Gaussian for large $N$ with 
\begin{equation}
    \langle \eta_i(t) \rangle = 0, \quad \langle \eta_i(t) \eta_j(s) \rangle = \upsilon(\varepsilon)(1-q(t)) \delta_{t,s} \delta_{i,j},
    \label{eq:DMFT_CLT}
\end{equation}
where $q(t)=C(t,t)$ as defined in Eq. \eqref{eq:Cdef}.
As required, in the noiseless limit $\beta \to \infty$ limit, one has $q(t) = 1 \, \forall t$ and the random variables $\eta_i$ are identically zero.  

Starting from the $N$ equations
\begin{equation}
    Q_i(t+1) = (1-\alpha) Q_i(t) + \alpha \sum_j J_{ij} m_j(t) + \alpha \eta_i(t) + \alpha h_i(t),
\end{equation}
where $h_i(t)$ is an arbitrary external field that will eventually be set to 0, the DMFT can be derived using path integral techniques or the cavity method, the latter being detailed in Appendix~\ref{appendix:DMFT}. Remaining in discrete time to explore the entire range of values of $\alpha$, one finds, in the $N \to \infty$ limit, 
\begin{equation}
\begin{aligned}
    Q(t+1) &= (1-\alpha) Q(t) \\
    &+ \alpha^2 (1-\varepsilon) \sum_{s < t} G(t,s) m(s) + \alpha \phi(t) + \alpha h(t),
\end{aligned}
    \label{eq:DMFT_discrete}
\end{equation}
with $\langle \phi(t) \rangle = 0$, and
\begin{equation}
    \langle \phi(t) \phi(s) \rangle = \upsilon( \varepsilon) \left[C(t,s) + (1-q(t)) \delta_{t,s}\right],
\end{equation}
The memory kernel $G$ and correlation function $C$ are then to be determined self-consistently,
\begin{equation}
    G(t,s) = \bigg\langle \frac{\partial m(t)}{\partial h(s)} \bigg\rvert_{h=0}\bigg\rangle, \qquad C(t,s) = \langle m(t) m(s) \rangle,
\end{equation}
where the averages $\langle \ldots \rangle$ are over the realisations of the random variable $\phi$.
These discrete time dynamics can first be integrated numerically with an iterative scheme to update both the memory kernel and correlation function until convergence, see \cite{roy2019numerical} or \cite{mignacco2020dynamical}. As detailed in the original work of Eissfeller and Opper \cite{eissfeller1992new,eissfeller1994mean}, one can also make use of Novikov's theorem to compute the response function with correlations, avoiding the unpleasant task of taking finite differences on noisy trajectories, at the cost of the inversion of the correlation matrix. Note that this inversion will however mean that very long trajectories become difficult to integrate.

While we shall see that this numerical resolution can provide precious intuition to understand the role of finite $N$ in the dynamics, a continuous description will be much more convenient to obtain analytical insights. In the $\alpha \ll 1$, $t \gg 1$ regime, we can rescale the time as $t \to t/\alpha$. Interestingly, doing so requires expanding $Q(t+1)$ to the second order if one is to keep an explicit dependence on~$\alpha$. The resulting continuous dynamics reads
\begin{equation}
\begin{aligned}
    \frac{\alpha}{2} \ddot{Q}(t) + \dot{Q}(t) = &-Q(t) + (1-\varepsilon) \int_0^t \dd s\, G(t,s) m(s)\\ 
    &+ \phi(t) + h(t)
    \label{eq:DMFT_continuous}
\end{aligned}
\end{equation}
with 
\begin{equation}
\label{eq:corr_phi}
    \langle \phi(t) \phi(s) \rangle = \upsilon(\varepsilon)  \big[C(t,s) + \alpha (1-q(t)) \delta(t-s) \big],
\end{equation}
and the memory kernel and correlation function are similarly defined self-consistently
\begin{equation}
    G(t,s) = \bigg\langle \frac{\delta m(t)}{\delta h(s)} \bigg\rvert_{h=0} \bigg\rangle, \qquad C(t,s) = \langle m(t) m(s) \rangle,
\end{equation}
with, we recall, $q(t) = C(t,t) = \langle m^2(t) \rangle$. Very importantly, note the rescaling in time introduces a prefactor $\alpha$ in the variance of the $\phi$, which stems from the noise in the learning process. Since $1 - q(t) \sim \beta^{-2}$ for large $\beta$, this extra term is of order $\alpha/\beta^2$, as anticipated above.

In the next sections, the DMFT equations will be used to shed light on the dynamical behaviour of the model in the limit $N \to \infty$.

\section{Noiseless Learning}
\label{sec:deterministic}

In this section, we use both the DMFT equations and the results on the complexity of fixed points and limit cycles to classify the different dynamical behaviours of the learning process in the noiseless case $\beta \to \infty$, where the realized and expected decisions are equal, $m_i(t) = S_i(t) = \sign(Q_i(t))$.

\subsection{The Memory-less Limit $\alpha=1$}

In this case, corresponding to Eq. \eqref{eq:TAP_0T}, both approaches (DMFT and complexity of limit cycles) seem to agree on the overall picture: as $\varepsilon$ increases from $0$ to $1$, the system transitions from $L=2$ cycles to over-damped oscillations and chaos -- see Fig. \ref{fig:correls_0T}. However, upon scrutiny, one realizes that the perfect agreement between DMFT and direct numerical simulations of the dynamics for finite $N$ is only valid in a region where $\varepsilon$ is small and $N$ large enough -- see Fig. \ref{fig:chaos_0T}. In particular, when $0.5 \lesssim \varepsilon \lesssim 0.8$, $L=2$ cycles do persist when $N$ is smaller than $\sim 200$. 
For larger $N$, the lag 2 autocorrelation function $C(\tau=2)$ is noticeably smaller than unity, and well predicted by DMFT as soon as $N \gtrsim 1000$.

What happens for $\varepsilon \lesssim 0.5$ when $N \to \infty$? The numerical solution of the DMFT equations suggest the following scenario: when $\varepsilon < \varepsilon_{\text{RM}} \approx 0.473$, the long time value $m_\infty$ of the correlation with the initial conditions $C(0,2n)$ at even time steps is strictly positive, hence the subscript ``RM'' for Remnant Magnetisation.\footnote{The convergence to $m_\infty$ is as slow a power law of $\tau$, which makes difficult its numerical determination. Obtaining the precise value of $\varepsilon_{\text{RM}}$ is therefore challenging \cite{eissfeller1994mean}. } It is only exactly equal to one for $\varepsilon = 0$ (permanent oscillations) and decreases to reach zero when $\varepsilon \to \varepsilon_{\text{RM}}$ \cite{eissfeller1994mean}.  Below $\varepsilon_{\text{RM}}$, we can conclude that the system is not ergodic, which will have important implications on the finite temperature dynamics. For asymmetries greater that $\varepsilon_{\text{RM}}$ on the other hand, the decorrelation becomes exponential and we enter a \textit{bona fide} chaotic, ergodic regime.

Although memory-less learning is clearly unrealistic, these results are rather instructive. The system is indeed unable to display aggregate coordination when interactions are mutually independent (chaotic region around $\varepsilon = 1$). Placing ourselves in the socio-economic setting, it seems evident that the number of players vastly exceeds the number of iterations, and the results above indicate that the chaotic region is in fact quite large. Perhaps more importantly, in the absence of learning, agents will see their decisions vary at a high frequency without ever reaching any static steady-state, not only when the game is close to zero sum ($\varepsilon \to 2$), but even when it is fully reciprocal ($\varepsilon \to 0$). Clearly, this last point underlines the importance of introducing memory to recover realistic learning dynamics.

\subsection{Memory Helps Convergence to Fixed Points}

\begin{figure}
    \centering
    \includegraphics[width=\linewidth]{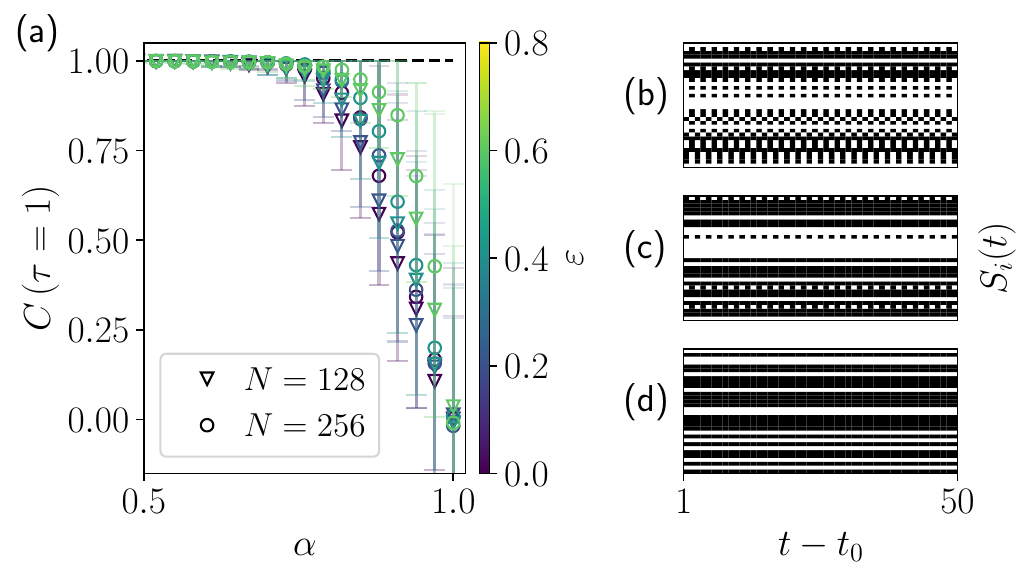}
    \caption{Convergence to fixed points with decreasing $\alpha$ and for different values of $\varepsilon < 0.8$, $\beta \to \infty$ and finite $N$. (a) Steady-state two-point correlation function between successive configurations from finite $N$ numerical simulations averaged over 200 samples of disorder and initial conditions, error-bars showing 95\% confidence intervals. (b), (c) and (d) Sample trajectories of 32 randomly chosen sites among $N = 256$ for $\varepsilon = 0.4$, $t_0 = 10^5/\alpha$, for $\alpha = \{1.0,0.88,0.7\}$ respectively.}
    \label{fig:FP_convergence}
\end{figure}

For $N$ not too large, we observe numerically that the fraction of ``frozen'' agents for which $S_i(t+1) = S_i(t)$ quickly tends to 1 as $\alpha$ decreases from $1$, as shown in Fig.~\ref{fig:FP_convergence}. This is somewhat consistent with intuition, as the learning dynamics average rewards over a period $\tau_\alpha \sim 1/\alpha$, meaning that high frequency cycles observed for $\alpha=1$ are expected to be ``washed out'' when $\alpha$ is sufficiently small. Since fixed points exist in large numbers, it appears natural that they are eventually reached given their abundance at zero temperature. However, as we have shown in the previous section, $L=2$ limit cycles are still much more numerous than fixed points for $\alpha \gtrsim 0.5$. The fact that $C(\tau=1)$ approaches unity as $\alpha$ is reduced much faster than in Fig.~\ref{fig:L2_cycles}(b) suggests that the basin of attraction of fixed points quickly expands, at the expense of $L=2$ limit cycles.\footnote{Note that $C(\tau=2)$ remains equal to $1$ in the whole range of parameters shown in Fig.~\ref{fig:FP_convergence}, i.e. we only observe fixed points or 2-cycles.}

\begin{figure}
    \centering
    \includegraphics[width=\linewidth]{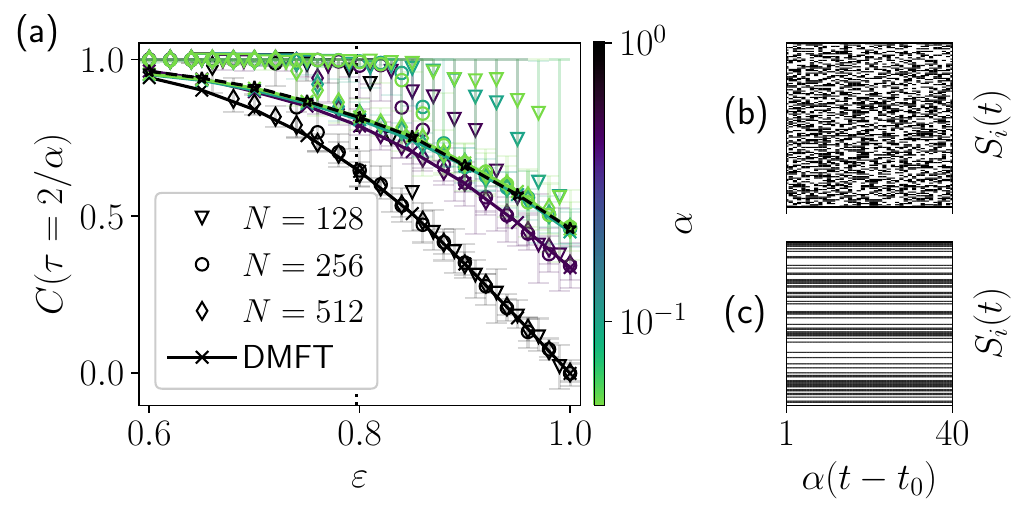}
    \caption{Influence of finite size $N$, non-reciprocity $\varepsilon$ and memory span $\alpha$ on the learning dynamics.  (a) Steady-state two-point correlation function shifted by $\tau = 2/\alpha$ in the $\beta \to \infty$ limit for different memory loss rates $\alpha$, from light green to black (color map on the right axis). Symbols correspond to direct simulations and plain lines to the solutions of the DMFT equations, while the dashed line is the solution to the DMFT equations for $\alpha \to 0$. The $N = 128$ simulations are initialized with $t_0 = 10^8$ time-steps, whereas $N = \{256,512\}$ have been simulated for $t_0 = 10^6$ iterations before taking measurements. Results are averaged over 32 samples, with error-bars showing 95\% confidence intervals. (b) and (c) Sample trajectories for all $N = 128$ sites for $\varepsilon = 1$ and $\alpha = \{1.0,0.1\}$ respectively, clearly reaching chaotic and fixed-point steady states.}
    \label{fig:chaos_0T_alpha}
\end{figure}

Our numerical results therefore indicate that for any finite size system which has enough time to reach a steady state, the effect of $\alpha$ is effectively to help the system find fixed points -- see  
Fig.~\ref{fig:chaos_0T_alpha}. Focusing for example on the points corresponding to $N = 128$ and $\alpha=0.1$, we indeed observe that the 95\% quantile includes $C(2/\alpha)=1$ even for $\varepsilon=1$, i.e. fixed points can be reached even in the chaotic regime with modest simulation times, which would be an overwhelmingly improbable scenario in the memory-less case, as illustrated by Figs.~\ref{fig:chaos_0T_alpha} (b) and (c).

As the number of agents $N$ increases, we enter the DMFT regime shown as plain lines in Fig.~\ref{fig:chaos_0T_alpha}. One finds that decreasing the value of $\alpha$ slows down the decorrelation of the system. However, for small $\alpha$, the evolution becomes a function of $\alpha \tau$ only, as suggested by Eq. \eqref{eq:DMFT_continuous} when $\alpha \to 0$: the dynamical slowdown is dominated by the long memory of learning itself.

Fig.~\ref{fig:chaos_0T_alpha} shows that when $\varepsilon \gtrsim 0.5$, sufficiently large systems (described by DMFT) decorrelate with time for all $\alpha$, and we expect $C(\tau\to \infty) \to 0$: learning leads to chaos in such cases.

When $\varepsilon \lesssim 0.5$, on the other hand, we found that there is ergodicity breaking, in the sense that $C(\tau \to \infty) > 0$, as we found above for cycles when $\alpha = 1$. More precisely, a numerical analysis of the DMFT equations suggests that when $\alpha \to 0$ and $\varepsilon$ small, $1 - C(\tau \to \infty)$ is extremely small but non zero. For example, when $\varepsilon=0.4$ we found $1 - C(\tau \to \infty) \approx 0.005$. This is compatible with the numerical results of \cite{eissfeller1994mean}. 

In other words, there seems to exist a critical value $\varepsilon_{\text{RM}}(\alpha)$ separating the ergodic, chaotic phase for $\varepsilon > \varepsilon_{\text{RM}}(\alpha)$ from the non-ergodic, quasi fixed-point behaviour for $\varepsilon < \varepsilon_{\text{RM}}(\alpha)$. However, our numerical results are not precise enough to ascertain the dependence of $\varepsilon_{\text{RM}}$ on $\alpha$, which seems to hover around the value $0.473$ found for $\alpha =1$. More work on this specific point would be needed to understand such a weak dependence on the memory length. 

\begin{figure}
    \centering
    \includegraphics[width=\linewidth]{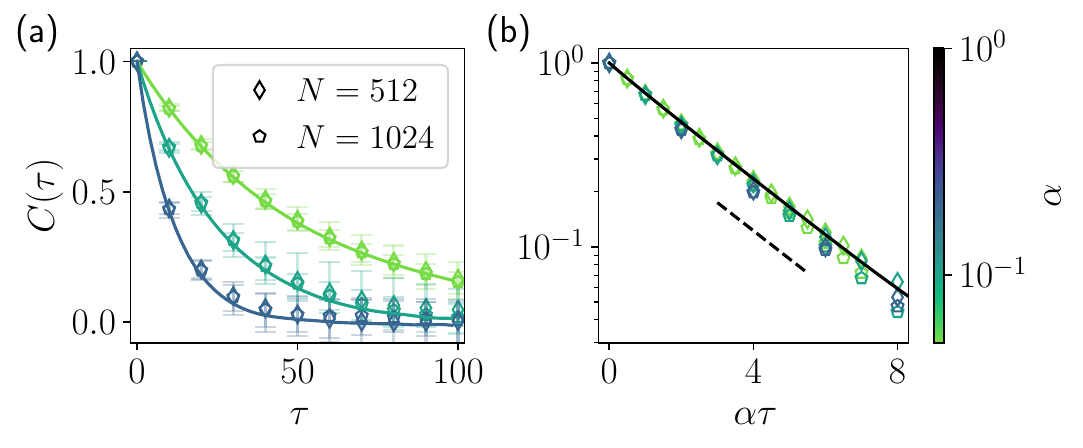}
    \caption{Evolution of the time shifted autocorrelation function in the non-symmetric case $\varepsilon = 1$, $\beta \to \infty$, for different system sizes and memory loss parameters averaged over 20 realizations. Left: lin-lin scale, errobars showing 95\% confidence intervals, continuous lines representing the numerically integrated full DMFT equations (Eq.~\eqref{eq:DMFT_nonsym}). Right: lin-log scale and rescaling of the time shift by $\alpha$ such that points collapse onto a single curve (errobars not shown), black continuous line representing the analytical solution found by solving the Sompolinsky \& Crisanti ODE Eq.~\eqref{eq:DMFT_Delta}, dashed line representing a pure exponential decay with characteristic time $\tau_1$.}
    \label{fig:0T_chaos}
\end{figure}

The precise dynamical behavior of the autocorrelation function $C(\tau)$ can be ascertained in the continuous limit $\alpha \to 0$ when $\varepsilon = 1$. Indeed, the influence of the memory kernel vanishes in this case where interactions are exactly non-symmetric, leaving us with
\begin{equation}
    \dot{Q}(t) = -Q(t) + \phi(t), \qquad (\alpha \to 0)
    \label{eq:DMFT_nonsym}
\end{equation}
where we emphasize that the time variable has been rescaled as $t \to \alpha t$. From there, the classical solution method proposed by Crisanti \& Sompolinsky \cite{sompolinsky1988chaos,crisanti1988dynamics} can be straightforwordly adapted with a small modification due to our parametrization of the interaction matrix that scales the variance of the entries by a factor $1/2$ for $\varepsilon=1$, see Appendix~\ref{appendix:Sompolinsky}. The two-point autocorrelation function is found to be given by
\begin{equation}
    C(\tau) = \frac{2}{\pi} \sin^{-1}\bigg( \frac{\Delta(\tau)}{\Delta(0)} \bigg), 
\end{equation}
where $\Delta(\tau) = \langle Q(t+\tau) Q(t) \rangle$ follows the second-order ordinary differential equation
\begin{equation}
    \ddot{\Delta}(\tau) = \Delta(\tau) - \frac{1}{2} C(\tau),
    \label{eq:DMFT_Delta}
\end{equation}
with $\Delta(0) = 1 - \frac{2}{\pi}$ \cite{crisanti2018path}. Very quickly, this means that the autocorrelation decays exponentially, $C(\tau) \propto \e^{-\frac{\tau}{\tau_1}}$ with
\begin{equation}
    \tau_1 = \sqrt{\frac{\pi - 2}{\pi - 3}} \approx 2.84. 
\end{equation}
Both the full solution, obtained by integrating the ODE numerically, as well as this exponential decay, are shown in Fig.~\ref{fig:0T_chaos}, displaying a very satisfactory match with numerical simulations.

\subsection{Anomalous Stretching of Cycles}

Having a better understanding of how the memory allows the system to find fixed points when they exist, a central question is what will happen if the memory loss rate is reduced in the region of parameter space where there are only limit cycles. As previously stated, averaging over a period $\tau_\alpha \sim 1/\alpha$ clearly suggests that the occurrence of short cycles (starting at $L=4$ for $\alpha = 1$) should gradually vanish.

\begin{figure}
    \centering
    \includegraphics[width=\linewidth]{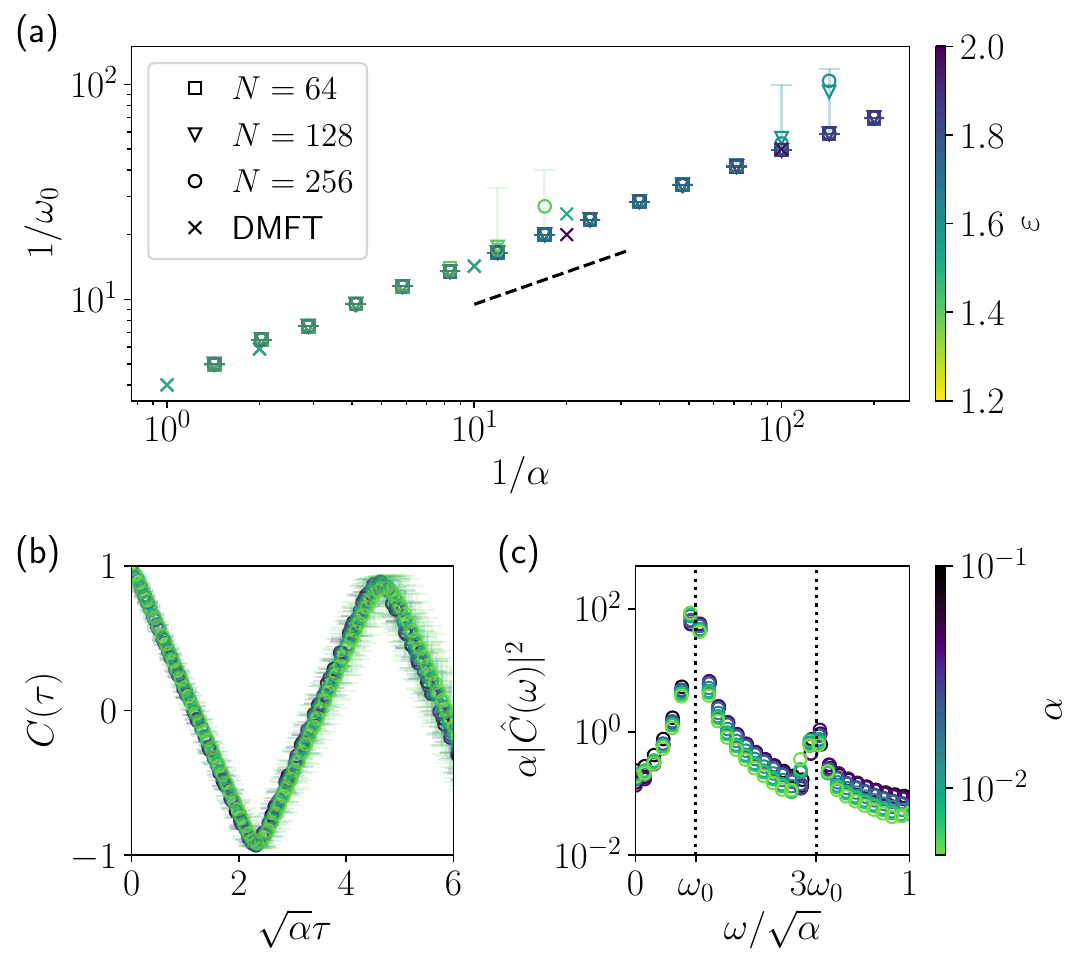}
    \caption{Evolution of the oscillation frequency $\omega_0$ with the memory loss rate $\alpha$ for $\beta \to \infty$, $\varepsilon > 2 - \varepsilon_c$ averaged over 96 samples of disorder and initial conditions, errorbars showing 95\% confidence interval. (a) Log-log plot highlighting the near square root dependency $\omega_0 \sim \sqrt{\alpha}$ the dashed line correponding to an exponent $1/2$. (b) Two-point autocorrelation for $\varepsilon = 1.8$, $N = 256$ as a function of the rescaled time lag for different values of $\alpha$, confirming the suitability of the scaling in particular at short time scales. (c) Power spectrum of the autocorrelation for the same parameters, displaying the maximum at $\omega_0$ and secondary peaks at odd multiples of this fundamental frequency as expected from the triangular aspect of the autocorrelation.}
    \label{fig:cycle_stretching}
\end{figure}

Naively, one might expect a simple rescaling in time $t \to t/\alpha$, yielding cycles -- when they exist -- of period inversely proportional to $\alpha$ itself. Looking at the numerical results from both the finite size game and the DMFT integrated numerically in Fig.~\ref{fig:cycle_stretching} (a), it quickly appears that such a simple rescaling in time does not provide the correct description. Indeed, the period of cycles is observed to be proportional to $1/\sqrt{\alpha}$, i.e. much shorter than $1/\alpha$ -- see Figs.~\ref{fig:cycle_stretching} (b) and (c).  

One important aspect to note is that there is some decorrelation, as the second peak of $C(\tau)$ does not quite reach unity (in Fig.~\ref{fig:cycle_stretching} (b)), meaning that we may see quasi-cycles and not exact limit cycles, complicating the analytical description of the phenomenon. Just as true fixed points in the $N \to \infty$ limit only exist only  $\varepsilon=0$, it appears that only the case $\varepsilon = 2$ does display true limit cycles. 

Another subtle point to consider is similar to the $\varepsilon < 1$ cases discussed, we expect the time taken to reach these cycles will depend on the system size and the relative distance to the chaotic region. This is confirmed by the DMFT solved for fixed trajectory times for $\varepsilon = 1.5$ (light crosses), which progressively departs from the $\omega_0 \sim \sqrt{\alpha}$ regime around $\alpha = 0.1$.

To understand how such non-trivial stretching occurs, we go back to the continuous DMFT equation,
\begin{equation*}
\begin{aligned}
    \frac{\alpha}{2} \ddot{Q}(t) = &-\dot{Q}(t) -Q(t) + (1-\varepsilon) \int_0^t \dd s\, G(t,s) m(s)\\
    &+ \phi(t) + h(t).
\end{aligned}
\end{equation*}
While the presence of the second order derivative $\ddot{Q}(t)$ appears natural to recover limit cycles, it should be noted that this term, being pre-factored by $\alpha$, is superficially subdominant relative to the dissipation represented by $\dot{Q}(t)$. While we have seen that there is some decorrelation, the fact that robust oscillations are present therefore suggests that the complicated self-consistent forcing terms almost exactly compensate dissipation over a period, allowing the system to periodically revisit quasi-identical configurations. In fact, the shape of these oscillations is far from sinusoidal, but rather of see-saw type, see Fig.~\ref{fig:cycle_stretching} (b). This suggests that in the limit $\alpha \to 0$, $\ddot{Q}(t)$ diverges each time $\dot{C}(\tau)$ changes sign, such that $\frac{\alpha}{2} \ddot{Q}(t)$ cannot be neglected and therefore sets the relevant time scale to $\alpha^{-1/2}$. We have however not been able to perform a more precise singular perturbation analysis of this phenomenon. 

Going beyond this rather loose argument, and precisely characterizing such see-saw patterns appears very challenging and is left for future work. A possible approach would be to first take the $\varepsilon = 2$ case where true cycles should exist, and to assume the correlation function is an exact triangular wave of frequency $\omega$ as suggested by Fig.~\ref{fig:cycle_stretching} (c). As a result, $m(s) = \sign(Q(s))$ is an exact square wave, and the convolution with $G$ can be written as a product in Fourier space. Enforcing the dissipation over a period to be zero, one could then perhaps find a closed equation for $Q$ and $\omega$ if appropriate ans\"atze for the response and forcing functions are taken. 

\section{Noisy Learning}
\label{sec:fluctuations}
While we have shown that the $\beta \to \infty$ deterministic limit can be relatively well understood with the analytical tools at our disposal, one of the key features of our model is the uncertainty in the decision occurring for boundedly rational agents. Besides, it is also in this situation that the \textit{online} learning dynamics differ significantly from the more widely studied \textit{offline} learning where the entire model can be understood in terms of deterministic mixed strategies parameterized by the coefficients $m_i(t)$ (compare Eqs. \eqref{eq:NMFE_general} and \eqref{eq:NMFE_offline}).

\begin{figure}
    \centering
    \includegraphics[width=\linewidth]{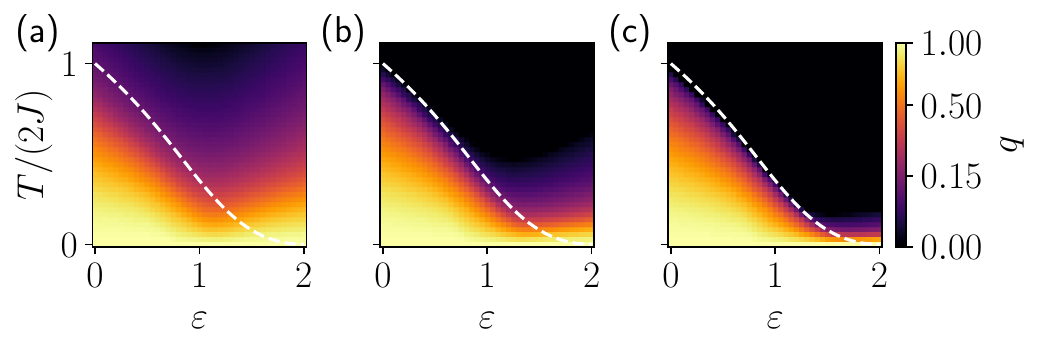}
    \caption{Heat map of $q = C(0)$ in $(T=1/\beta,\varepsilon)$ space from numerical simulations for $N = 256$, $t_0 = 10^6$ averaged over 32 samples of disorder and initial conditions and (a), (b) and (c) corresponding to $\alpha = \{0.5,0.1,0.01\}$ respectively. The white dashed line represents the critical temperature $T_c(\varepsilon)$ where the paramagnetic solution ($q = 0$) becomes linearly unstable (Sec.~\ref{sec:statics}).}
    \label{fig:q_phase_diag}
\end{figure}

When $\alpha$ is close to unity and $\beta$ becomes small, the fluctuations are too large for coordination to occur. Taking for instance $\alpha = 1$, it is indeed clear that the iteration
\begin{equation*}
    m_i(t+1) = \tanh\left( \beta \sum_J J_{ij}S_j(t) \right)
\end{equation*}
will have extremely large fluctuation in the argument on the right hand side. As a result, we expect to lose the sharp transition as a function of $T$ that can be observed for the NMFE (see Fig.~\ref{fig:q_phase_diag}). The order parameter $q$ instead continuously tends to 0 with $T$, regardless of the asymmetry $\varepsilon$. This regime is shown in Fig.~\ref{fig:q_phase_diag} (a), representing the heat map of $q$ for $\alpha = 0.5$. Clearly, the linear stability analysis of the paramagnetic fixed point presented in Sec.~\ref{sec:statics} cannot hold when the thermal fluctuations are not averaged on large periods of time. To find a richer phenomenology, we will therefore focus on the $\alpha \ll 1$ regime where more complex dynamics can be observed.

\subsection{Fixed Points for Intentions}
\label{subsec:DMFT_FP}
In Sec.~\ref{sec:statics}, we studied the fixed points of the NMFE that the game may reach if the fluctuations from imperfect learning can be neglected, i.e. if $\alpha \to 0$. Now, we have seen that the DMFT equations proved effective in the zero ``temperature'' limit, and importantly established that true fixed points or two-cycles are only reached in finite time in the fully reciprocal case $\varepsilon = 0$ when the system size diverges (see Fig.~\ref{fig:chaos_0T}). The same equations can also be used to revisit this finite $\beta$ regime.

Going back to Eq.~\eqref{eq:DMFT_continuous} and neglecting the term in $\sqrt{\alpha}$ from the correlation function as we did in the static setup, fixed point solutions require
\begin{equation}
    Q =  (1-\varepsilon) m \chi + J \sqrt{q\upsilon(\varepsilon)} z,
    \label{eq:DMFT_FP}
\end{equation}
where $z$ is now a \textit{static} white noise of unit variance, $q$ is simply the now constant autocorrelation and $\chi$ is the integrated response function that we assume to be time-translation invariant,
\begin{equation}
    \chi = \int_0^\infty \dd \tau \, G(\tau).
\end{equation}
The averages on the effective process can now be taken on $z$ to self-consistently solve for $q$ and $\chi$ (see e.g. \cite{galla2006random} for a more detailed description). The resulting set of equations are then
\begin{equation}
    q = \langle m^2(z) \rangle_z,
    \label{eq:q}
\end{equation}
to be solved simultaneously with
\begin{equation}
\begin{aligned}
    {\chi} = \bigg\langle  \frac{\beta(1 - m^2(z))}{1 - \beta (1-\varepsilon) {\chi} (1-m^2(z))} \bigg\rangle_z
\end{aligned}
\label{eq:chi}
\end{equation}
where $m(z)$ is the solution to
\begin{equation}
    m(z) = \tanh(\beta (1-\varepsilon) {\chi} m(z) + \beta  \sqrt{q\upsilon(\varepsilon)} z).
    \label{eq:m_z}
\end{equation}
Although our model is entirely built on a dynamical evolution equation, and not on a notion of thermal equilibrium, this set of self-consistent equations coincides with the replica-symmetric solution of the NMFE model found by Bray, Sompolinsky and Yu \cite{bray1986naive} for $\varepsilon = 0$. Since replica symmetry is broken in the whole low temperature phase of the NMFE model, we expect that these static solutions of the DMFT cannot correctly describe the long time limit of the dynamics, as we now show.

\begin{figure}
    \centering
    \includegraphics[width=\linewidth]{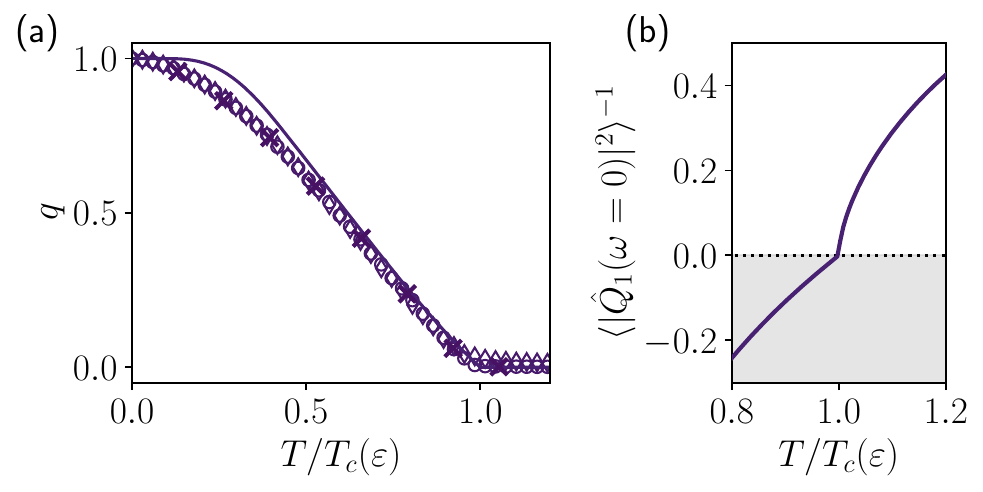}
    \caption{(a) Order parameter $q = C(t,t)$ averaged in time in the (quasi) stationary regime vs. rescaled temperature for $\varepsilon = 0.1$. Circular and diamond markers correspond direct, finite size simulations at $N = 256$ for $\alpha = 0.01$ and $\alpha = 0.1$ respectively, whereas crosses represent the (dynamical) numerical solution to the complete set of $N \to \infty$ DMFT equations for $\alpha = 0.1$. Continuous lines show the solution to the static DMFT fixed point equations (Eq.~\eqref{eq:q}-\eqref{eq:m_z}) for $\varepsilon=0.1$. (b) Spectral density of a small perturbation $Q_1$ to the fixed point solution of the DMFT close to the critical temperature. As the quantity is necessarily positive for a valid solution, the grey region corresponds to instability.}
    \label{fig:FP_q}
\end{figure}

The numerical solutions for the DMFT fixed point equations are shown in Fig.~\ref{fig:FP_q} (a) and compared to numerical results of the game for small $\alpha$ and for $\varepsilon = 0.1$ (similar results, not shown, are obtained for other values of $\varepsilon \lesssim 0.8$).\footnote{We have in fact noted that in this regime, all results seem to collapse when plotted vs. $T/T_c(\varepsilon)$, as in Fig. \ref{fig:FP_q} (a).} We find that the long time behaviour of $q$ for the direct simulation of the SK-game (circles and squares) and for long time dynamical solution of the DMFT equations match very well, but differ from the value of $q$ inferred from the set of self-consistent equations established above. This is expected since with such solution the order parameter $q$ approaches unity exponentially fast as $T \to 0$, whereas the fact that the probability of small local fields (i.e. rewards in the game analogy) vanish linearly (see Fig. \ref{fig:distrib_ei}) suggest that $q = 1 - \kappa T^2$, as for the full RSB solution of \cite{bray1986naive} but with presumably a different value of $\kappa$.  

To ascertain the range over which this non-trivial mean-field solution should be valid, we can study the stability of the DMFT fixed point close to the critical temperature $1/\beta_c$, following the procedure first detailed in \cite{opper1992phase}. Considering a random perturbation to the fixed point $\epsilon \xi(t)$, with $\xi(t)$ a Gaussian white noise and $\epsilon \ll 1$, we study the perturbed solution
\begin{equation}
    Q(t) = Q_0 + \epsilon Q_1(t), 
\end{equation}
with $Q_0$ the fixed point given in Eq.\eqref{eq:DMFT_FP}, where the noise is no longer static but similarly given by $\phi(t) = \sqrt{q\upsilon(\varepsilon)} z + \epsilon \phi_1(t)$. Replacing in the DMFT continuous dynamics for $\alpha \to 0^+$ and collecting terms of order $\epsilon$, we find that the perturbation evolves as
\begin{equation}
\begin{aligned}
    \dot{Q}_1(t) &= - Q_1(t) + \beta (1-\varepsilon) (1-m^2(z)) \int_{0}^t G(t,s) Q_1(s)\\
    &\quad + \phi_1(t) + \xi(t), 
\end{aligned}
\end{equation}
where we have used $\sech^2(\beta Q_0) = 1 - m^2(z)$ from Eq.~\eqref{eq:m_z}, giving in Fourier space
\begin{equation}
    \hat{Q}_1(\omega) = \frac{\hat{\phi}_1(\omega) + \hat{\xi}(\omega)}{i\omega + 1 - \beta  (1-\varepsilon) (1-m^2(z)) \hat{G}(\omega)},
\end{equation}
where we have again assumed that the memory kernel is time-translation invariant.

Now, in the limit $\beta Q_1 \ll 1$, i.e. close to the critical temperature, one can we linearize the hyperbolic tangent  $\tanh(\beta Q_1(t))$ and write a closed equation for the spectral density of $Q_1$ at order $\beta^2 Q_1^2$,
\begin{equation}
\begin{aligned}
    \frac{1}{\langle \lvert \hat{Q}_1(\omega) \rvert^2 \rangle} &= \lvert i\omega + 1 - \beta  (1-\varepsilon)\hat{G}(\omega)\rvert^2 - \beta^2 \upsilon(\varepsilon).
\end{aligned}
\end{equation}
As a result, we have the criterion for the onset of instability for $\omega = 0$:
\begin{equation}
    (1-\varepsilon){\chi} = 1 - \beta \sqrt{\upsilon(\varepsilon)},
    \label{eq:stability_condition}
\end{equation}
where we noticed $\hat{G}(\omega = 0) = {\chi}$, given, close to $\beta_c$, by
\begin{equation}
    {\chi} = \frac{1}{2\beta (1-\varepsilon)} \left( 1 - \sqrt{1 - 4\beta^2 (1-\varepsilon)}\right).
    \label{eq:chi_paramag}
\end{equation}
Taking $\varepsilon = 0$, we recover the criterion found by Bray \textit{et al.} \cite{bray1986naive} for the critical temperature, giving $T_c = 1/\beta_c = 2$ in their case.

For non-zero $\varepsilon$, we can also find the critical temperature by replacing Eq.~\eqref{eq:chi_paramag} in \eqref{eq:stability_condition}, to  recover yet again the critical value given by Eq.~\eqref{eq:critical_temp}. The invalidity of the fixed point solution is illustrated in Fig.~\ref{fig:FP_q} (b), where the spectral density evaluated at $\omega = 0$ can be observed to become negative for $T < T_c(\varepsilon)$. While the linearization used to obtain the relation is expected to become invalid, we understand this negativity as a strong sign that the solution is unstable and thus likely invalid throughout the low ``temperature'' phase, consistently with the discrepancy observed between the static prediction and the direct simulations and the dynamic solution of the DMFT show in Fig.~\ref{fig:FP_q} (a). In this finite $\beta$ regime, we therefore effectively only have what we previously referred to as quasi-fixed points at best (including when $\varepsilon = 0$), even when neglecting the vanishing fluctuations caused by the online learning dynamics as $\alpha \to 0$. 

\begin{figure*}
    \centering
    \includegraphics[width=\textwidth]{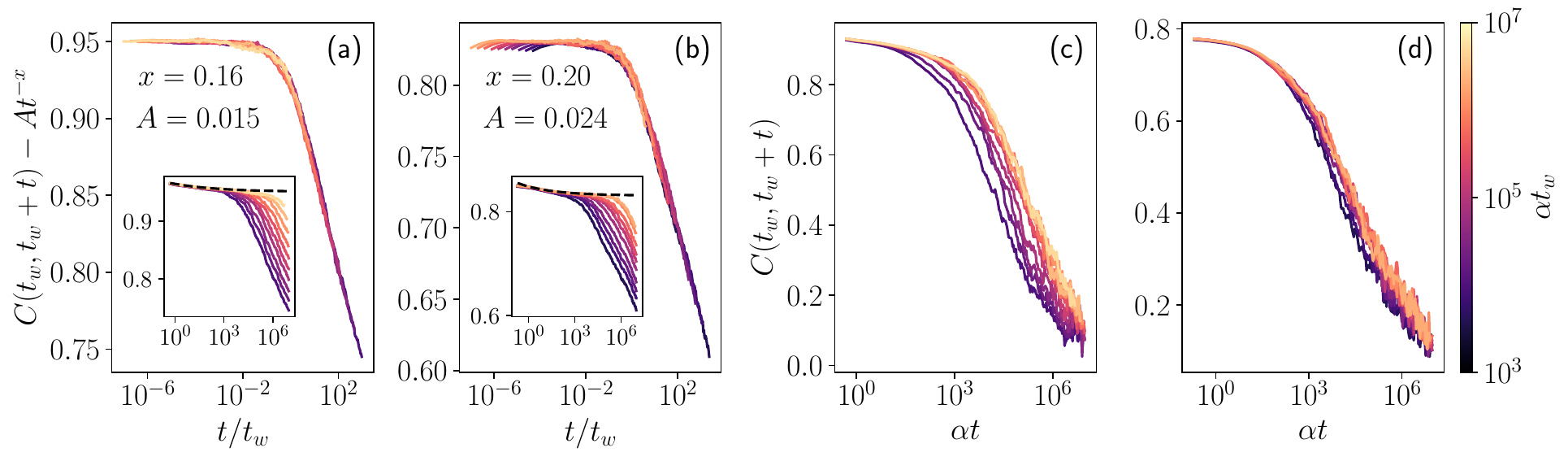}
    \caption{Aging behavior of the system for $N = 256$, averages performed over 576 realizations. Color indicates the value of $\alpha t_w$, see scale on the far right. (a) and (b): Aging two-point correlation functions with the initial power law decay removed to isolate the aging component plotted as a function of $t/t_w$, inset showing the entire correlation function as a function of $\alpha t$, dashed line representing the power law fit $At^{-x}$ of the first relaxation. (a) $\alpha = 0.5$, $\beta = 4$, $\varepsilon = 0$, (b) $\alpha = 0.2$, $\beta = 2$, $\varepsilon = 0.25$. (c) and (d): Partially aging two-point correlation functions plotted as a function of $\alpha t$. (c) $\alpha = 0.5$, $\beta = 4$, $\varepsilon = 0.5$, (d) $\alpha = 0.2$, $\beta = 2$, $\varepsilon = 0.5$. }
    \label{fig:aging_plots}
\end{figure*}

\subsection{Aging}

The inadequacy of the static solution of the DMFT equations to describe the long term dynamics of the system is a well known symptom associated with the ``aging'' phenomenon \cite{cugliandolo1994out}, i.e. the fact that equilibrium is never reached and all correlation functions depend on the ``age'' of the system \cite{bouchaud1998out}: 
\begin{equation} \label{eq:Caging}
    C(t_w,t_w + t) = C_\mathrm{relax}(t) + C_\mathrm{aging}(t,t_w),
\end{equation}
where $t_w$ is the waiting time, or age of the system, $t_w=0$ corresponding to a random initial condition. 

Aging typically arises in complex systems in the low temperature (low noise) limit. Pictorially, the energy landscape of such systems (like spin-glasses) are highly non convex and ``rugged'', with a very large number of local minima or quasi-stable saddle points in which the dynamics gets stuck for extended periods of time \cite{bouchaud1998out,vincent2007slow}. The consequence is then that the time required to exit a local minimum is a function of the age of the system, i.e. the time taken for the system to reach this configuration. Intuitively, the deeper in the energy landscape the solution is, the longer it will take for a sufficiently large random fluctuation to occur and allow the system to resume its exploration of the landscape. Such \textit{aging} phenomena are known to occur in a wide range of complex systems with reciprocal interactions, such as glassy systems,  populations dynamics \cite{altieri2021properties} or neural networks that are described by very similar mean-field dynamics \cite{marti2018correlations}. Aging dynamics was also recently found in a ``habit formation'' model, see \cite{moran2020force}.

Not surprisingly in view of its similarity with usual spin-glasses, the SK-game displays aging for reciprocal interactions ($\varepsilon=0$) and sufficiently low temperatures $\beta > \beta_c$, see Fig.~\ref{fig:aging_plots} (a), for which Eq. \eqref{eq:Caging} accurately describes the data with the initial relaxation component $C_\mathrm{relax}(t)$ well fitted by a power law $t^{-x}$ and 
\begin{equation}
    C_\mathrm{aging}(t,t_w) = \mathcal{C}\left(\frac{t}{t_w}\right),
\end{equation}
corresponding to the aging behaviour found in a wide range of glassy models \cite{bouchaud1992weak,rieger1993nonequilibrium,cugliandolo1994evidence,yoshino1996off,bouchaud1998out,berthier2002geometrical,de2023aging}. Interestingly, this is not the behavior observed in the ``physical'' SK model \cite{marinari1998numerical,baldassarri1998numerical}, which is typically found to display ``sub-aging'', although the precise scaling of the aging correlation function remains unclear. It is however important to emphasize that the learning dynamics of the SK-game is markedly different from the physical dynamics of the original SK model, so there is \textit{a priori} no reason to expect their complicated out-of-equilibrium relaxation to be directly comparable.

What happens in our model when $\varepsilon > 0$? It is known from previous work (in somewhat different contexts), that aging is interrupted at long times in the presence of non reciprocal interactions, but survives for finite times provided the asymmetry is not to large \cite{cugliandolo1997glassy,marinari1998off,berthier2013non}. If the asymmetry strength is further increased, we expect the amount of mixing in the system to eventually be large enough for the dynamics to no longer get stuck \cite{iori1997stability}. From our numerical simulations, shown in Fig.~\ref{fig:aging_plots} (b)-(d), it appears that aging (as described by Eq. \eqref{eq:Caging}) still holds when $\varepsilon < \varepsilon^\star$, but that the dynamics becomes time translation invariant when $\varepsilon > \varepsilon^\star$. It is tempting to conjecture that aging disappears exactly when the dynamics becomes ergodic, i.e. when the correlation with a random initial condition decays to zero. This suggests that $\varepsilon^\star = \varepsilon_{\mathrm{RM}}$, which is roughly in line with our numerical data. Note however that the transition between aging dynamics and time translation invariant correlations seems to occur somewhat progressively, hence it may well be that we in fact observe interrupted aging beyond a time that decreases not only with $\varepsilon$ but also with temperature $1/\beta$. More analytical work is needed to clarify the situation.  

A particularity of the aging dynamics of the SK-game is related to its \textit{online} learning dynamics. As clearly visible in both the DMFT equation and the derivation of the Naive Mean-Field equation detailed in Appendix~\ref{appendix:NMFE}, there will inevitably be some decorrelation in time of the \textit{expected} decisions if $\alpha(1-q)$ becomes significant. We therefore naturally expect the region where time translation invariance breaks down to be dependent on all three parameters $\alpha$, $\beta$ and $\varepsilon$.

The interpretation of aging in the socio-economic context is quite interesting and has been discussed in Sec.~\ref{sec:summary_aging}. In a nutshell, it means that as time goes on, agents get stuck in locally satisficing strategies for longer and longer, but after a time proportional to the total time the game has already been played, the system eventually evolves and individual strategies $m_i$ reach an altogether different configuration. This process goes on forever, but becomes slower and slower with time: the notion of quasi-equilibrium therefore makes sense at long times, for small enough noise and small enough non-reciprocity.   

\subsection{Chaos and (quasi-)Limit cycles}

When the non-reciprocity of interactions is sufficiently small and quasi-fixed points exist, we have established that boundedly rational systems displays complicated aging dynamics when learning noise, parameterised by the value of $\beta$, is present. The immediate question is now how such noise influences the complex dynamics, chaos and (quasi) limit cycles that we have found in the $\beta \to \infty$, $\alpha \ll 1$ regime (see Sec.~\ref{sec:deterministic}). 

\begin{figure*}
    \centering
    \includegraphics[width=\textwidth]{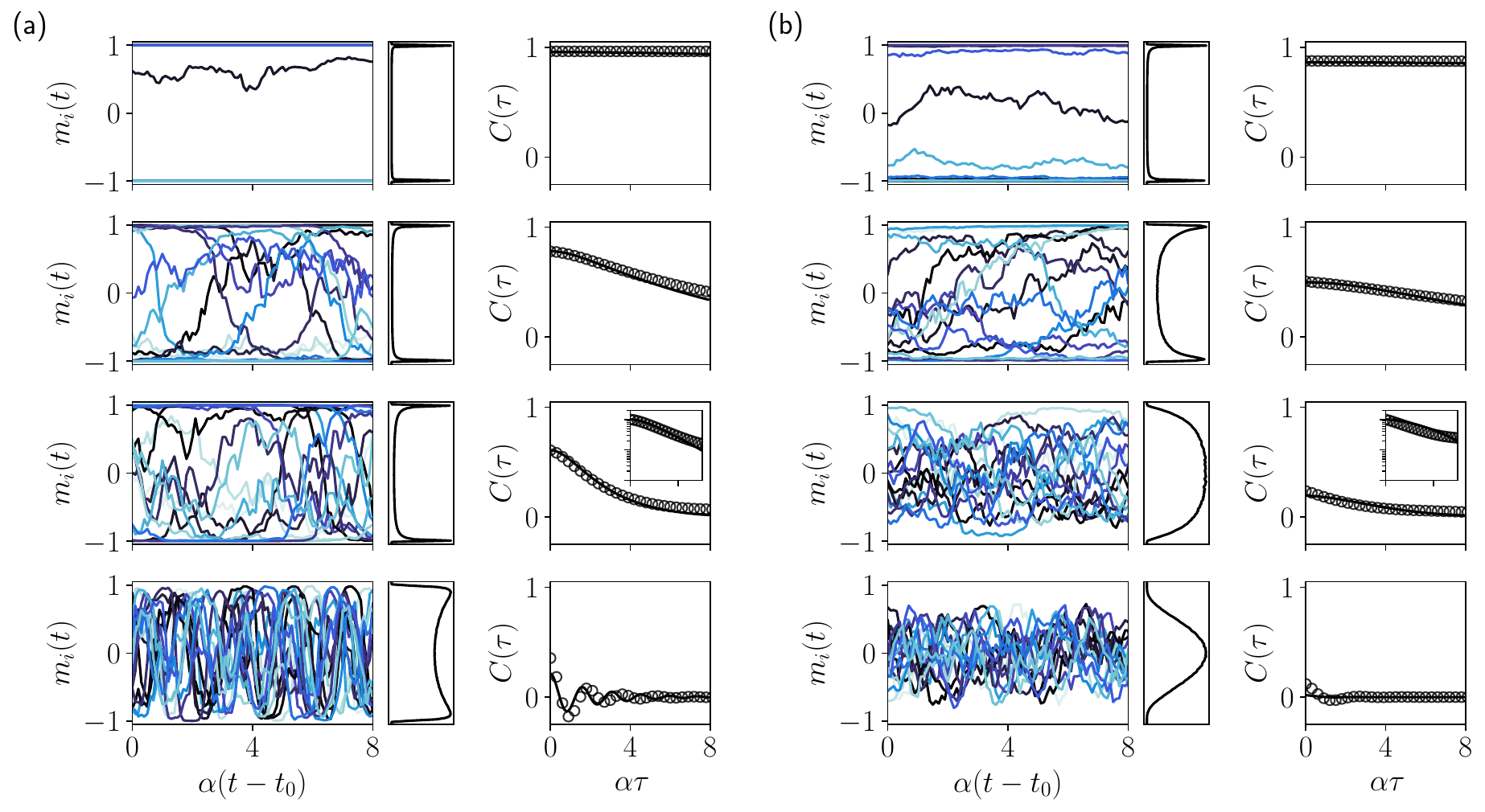}
    \caption{Finite $\beta$ trajectories obtained from numerical simulations at $\alpha = 0.1$, $N = 256$, $t_0 = 10^8$ averaged over 96 samples of disorder and initial conditions for (a) $\beta = 4$, (b) $\beta = 2$ and from top to bottom $\varepsilon = \{0.1,0.85,1.05,1.5\}$. Each line displays (from left to right) the evolution of 16 randomly selected agents, the associated histogram of $m_i$ over both agents, time and realizations and the autocorrelation function assumed to be time translation invariant on short time scales. Third row: insets representing the evolution of the normalized autocorrelation $C(\tau)/C(0)$ with a logarithmic vertical scale.}
    \label{fig:finiteT_traj}
\end{figure*}

To qualitatively illustrate the effect of non-zero noise, we have run simulations for $\alpha = 0.1$ and different values of $\varepsilon$ and $\beta$. Fig.~\ref{fig:finiteT_traj} displays individual trajectories of the intentions $m_i(t)$, as well as the distribution of the values of individual $m_i$ over all agents, realizations and time-steps. We also show the auto-correlation function $C(\tau)$ (assumed to be time-translation invariant over the short time scales considered), which is also compared to the DMFT solved numerically. Note that for the smallest value $\varepsilon = 0.1$, the trajectories illustrate the previously discussed quasi-fixed points emerging from the online dynamics. Clearly, while the correlation function remains close to constant, individual intentions are not exactly frozen (notice the wiggles in the top row of Fig.~\ref{fig:finiteT_traj}, specially for $\beta=2$), explaining how the system as a whole eventually decorrelates and displays aging, as discussed in the previous section.

In the chaotic regime around $\varepsilon = 1$, it is clear that decreasing $\beta$ (increasing noise) spreads the distribution of the $m_i$, which is less and less concentrated around $\pm 1$ as an immediate consequence of the smoothed out hyperbolic tangent. As a result, the equal-time autocorrelation $C(0) = q$ naturally decreases when $\beta$ decreases. It is furthermore interesting to note that its value significantly decreases as the asymmetry parameter $\varepsilon$ increases. We expect a decrease in  $\alpha$ to have a similar role, as suggested by the phase diagrams presented in Fig.~\ref{fig:q_phase_diag} (b) and (c). 

Dynamically, the decay of the autocorrelation in this chaotic regime appears to be more or less independent of the strength of the noise, which can be seen by comparing the insets of the third row of Fig.~\ref{fig:finiteT_traj}. While it is known that external noise kills deterministic chaos in neural networks with uncorrelated couplings \cite{molgedey1992suppressing}, what is interesting in our case is that both the non-linearity of the hyperbolic tangent (governed by $\beta$), and the strength of the effective noise (which scales as $1-q$, see Eq. \eqref{eq:corr_phi}) are varied simultaneously and, in a sense, self-consistently. Determining the way the decorrelation rate evolves with both $\beta$ and $\varepsilon$ is therefore quite non trivial and would be a very interesting endeavor. 

Where we previously had limit cycles, for $\varepsilon = 1.5$ for instance (last row of Fig.~\ref{fig:finiteT_traj}), it appears that oscillations survive for large values of $\beta$. Note however that in this case the value $q=C(0)$ appears to very quickly vanish when $\beta$ decreases. This is consistent with the linear stability analysis of the paramagnetic solution that should be valid for vanishingly small $\alpha$. As visible in Fig.~\ref{fig:q_phase_diag} (c) we indeed expect that the region in which the system displays any form of aggregate coordination becomes increasingly narrow as $\varepsilon$ gets closer to its maximum value of 2. Precisely for $\varepsilon = 2$, the system likely becomes fully disordered ($q=0$) for any finite values of $\beta$ when $\alpha \to 0$. For small but finite values of $\alpha$ as those presented here, this is not quite the case however, and large asymmetries $\varepsilon > 2-\varepsilon_c$ do give rise to clear oscillations, both in individual trajectories and the correlation function (for instance Fig.~\ref{fig:finiteT_traj}, bottom row shows for $\varepsilon = 1.5$ that some oscillations can be somewhat sustained).

\subsection{Role of the Noise: Recap}

In the presence of noise, the ``conviction'' of agents naturally goes down, in the sense that individual $m_i$'s become smaller in absolute value, i.e. less polarized around $\pm 1$. For small $\alpha$ (long memory time), there exists a well defined transition line in the plane $\varepsilon$ (asymmetry), $1/\beta$ (amplitude of the noise) above which agents start playing randomly at each round (i.e. $m_i = 0$), but below which some instantaneous propensity to overplay $+1$ or $-1$ appears. The time evolution of this propensity depends on the value of $\varepsilon$: for small asymmetries, the fixed points that are reached in the absence of noise become quasi-fixed points around which the system settles for longer and longer periods of time, after eventually moving on to a completely different configuration (aging). For $\varepsilon \sim 1$ (uncorrelated influence from $i$ to $j$ and from $j$ to $i$), deterministic chaos when noise is absent becomes noisy chaos, with not much changes. In the highly competitive region $\varepsilon \to 2$, periodic cycles progressively become over-damped, as expected since noise does not allow synchronisation to survive at long times. In terms of individual rewards, not surprisingly, noise tends to be detrimental, except in the reciprocal region ($\varepsilon$ small) where weak noise actually {\it helps} agents coordinating around mutually beneficial actions.

While the precise characterization of this very rich ecology of dynamical behaviors is left for future work, we emphasize that the decorrelation induced by the fact that incentives $Q_i$ are themselves random variables is a key difference between the \textit{online} learning presented here and their \textit{offline} counterpart that are often considered \cite{galla2011cycles,galla2013complex}. While the phenomenology between these two types of learning is somewhat similar for $\alpha \ll 1$, $\beta \gg 1$, there are clearly key differences whenever the ratio $\sqrt{\alpha}/\beta$ ceases to be negligible. 

\section{Summary \& Discussion}
\label{sec:conclusion}

\subsection{Blindsided by Complexity}

Let us summarize our main conceptual 
assertions. As a schematic model of the complexity economic agents are confronted with, we introduced the ``SK-game'', a discrete time binary choice model with $N$ interacting agents and three parameters: $\alpha$ (memory loss rate), $\beta$ (inverse amplitude of noise in the learning process or \textit{intensity of choice}) and $\varepsilon$ (non reciprocity of interactions). 

We have shown that even in a completely static environment where the pay-off matrix does not evolve, agents are unable to learn collectively optimal strategies. This is either because the learning process gets trapped by a sub-optimal fixed point (or remains around one for very long times), or because learning never converges and leads to a never ending (chaotic or quasi-periodic) evolution of agents intentions. 

Hence, contrarily to the hope  that learning might save the ``rational expectation'' framework \cite{evans2013learning}, which still holds the upper hand in macroeconomics textbooks, we argue that complex situations are generically {\it unlearnable}. Agents, therefore, must do with {\it satisficing} solutions, as argued long ago by H. Simon \cite{simon1955behavioral}, an idea embodied by our model in a concrete and tangible way. 

Only a centralized, omniscient agent may be able to ascribe an optimal strategy to all agents -- which incidentally raises the question of trust: would agents even agree to follow the central planner advice? Would they even believe in her ability to solve complex problems, knowing that their solution sensitively depends on all parameters of the model? If a finite fraction of all agents fail to comply, the resulting average reward will drop precipitously below the optimal value and not be much better than the result obtained through individual learning.

As general ideas of interest in a socio-economic context, we have established that 
\begin{enumerate} 
\item  long memory of past rewards is beneficial to learning whereas over-reaction to recent past is detrimental; \item increased competition generically destabilizes fixed points and leads first to chaos and, in the high competition limit, to quasi-cycles; 
\item some amount of noise in the learning process, quite paradoxically, allows the system to reach better collective decisions, in the sense that the average reward is increased; 
\item non-ergodic behaviour spontaneously appear (in the form of ``aging'') in a large swath of parameter space, when $\alpha$, $1/\beta$ and $\varepsilon$ are small.
\end{enumerate}
On the positive side, we have shown that learning is far from useless: instead of getting stuck among one of the most numerous fixed points with low average reward, the learning process does allow the system to coordinate around {\it satisficing} solutions with rather high (but not optimal) average reward. Numerically, the average reward at the end of the learning process is, for $\varepsilon=0$, $\approx 8 \%$ below the optimal value, when the majority of fixed points lead to a much worse average reward  $\approx 33 \%$ below the optimal value \cite{bray1980metastable}.   

\subsection{Technical Results \& Conjectures} 

From a statistical mechanics perspective, our model is next of kin to, but different from several well studied models; a synthesis of our original results can be found in Figs. \ref{fig:qualitative_phase_diag}, \ref{fig:avg_reward}, \ref{fig:NMFE_FP}, \ref{fig:q_phase_diag}. 

For example, when $\alpha = 1$, $\beta \to \infty$, the dynamics is  equivalent to a Hopfield model of learning with non-symmetric Gaussian synaptic couplings, for which many results are known, in particular on the number of fixed points and $L$-cycles. Introducing some memory with $\alpha < 1$, we found that previously dynamically unattainable fixed points become typical solutions, replacing the short limit cycles in which the $\alpha=1$ parallel dynamics get stuck. We also showed how the number of $L$-cycles can be calculated for all values of $\alpha < 1$.  
 
The chaotic region that is known to exist when interactions are mostly non-symmetric ($\varepsilon \approx 1$) also appears to be reduced by memory. When couplings are mostly non-reciprocal $\varepsilon \lesssim 2$, periodic oscillations survive but we found that decreasing $\alpha$ non-trivially increases the cycle length, as $\alpha^{-1/2}$. 

When $\beta$ is finite, the fixed point solutions to the dynamics correspond to the so-called Naive Mean-field Equation of spin glasses, another model that has been studied in detail \cite{bray1986naive}. One knows in particular that such solutions become exponentially abundant for small enough noise $1/\beta$ and for  $\varepsilon=0$, a result that we have extended to all $\varepsilon < 1$. When $\varepsilon > 1$, on the other hand, the only fixed point (or Nash equilibrium) is $m_i =0$, $\forall i$, i.e. completely random decisions at each time step.

The Dynamical Mean-Field Theory (DMFT) is a tool of choice for investigating the dynamics of the model when $N \to \infty$. DMFT is however frustratingly difficult to exploit analytically in the general case, so we are left with numerical solutions of our DMFT equations that accurately match direct numerical simulations of the model when $N$ is large, but fails to capture some specific features arising when $N$ is small. From our numerical results, we conjectured that quasi-fixed points (and correspondingly, aging dynamics) persist for small noise and when $\varepsilon \leq \varepsilon^\star$, where the value of $\varepsilon^\star$ is difficult to ascertain but could be as high as $\varepsilon_{\mathrm{RM}} = 0.47$, perhaps related to the remnant magnetisation transition found in \cite{eissfeller1994mean}. 

For $\beta = \infty$, the long-time, zero temperature autocorrelation $C(\infty)$ appears to drop extremely slowly with $\varepsilon$, but we have not been able to get an  analytical result. DMFT equations also clearly lead to the anomalous $\alpha^{-1/2}$ stretching of the cycles mentioned above, but an analytic solution again eluded us. 

One of the reasons analytical progress with DMFT is difficult is presumably related to the phenomenon of ``Replica Symmetry Breaking'' and its avatar in the present context of dynamical learning. Indeed, any attempt to expand around a static solution of the DMFT equations leads to inconsistencies in the interesting situation $\beta > \beta_c(\varepsilon)$ when decisions are not purely random, see Fig. \ref{fig:FP_q}(b). In fact, as shown in Fig. \ref{fig:FP_q}(a), the value of the order parameter $q$ predicted by such static solutions is substantially off the value found from the long time, numerical solution of the DMFT equations, which itself coincides with direct simulations of the SK-game. {\it En passant}, we noticed that the value of $q$ seems to be given by a universal function of $\beta_c(\varepsilon)/\beta$, independently of the value of $\varepsilon$. Again, we have not been able to understand why this should be the case. 

Finally, we have numerically established several interesting results concerning average and individual rewards, that would deserve further investigations. For example, the average reward seems to converge towards its asymptotic value as $N^{-2/3}$, exactly as for the SK model, although, as already noted above, this asymptotic value is $\approx 8 \%$ below the optimal SK value. Is is possible to characterize more precisely the ensemble of configurations reached after learning in the long memory limit $\alpha \to 0$? Can one, in particular, understand analytically the distribution of individual rewards shown in Fig. \ref{fig:distrib_ei} and the corresponding asymptotic value of the average reward, as well as its non monotonic behaviour as a function of the noise parameter $\beta$? These are, in our opinion, quite interesting theoretical questions left for future investigation.

\subsection{Extensions and Final Remarks}

Many extensions of the very simple framework presented here can be imagined for the model to be more representative of real socio-economic systems. For example, by analogy with spin-glasses, going beyond the fully connected interactions and towards a more realistic network structure should not change the overall phenomenology although some subtle differences may show up. While analytical predictions become even more challenging, recent works on dynamical mean-field theories with finite connectivity, so far developed for Lotka-Volterra type systems, could perhaps be adapted to the learning dynamics of our model.

Allowing the interaction network to evolve with time would of course also make the model more realistic, as in \cite{colon2022radical}; in this case one would have to distinguish the case where the learning time is much longer or much shorter than the reshuffling time of the network. 

Other interesting additions to Ising games could also include the introduction of self-excitation \cite{antonov2021self} or of alternative decision rules that might be less statistical mechanics-friendly \cite{bouchaud2013crises}. Extension to multinary decisions, beyond the binary case considered here, as well as higher-order interactions (i.e. a ``$p$-spin game''), would obviously be interesting as well, specially as higher order interactions are known to change the phenomenology of the SK model (see e.g. \cite{bouchaud1998out}). In particular, we expect that in the $p$-spin case with $p \geq 3$ a much larger gap would develop between the optimal average reward and the one reached by learning. 

Finally, whereas temperature in physics is the same for all spins, there is no reason to believe that the amount of noise $\beta$ or the memory span $\alpha$ should be the same for all agents. Introducing such heterogeneities might be worth exploring, as some agents with longer memory may fare systematically better than others, like in Minority Games, see \cite{challet2004minority}.  

Beyond the socio-economic context that was our initial motivation for its design, we believe that the simplicity and generality of our model makes it a suitable candidate to describe a much wider range of complex systems. In the context of biological neural networks, the parameter $\alpha$ indeed allows one to interpolate between simple discrete-time Hopfield network \cite{hopfield1982neural}, and continuous-time models where $Q_i$ is an activation variable for the firing rates $m_i$ \cite{van1996chaos, brunel2000dynamics,rajan2010stimulus,stern2014dynamics,kadmon2015transition,clark2023theory}. Although in our case the influence of $\beta$ introduces some perhaps unwanted stochasticity, these fluctuations can in principle be suppressed (at least partially) with sufficiently small $\alpha$. The memory loss parameter could also represent an interesting way to tune the effective slowing down of the dynamics caused by symmetry and described in \cite{marti2018correlations}. Here, the description of real neural networks would likely require much more sparse interactions, but also perhaps the introduction of some dedicated dynamics for the interactions themselves, see e.g. \cite{pereira2023forgetting} for recent ideas. 

Closer to our original motivation, more applied socio-economic problems might benefit from the introduction of this type of reinforcement learning. In macroeconomics for instance, some form of ``habit formation'' could perhaps be relevant to extend existing descriptions of input/output networks \cite{dessertaine2022out, colon2022radical}, where client/supplier relationships are probably strongly affected by history (on this point, see Kirman's classic study on Marseilles' fish market \cite{kirman2000learning}, see also \cite{dosi2023foundations}). Finally, while it is an aspect of our model we have not investigated here, previous works have reported that similar dynamics yield interesting volatility clusters and heavy tails, corresponding to sudden changes of quasi-fixed points. Such effects might be relevant to describe financial time series \cite{galla2013complex,sanders2018prevalence}.

In the context of financial markets, our model could also be used to challenge the idea that efficiency is reached by evolution. Indeed, a theory that has been proposed to explain empirical observations going against the Efficient Market Hypothesis is the so-called ``Adaptive Markets Hypothesis'', stating that inefficiencies stem from the (transient) evolution of a market towards true efficiency \cite{lo2004adaptive,lo2005reconciling}. However, reinterpreting the $Q_i$ in our model as some form of fitness measure, it appears unlikely that simple evolutionary dynamics (akin to the learning dynamics) could overcome the type of radical complexity discussed here. As a matter of fact, such an evolutionary twist to the ``SK-game'' could also be used to conjecture that simple Darwinian dynamics is unlikely to lead to a global optimum in a complex biological setting \cite{brotto2017model}.

Last but not least, we believe that the learning dynamics presented here may be useful from a purely algorithmic point of view in the study of spin-glasses and so-called TAP states. Indeed, in the $\beta \to \infty$ limit, we have seen that our iteration relatively frequently finds fixed points in regions where their abundance is known to be sub-exponential (close to $\varepsilon=1$ in particular), and this even for relatively large values of $\alpha$. Interestingly, similar exponentially weighted moving averages have been employed in past numerical studies of TAP states for symmetric interactions \cite{aspelmeier2006free,aspelmeier2019realizable}, but on the magnetizations $m_i$ themselves and not on the local fields $Q_i$ like is the case above. 

Using an offline version of our learning procedure could then be of use to effectively converge to fixed points of the TAP equations or Naive Mean-Field Equations, and study their properties. Perhaps even more interestingly, the online dynamics and the resulting fluctuations of the $m_i$ themselves could prove to be extremely valuable to probe hardly accessible regions of the solution space. In some sense, the fluctuations related to finite values of $\alpha$ and $\beta$ could allow to define ``meta-TAP'' states, in the sense of closely related TAP states mutually accessible thanks to such extra fluctuations, in the same spirit as standard Langevin dynamics in an energy landscape. 

Finally, as mentioned in Sec.~\ref{sec:self-reinforcement}, in the zero temperature limit and for $\varepsilon = 1$, it has recently been reported that for a certain range of self-interaction strengths $J_{\rm d}$, there appears an exponential number of accessible solutions to the TAP equations that are seemingly not reachable with standard Hopfield dynamics \cite{ZecchinaLesHouches}. Preliminary numerical experiments seem to suggest that our learning dynamics find such fixed points, as suggested by their effectiveness for $\varepsilon=1$ without any form of self-interaction. Beyond existing interest around neural networks,  such a self-reinforcement, ``habit formation'' term could also be, as stated above, interesting to study from the socio-economic perspective \cite{pemantle2007survey, moran2020force}.

\begin{acknowledgments}
The authors are indebted to T. Galla and S. Hwang for their insights on the DMFT and the enumeration of cycles respectively. We also thank C. Aubrun, F. Aguirre-Lopez, G. Biroli, C. Colon, J. D. Farmer, R. Farmer, A. Kirman, S. Lakhal, M. Mézard, P. Mergny, A. Montanari, J. Moran, N. Patil, V. Ros, P.-F. Urbani, R. Zakine and F. Zamponi for fruitful discussions on these topics. J.G.-B. would finally like to thank F. Mignacco for precious tips on the numerical resolution of the DMFT equations. This research was conducted within the Econophysics \& Complex Systems Research Chair, under the aegis of the Fondation du Risque, the Fondation de l’Ecole polytechnique, the Ecole polytechnique and Capital Fund Management.
\end{acknowledgments}

\clearpage

\onecolumngrid

\small

\appendix
\addcontentsline{toc}{section}{Appendices}

\section{Static NMFE}
\label{appendix:NMFE}
We claim that for $\alpha \ll 1$
\begin{equation}
    \tilde{m}^\alpha_j(t) \simeq \alpha \sum_{t' = 1}^{t} (1-\alpha)^{t-t'} m_j(t').
\end{equation}
Indeed, given the assumption of independence in time, i.e.  $\E{\left(S_j(t') - m_j(t')\right)\left(S_j(t'') - m_j(t'')\right)} = \delta_{t,t'}\left( 1 - m_(t) \right)$, we have
\begin{equation}
\begin{aligned}
    \E{\left(\tilde{m}^\alpha_j(t) - \alpha \sum_{t'\leq t} (1-\alpha)^{t-t'} m_j(t') \right)^2 } &= \alpha^2 \sum_{t'\leq t} (1-\alpha)^{2(t-t')} \E{ \left(S_j(t') - m_j(t')\right)^2} \\
    & \quad + \alpha^2 \sum_{t' \leq t} \sum_{t'' \neq t'} (1-\alpha)^{t-t'} (1-\alpha)^{t-t''} \E{\left(S_j(t') - m_j(t')\right)\left(S_j(t'') - m_j(t'')\right)}\\
    &= \alpha^2 \sum_{t'\leq t} (1-\alpha)^{2(t-t')} \left(1 - (m_j(t'))^2 \right)\\
    &\leq \alpha^2 \sum_{t'\leq t} (1-\alpha)^{2(t-t')} = \frac{\alpha}{2 - \alpha} \xrightarrow[\alpha \to 0]{} 0.
\end{aligned}
\end{equation}
We can then make the \textit{ansatz} that the expected decision reaches a fixed point $m_j^\star$ after some time. For sufficiently large $t$ and small but finite values of $\alpha$, we will therefore have 
\begin{equation}
    \E{\tilde{m}^\alpha_j(t)} \simeq m_j^\star
\end{equation}
with fluctuations characterized by
\begin{equation}
    \E{\left(\tilde{m}^\alpha_j(t) - m_j^\star \right)^2} = \frac{\alpha}{2}\left(1 - (m_j^\star)^2 \right) + O(\alpha^2).
\end{equation}

\section{Limit Cycle Complexity with Memory}
\label{appendix:LC_complexity}

To study the influence of the memory loss rate $\alpha < 1$, we may adapt the method of Hwang \textit{et al.}, although this requires the introduction of either a strong nonlinearity in the exponent and subsequent saddle equations, or of new variables. We opt for the latter, and write the number of cycles of length $L$ as
\begin{equation}
\begin{aligned}
    {\mathcal{N}}_L(N,\alpha,\varepsilon) &= \sum_{\{S_i(t) \} } \int_{-\infty}^{\infty} \bigg( \prod_{i=1}^N \prod_{t=1}^L \dd Q_i(t) \bigg) \left\lvert \det {\boldsymbol{\mathcal{J}}}^\alpha \right\rvert^N \prod_{i,t} \delta \Big( Q_i(t+1) - (1-\alpha) Q_i(t) - \alpha \sum_j J_{ij} S_j(t) \Big) \Theta(Q_i(t) S_i(t)),
\end{aligned}
    \label{eq:N_L_count}
\end{equation}
where the Dirac $\delta$ ensures that the dynamics are satisfied at each step, while the second enforces $S_i(t) = \sign(Q_i(t)) \; \forall t$. The $L \times L$ matrix $\bm{\mathcal{J}}^\alpha$ is the $\alpha$-dependent Jacobian ensuring that the zeros of the $\delta$ function are correctly weighted, i.e. for $L > 1$
\begin{equation}
    \mathcal{J}^\alpha_{ts} =
    \begin{cases}
        -(1-\alpha), &\qquad \text{if} \quad {s = t}{\pmod L}\\
        1, &\qquad \text{if} \quad {s = t + 1}{\pmod L}\\
        0 &\qquad \text{otherwise}.
    \end{cases}
\end{equation}
The first step is, as usual, to perform the average over the disorder after introducing the integral representation of the Dirac $\delta$,
\begin{equation}
\begin{aligned}
    \overline{\mathcal{N}_L}(N,\alpha,\varepsilon) &= \left\lvert \det \bm{\mathcal{J}}^\alpha \right\rvert^N \sum_{\{S_i(t) \} } \int_{-\infty}^{\infty}  \bigg( \prod_{i=1}^N \prod_{t=1}^L \dd Q_i(t)  \frac{\dd \lambda_i(t)}{2\pi} \bigg) \bigg( \prod_{i,t} \Theta(S_i(t) Q_i(t)) \bigg) \exp \bigg[ -i\sum_{i,t} \lambda_i(t) Q_i(t+1) + i(1-\alpha)\sum_{i,t} \lambda_i(t) Q_i(t) \\
    & - \frac{1}{2N} \alpha^2  \Big( 1- \frac{\varepsilon}{2}\Big)^2 \sum_{i<j} \Big( \sum_t [\lambda_i(t) S_j(t) + \lambda_j(t) S_i(t) ] \Big)^2 - \frac{1}{2N} \alpha^2  \Big(\frac{\varepsilon}{2}\Big)^2 \sum_{i<j} \Big( \sum_t [\lambda_i(t) S_j(t) - \lambda_j(t) S_i(t) ] \Big)^2\bigg].
\end{aligned}
\end{equation}
The last two terms, resulting from the average on disorder, may be rearranged to give
\begin{equation*}
    -\frac{\alpha^2 }{2N} \sum_{t,s} \bigg( \sum_{i,j} \Big[ \upsilon(\varepsilon)  \lambda_i(t) S_j(t) \lambda_i(s) S_j(s) + (1-\varepsilon) \lambda_i(t) S_j(t) \lambda_j(s) S_i(s)  \Big] - \frac12 (\varepsilon - 2)^2 \sum_i \lambda_i(t) S_i(t) \lambda_i(s) S_i(s) \bigg).
\end{equation*}
Similar to ref. \cite{hwang2019number}, we introduce a set of auxiliary functions,
\begin{alignat}{4}
U(t,s) &= \frac{1}{N} \sum_i \lambda_i(t) S_i(t) \lambda_i(s) S_i(s),\\
V(t,s) &= \frac{1}{N} \sum_i \lambda_i(t) S_i(s),\\
R(t,s) &= \frac{1}{N} \sum_i \lambda_i(t) \lambda_i(s),\\
K(t,s) &= \frac{1}{N} \sum_i S_i(t) S_i(s),
\end{alignat}
such that the last term gives
\begin{equation*}
    -\frac12 N \alpha^2  \sum_{t,s} \Big[ \upsilon(\varepsilon) R(t,s) K(t,s) + (1-\varepsilon) V(t,s) V(s,t) - \frac{(2-\varepsilon)^2}{2N} U(t,s) \Big].
\end{equation*}
In the limit $N\to \infty$, the $O(N^{-1})$ term is insignificant and can thus be neglected. The complete expression is then,
\begin{equation}
\begin{aligned}
    \overline{\mathcal{N}_L}(N,\alpha,\varepsilon) &= \left\lvert \det \bm{\mathcal{J}}^\alpha \right\rvert^N \sum_{\{S_i(t) \} } \int_{-\infty}^{\infty}  \bigg( \prod_{i,t} \dd Q_i(t)  \frac{\dd \lambda_i(t)}{2\pi} \bigg) \bigg( \prod_{i,t} \Theta(S_i(t) Q_i(t)) \bigg)\bigg( \prod_{t,s} \dd K(t,s)  \frac{\dd \hat{K}(t,s)}{2\pi} \dd R(t,s)  \frac{\dd \hat{R}(t,s)}{2\pi} \dd V(t,s)  \frac{\dd \hat{V}(t,s)}{2\pi} \bigg) \\
    & \exp \bigg[ -i\sum_{i,t} \lambda_i(t) Q_i(t+1) + i(1-\alpha)\sum_{i,t} \lambda_i(t) Q_i(t) -\frac12 N \alpha^2  \sum_{t,s} \bigg[ \upsilon(\varepsilon) R(t,s) K(t,s) + (1-\varepsilon) V(t,s) V(s,t) \bigg]\\
    & - i\sum_{t,s} \hat{K}(t,s) \Big(N K(t,s) - \sum_i S_i(t) S_i(s)\Big) - i\sum_{t,s} \hat{R}(t,s) \Big(N R(t,s) - \sum_i \lambda_i(t) \lambda_i(s)\Big) \\
    & - i\sum_{t,s} \hat{V}(t,s) \Big(N V(t,s) - \sum_i \lambda_i(t) S_i(s) \Big) \bigg].
\end{aligned}
\end{equation}
Now, one can notice that
\begin{equation}
    \int_{-\infty}^\infty \bigg( \prod_{t,s} \frac{\dd R(t,s)}{2\pi} \bigg) \, \e^{- \sum_{t,s} i R(t,s)\big[ N\hat{R}(t,s) - \frac{1}{2}iN \alpha^2 \upsilon(\varepsilon) K(t,s) \big]} = \prod_{t,s}\delta \bigg(N\hat{R}(t,s) - \frac{1}{2}i N \alpha^2 \upsilon(\varepsilon) K(t,s)\bigg),
\end{equation}
which means the integral over $K(t,s)$ now reads,
\begin{equation}
    \int_{-\infty}^\infty \bigg( \prod_{t,s} \frac{\dd R(t,s)}{2\pi} \bigg) \, \prod_{t,s} \e^{-i N \hat{R}(t,s) R(t,s)} \delta \bigg(N\hat{R}(t,s) - \frac{1}{2}i N \alpha^2\upsilon(\varepsilon) K(t,s)\bigg) = \e^{-\frac{2N}{\alpha^2 \upsilon(\varepsilon)}\sum_{t,s} \hat{R}(t,s) \hat{K}(t,s)}
\end{equation}
For $V(t,s)$, we have to be a bit more careful, as the expression is not linear for all terms, as the diagonal for $t = s$ gives a quadratic term that will have to be treated separately. Noticing that the product $V(t,s) V(s,t)$ is symmetric, we start by considering the $t < s$,
\begin{equation}
    \int_{-\infty}^\infty \bigg( \prod_{t < s} \frac{\dd V(t,s)}{2\pi} \bigg) \, \e^{-\sum_{t < s} i V(t,s)\big[ N \hat{V}(t,s) - iN \alpha^2  (1-\varepsilon) V(s,t) \big]} = \prod_{t < s} \delta \Big( N \hat{V}(t,s) - iN \alpha^2  (1-\varepsilon) V(s,t) \Big).
\end{equation}
Now integrating over the $V(t,s)$ for $t > s$,
\begin{equation}
    \int_{-\infty}^\infty \bigg( \prod_{t > s} \frac{\dd V(t,s)}{2\pi} \bigg) \prod_{t>s} \e^{-iN \hat{V}(t,s) V(t,s)} \delta \Big( -i \alpha^2  (1-\varepsilon) V(s,t) + 2\hat{V}(t,s) \Big) = \e^{-\frac{N}{\alpha^2  (1-\varepsilon)}\sum_{t > s} \hat{V}(t,s) \hat{V}(s,t)}.
\end{equation}
Finally, the diagonal $V(t,t)$ for which the expression is quadratic can be computed with a Gaussian integral,
\begin{equation}
    \int_{-\infty}^\infty  \bigg(\prod_t \dd V(t,t) \bigg) \, \e^{-\frac{1}{2}N\alpha^2(1-\varepsilon)\sum_{t} V(t,t)^2 - iN\sum_t V(t,t) \hat{V}(t,t)} \sim \e^{-\frac{N}{2\alpha^2  (1-\varepsilon)}\sum_t \hat{V}(t,t)^2}
\end{equation}
Up to an $O(1)$ multiplicative constant, the complete expression now reads
\begin{equation}
\begin{aligned}
    \overline{\mathcal{N}_L}(N,\alpha,\varepsilon) \sim  & \left\lvert \det \bm{\mathcal{J}}^\alpha \right\rvert^N \sum_{\{S_i(t) \} } \int_{-\infty}^{\infty}  \bigg( \prod_{i,t} \dd Q_i(t)  \frac{\dd \lambda_i(t)}{2\pi} \bigg)\bigg(\prod_{i,t} \Theta(S_i(t) Q_i(t)) \bigg) \bigg( \prod_{t,s} \dd \hat{K}(t,s) \dd \hat{R}(t,s) \dd \hat{V}(t,s) \bigg)\\
    & \exp \bigg( -i\sum_{i,t} \lambda_i(t) Q_i(t+1) + i(1-\alpha)\sum_{i,t} \lambda_i(t) Q_i(t) + i \sum_{i,t} \nu_i(t) Q_i(t) S_i(t)\\
    & - \frac{2N}{\alpha^2 \upsilon(\varepsilon)} \sum_{t,s} \hat{R}(t,s) \hat{K}(t,s) - \frac{N}{2\alpha^2  (1-\varepsilon)} \sum_{t,s}\hat{V}(t,s) \hat{V}(s,t)\\
    & + i\sum_{t,s} \hat{K}(t,s) \sum_i S_i(t) S_i(s) + i\sum_{t,s} \hat{R}(t,s) \sum_i \lambda_i(t) \lambda_i(s) + i\sum_{t,s} \hat{V}(t,s) \sum_i \lambda_i(t) S_i(s) \bigg).
\end{aligned}
\end{equation}
Taking the changes of variable
\begin{equation}
\begin{aligned}
    \hat{R}(t,s) &\to \frac{1}{2}\alpha^2  \upsilon(\varepsilon) \hat{R}(t,s)\\
    \hat{K}(t,s) &\to \frac{1}{2}\alpha^2  \upsilon(\varepsilon) \hat{K}(t,s)\\
    \hat{V}(t,s) &\to \alpha^2  (1-\varepsilon) \hat{V}(t,s)
\end{aligned}
\end{equation}
for convenience, we can entirely factorize the problem in $N$, finally giving
\begin{equation}
\begin{aligned}
    \overline{\mathcal{N}_L}(N,\alpha,\varepsilon) \sim  \int_{-\infty}^{\infty}  \bigg( \prod_{t,s} \dd \hat{K}(t,s) \dd \hat{R}(t,s) \dd \hat{V}(t,s) \bigg) \exp\bigg( &N\bigg[ -\frac{1}{2} \alpha^2  \upsilon(\varepsilon) \sum_{t,s} \hat{R}(t,s) \hat{K}(t,s)\\
    & -\frac{1}{2} \alpha^2  (1-\varepsilon) \sum_{t,s} \hat{V}(t,s) \hat{V}(s,t) + \log {\mathcal{I}_L}  + \log \left\lvert \det \bm{\mathcal{J}}^\alpha \right\rvert \bigg] \bigg).
\end{aligned}
\label{eq:N_L_almost_done}
\end{equation}
where
\begin{equation}
\begin{aligned}
    \mathcal{I}_L = \sum_{\{S(t)\}} \int_{-\infty}^{\infty} \bigg(\prod_t \dd Q(t) \frac{\dd \lambda(t)}{2\pi} \bigg)\, & \bigg(\prod_t \Theta(S(t) Q(t))\bigg) \exp \bigg[ -i \sum_t \lambda(t) Q(t+1) + i (1-\alpha) \sum_t \lambda(t) Q(t) \\
    &+ \frac{1}{2} \alpha^2  \upsilon(\varepsilon) \sum_{t,s} (S(t) S(s) i \hat{K}(t,s) + \lambda(t) \lambda(s) i\hat{R}(t,s)) + \alpha^2  (1-\varepsilon) \sum_{t,s} \lambda(t) S(s) i \hat{V}(t,s) \bigg].
\end{aligned}
\end{equation}
At this stage, we may notice that $S(t)^2 = 1 \, \forall t$ and as such that the diagonal part of the sum over $K(t,s)$ can be taken out of $\mathcal{I}_L$. Doing so and combining this contribution to the first term of the exponent of Eq.~\ref{eq:N_L_almost_done}, we have
\begin{equation}
\begin{aligned}
    \overline{\mathcal{N}_L}(N,\alpha,\varepsilon) \sim  &\int_{-\infty}^{\infty}  \bigg( \prod_{t,s} \dd \hat{K}(t,s) \dd \hat{R}(t,s) \dd \hat{V}(t,s) \bigg) \exp\bigg( N\bigg[ \frac{1}{2} \alpha^2  \upsilon(\varepsilon) \sum_{t} (i\hat{R}(t,t) + 1) i\hat{K}(t,t)\\
    & + \frac{1}{2} \alpha^2  \bigg(1-\varepsilon+\frac{\varepsilon^2}{2}\bigg) \sum_{s\neq t} i\hat{R}(t,s) i\hat{K}(t,s) -\frac{1}{2} \alpha^2  (1-\varepsilon) \sum_{t,s} \hat{V}(t,s) \hat{V}(s,t) + \log \mathcal{I}_L + \log \left\lvert \det \bm{\mathcal{J}}^\alpha \right\rvert \bigg] \bigg).
\end{aligned}
\end{equation}
Now, the first term in the exponent can be integrated over the $\hat{K}(t,t)$ exactly, yieding a product of $\delta$ functions fixing
\begin{equation}
    i\hat{R}(t,t) = -1 \quad \forall t,
\end{equation}
as expected and introducing only a sub-dominant correction $O\left(\frac{\log N}{N}\right)$ in the complexity.

In the $N\to \infty$ limit, the complexity is then finally given 
\begin{equation}
\begin{aligned}
    \Sigma_L(\alpha,\varepsilon) = \saddle_{\hat{R},\hat{K},\hat{V}} \bigg\{ \frac{1}{2} \alpha^2  \upsilon(\varepsilon) \sum_{s\neq t} i\hat{R}(t,s) i\hat{K}(t,s) - \frac{1}{2} \alpha^2  (1-\varepsilon) \sum_{t,s} \hat{V}(t,s) \hat{V}(s,t) + \log \mathcal{I}_L +& \log \left\lvert \det \bm{\mathcal{J}}^\alpha \right\rvert\bigg\},
\end{aligned}
\end{equation}
with now
\begin{equation}
\begin{aligned}
    \mathcal{I}_L = & \sum_{\{S(t)\}} \int_{-\infty}^{\infty} \bigg(\prod_t \dd Q(t) \frac{\dd \lambda(t)}{2\pi} \bigg)\, \prod_t \Theta(S(t) Q(t)) \exp \bigg[ -i \sum_t \lambda(t) Q(t+1) + i (1-\alpha) \sum_t \lambda(t) Q(t) \\
    &- \frac{1}{2} \alpha^2  \upsilon(\varepsilon) \Big[ \sum_t \lambda(t)^2 - \sum_{s\neq t} (S(t) S(s) i \hat{K}(t,s) - \lambda(t) \lambda(s) i\hat{R}(t,s)) \Big]+ \alpha^2  (1-\varepsilon) \sum_{t,s} \lambda(t) S(s) i \hat{V}(t,s) \bigg].
\end{aligned}
\end{equation}
This integral may be rewritten as
\begin{equation}
\begin{aligned}
    \mathcal{I}_L &= \sum_{\{S(t)\}} \e^{\frac12 \alpha^2 \upsilon(\varepsilon) \sum_{s\neq t} S(t) S(s) i \hat{K}(t,s) } \int_{-\infty}^{\infty} \bigg(\prod_t \frac{\dd Q(t)}{\sqrt{2\pi}} \frac{\dd \lambda(t)}{\sqrt{2\pi}} \bigg)\, \prod_t \Theta(S(t) Q(t)) \, \exp\left( -\frac12 \bm{\lambda}^\top \mathbf{A} \bm{\lambda} - i \mathbf{b}^\top \bm{\lambda} \right).
\end{aligned}
\end{equation}
with the $L\times L$ matrix $\mathbf{A}$ constituted of 
\begin{equation}
    A(t,s) = \alpha^2 \upsilon(\varepsilon)(\delta_{ts} - (1-\delta_{ts}) i\hat{R}(t,s)),
\end{equation}
and $\mathbf{b} \in \mathbb{R}^L$,
\begin{equation}
    b(t) = Q(t+1) - (1-\alpha)Q(t) - \alpha^2(1-\varepsilon) \sum_s \hat{V}(t,s) S(s),
\end{equation}
such that we in fact have
\begin{equation}
    \mathbf{b} = \bm{\mathcal{J}}^{\alpha} \mathbf{Q} + \mathbf{c}, \qquad c(t) = - \alpha^2(1-\varepsilon) \sum_s \hat{V}(t,s) S(s).
\end{equation}
Computing the Gaussian integral on $\bm{\lambda}$, 
\begin{equation}
\begin{aligned}
    \mathcal{I}_L &= \sum_{\{S(t)\}} \frac{\e^{\frac12 \alpha^2 \upsilon(\varepsilon) \sum_{s\neq t} S(t) S(s) i \hat{K}(t,s) }}{\sqrt{\det \mathbf{A} }}\int_{-\infty}^{\infty} \bigg(\prod_t \frac{\dd Q(t)}{\sqrt{2\pi}} \bigg)\, \prod_t \Theta(S(t) Q(t)) \, \exp\left( -\frac12 (\bm{\mathcal{J}}^\alpha \mathbf{Q} + \mathbf{c})^\top \mathbf{A}^{-1} (\bm{\mathcal{J}}^\alpha \mathbf{Q} + \mathbf{c})\right).
\end{aligned}
\end{equation}
Taking the change of variable $\mathbf{u} = \bm{\mathcal{J}}^\alpha \mathbf{Q} + \mathbf{c}$, the above becomes
\begin{equation}
\begin{aligned}
    \mathcal{I}_L &= \sum_{\{S(t)\}} \frac{\e^{\frac12 \alpha^2 \upsilon(\varepsilon) \sum_{s\neq t} S(t) S(s) i \hat{K}(t,s) }}{\left\lvert \det \bm{\mathcal{J}}^{\alpha} \right\rvert \sqrt{\det \tilde{\mathbf{A}}}}\int_{-\infty}^{\infty} \bigg(\prod_t \frac{\dd Q(t)}{\sqrt{2\pi}} \bigg)\, \prod_t \Theta(S(t)(\mathbf{Q} - (\bm{\mathcal{J}}^\alpha)^{-1} \mathbf{c})(t)) \, \exp\left( -\frac12 \mathbf{Q}^\top \tilde{\mathbf{A}}^{-1} \mathbf{Q}\right),
\end{aligned}
\end{equation}
where $\tilde{\mathbf{A}} = (\bm{\mathcal{J}}^\alpha)^{-1} \mathbf{A} (\bm{\mathcal{J}}^\alpha)^{-1}$ As a result, the $\left\lvert \det \bm{\mathcal{J}}^\alpha \right\rvert$ contributions in the complexity cancel out, and we find ourselves with 
\begin{equation}
\begin{aligned}
    \Sigma_L(\alpha,\varepsilon) = \saddle_{\hat{R},\hat{K},\hat{V}} \bigg\{ \frac{1}{2} \alpha^2  \upsilon(\varepsilon) \sum_{s\neq t} i\hat{R}(t,s) i\hat{K}(t,s) - \frac{1}{2} \alpha^2  (1-\varepsilon) \sum_{t,s} \hat{V}(t,s) \hat{V}(s,t) + \log \mathcal{I}_L \bigg\},
\end{aligned}
\end{equation}
with now
\begin{equation}
\begin{aligned}
    \mathcal{I}_L &= \sum_{\{S(t)\}} \frac{\e^{\frac12 \alpha^2 \upsilon(\varepsilon) \sum_{s\neq t} S(t) S(s) i \hat{K}(t,s) }}{\sqrt{\det \tilde{\mathbf{A}} }}\int_{-\infty}^{\infty} \bigg(\prod_t \frac{\dd Q(t)}{\sqrt{2\pi}} \bigg)\, \prod_t \Theta(S(t)(\mathbf{Q} - (\bm{\mathcal{J}}^\alpha)^{-1} \mathbf{c})(t)) \, \exp\left( -\frac12 \mathbf{Q}^\top \tilde{\mathbf{A}}^{-1} \mathbf{Q}\right).
\end{aligned}
\end{equation}
The integral is challenging to study in generality, as the non-diagonal nature of $\bm{\mathcal{J}}^\alpha$ and $\mathbf{A}$ means that we cannot factorize the integrand. We now move on to specific cases of interest. The equations can be further simplified through the rescalings $\alpha^2 \upsilon(\varepsilon) \hat{K}(t,s) \to \hat{K}(t,s)$, $\alpha \sqrt{\upsilon(\varepsilon)} \hat{V}(t,s) \to \hat{V}(t,s)$ and $Q(t) \to \alpha \sqrt{\upsilon(\varepsilon)}$, in which case we finally get
\begin{equation}
\begin{aligned}
    \Sigma_L(\alpha,\eta) = \saddle_{\hat{R},\hat{K},\hat{V}} \bigg\{ \sum_{s < t} i\hat{R}(t,s) i\hat{K}(t,s) - \frac{\eta}{2} \sum_{t,s} \hat{V}(t,s) \hat{V}(s,t) + \log \mathcal{I}_L \bigg\},
\end{aligned}
\end{equation}
\begin{equation}
\begin{aligned}
    \mathcal{I}_L &= \sum_{\{S(t)\}} 
    \e^{\sum_{s < t} S(t) S(s) i \hat{K}(t,s) } \Psi_L(\Gamma_1(\alpha,\eta),\dots,\Gamma_L(\alpha,\eta);\mathbf{C}),
\end{aligned}
\end{equation}
where $\Psi_L(x_1,\dots,x_L;\mathbf{C})$ is the cumulative distribution of an $L$-dimensional Gaussian with a zero mean vector and covariance matrix $\mathbf{C}$ evaluated up to $x_t$, $t \in \{1,\dots,L\}$. In our case, the upper bounds of integration are given by $\bm{\Gamma} = \mathbf{S} \circ (\bm{\mathcal{J}}^\alpha)^{-1} \mathbf{c}$, with now
\begin{equation}
    c(t) = \eta \sum_s \hat{V}(t,s) S(s),
\end{equation}
while the covariance matrix must be taken with care as the off-diagonal elements must be adapted to the presence of $S(t)$ in the Heaviside step function. As a result, off-diagonal elements must be multiplied by $S(t) S(s)$. As a result, we have 
\begin{equation}
    C(t,s) = 
    \begin{cases}
        \tilde{A}(t,t) \quad & \quad \text{for} \; t = s\\
        S(t) S(s) \tilde{A}(t,s) \quad & \quad \text{for} \; t \neq s\\
    \end{cases}
\end{equation}
with $\tilde{\mathbf{A}} = (\bm{\mathcal{J}}^\alpha)^{-1} \mathbf{A} (\bm{\mathcal{J}}^\alpha)^{-1}$ and
\begin{equation}
    A(t,s) = \delta_{ts} - (1-\delta_{ts}) i\hat{R}(t,s).
\end{equation}

\subsection{Fixed Points}
\label{appendix:LC_complexity_FP}

We can start by checking that we recover the known result for $L = 1$. In this case, $\mathcal{J}^\alpha = \alpha$, and we only have to solve for a scalar $\hat{V}(1,1) = x$. Here, 
\begin{equation}
    \mathcal{I}_L = 2 \Psi_1\left(\frac{\eta}{\alpha} x; \frac{1}{\alpha^2} \right) = 2 \Phi(\eta x),
\end{equation}
therefore the saddle point equation becomes
\begin{equation}
    \Sigma_1(\alpha,\eta) := \Sigma_{\mathrm{FP}}(\eta) = \max_x \left\{ -\frac12 x^2 + \log 2 + \log \Phi(\eta x) \right\},
\end{equation}
from which the expression in the main text can immediately be recovered.

\subsection{Two-cycles}
There are now six variables to solve for: $i\hat{R}$, $i\hat{K}$, $\hat{V}_1 = \hat{V}(1,1)$, $\hat{V}_2 = \hat{V}(2,2)$, $\hat{V}_{12} = \hat{V}(1,2)$ and $\hat{V}_{21} = \hat{V}(2,1)$. In this case, the Jacobian matrix is given by
\begin{equation}
    \bm{\mathcal{J}}^\alpha = 
    \begin{bmatrix}
        -(1-\alpha) & 1 \\
        1 & -(1-\alpha)
    \end{bmatrix}
    \qquad \Rightarrow \qquad 
    (\bm{\mathcal{J}}^\alpha)^{-1} = \frac{1}{1-(1-\alpha)^2}
    \begin{bmatrix}
        1-\alpha & 1 \\
        1 & 1-\alpha
    \end{bmatrix},
\end{equation}
giving the bounds of integration
\begin{equation}
    \bm{\Gamma} = \frac{\eta}{1-(1-\alpha)^2} 
    \begin{bmatrix}
        \hat{V}_{21} + S_1 S_2 \hat{V}_2 + (1-\alpha)(\hat{V}_1 + S_1 S_2 \hat{V}_{12})\\
        \hat{V}_{12} + S_1 S_2 \hat{V}_1 + (1-\alpha)(\hat{V}_2 + S_1 S_2 \hat{V}_{21})\\
    \end{bmatrix}
\end{equation}
and the covariance matrix
\begin{equation}
    \mathbf{C} = \frac{1}{(1-(1-\alpha)^2)^2} 
    \begin{bmatrix}
        1+(1-\alpha)^2 -2(1-\alpha)i\hat{R} & S_1S_2(2(1-\alpha) - (1+(1-\alpha)^2)i\hat{R} )\\ 
        S_1S_2(2(1-\alpha) - (1+(1-\alpha)^2)i\hat{R} ) & 1+(1-\alpha)^2 -2(1-\alpha)i\hat{R}
    \end{bmatrix}.
\end{equation}
Now, it is immediately apparent that $\mathcal{I}_L$ is a function of the product $S_1 S_2 = \pm 1$ rather than $S_1$ and $S_2$. As a result, the complexity is given by
\begin{equation}
   \Sigma_2(\alpha,\eta) = \saddle_{\hat{R},\hat{K},\hat{V}_1,\hat{V}_2,\hat{V}_{12},\hat{V}_{21}} \left\{ i\hat{R} i\hat{K} - \frac{\eta}{2}(\hat{V}_1^2 + \hat{V}_2^2 + 2\hat{V}_{12}\hat{V}_{21}) + \log 2 + \log \sum_{S=\pm 1} \e^{S i\hat{K}} \Psi_2(\Gamma_1, \Gamma_2;\mathbf{C}) \right\}.
\end{equation}
The six saddle point equations read
\begin{equation}
    i\hat{R} \sum_{S = \pm 1} \e^{ S i\hat{K}} \Psi_2(\Gamma_1,\Gamma_2;\mathbf{C}) + \sum_{S = \pm 1} S \e^{ S i\hat{K}} \Psi_2(\Gamma_1,\Gamma_2;\mathbf{C}) = 0,
    \label{eq:saddle_R}
\end{equation}
\begin{equation}
    \left(i\hat{K} + \frac{i\hat{R}}{1 - (i\hat{R})^2} \right) \sum_{S = \pm 1} \e^{ S i\hat{K}} \Psi_2(\Gamma_1,\Gamma_2;\mathbf{C}) - \frac{1}{2} \sum_{S = \pm 1} \e^{ S i\hat{K}} \int_{-\infty}^{\Gamma_1} \dd Q_1 \int_{-\infty}^{\Gamma_2} \dd Q_2 \, \mathbf{Q}^\top \frac{\partial \mathbf{C}^{-1}}{\partial i\hat{R}} \mathbf{Q} \frac{\e^{-\frac{1}{2} \mathbf{Q}^\top \mathbf{C}^{-1} \mathbf{Q}}}{2\pi \sqrt{\det \mathbf{C}}} = 0 
    \label{eq:saddle_K}
\end{equation}
\begin{equation}
\begin{aligned}
    -\eta \hat{V}_1 \sum_{S = \pm 1} \e^{ S i\hat{K}} \Psi_2(\Gamma_1,\Gamma_2;\mathbf{C}) & + \frac{\eta (1-\alpha)}{1-(1-\alpha)^2}\sum_{S = \pm 1} \e^{ S i\hat{K}} \int_{-\infty}^{\Gamma_2} \dd Q_2 \frac{\e^{-\frac{1}{2} {\mathbf{Q}_2^\star}^\top \mathbf{C}^{-1} \mathbf{Q}_2^\star}}{2\pi \sqrt{\det \mathbf{C}}} \\
    & + \frac{\eta}{1-(1-\alpha)^2}\sum_{S = \pm 1} S \e^{ S i\hat{K}} \int_{-\infty}^{\Gamma_1} \dd Q_1 \frac{\e^{-\frac{1}{2} {\mathbf{Q}_1^\star}^\top \mathbf{C}^{-1} \mathbf{Q}_1^\star}}{2\pi \sqrt{\det \mathbf{C}}} = 0
    \label{eq:saddle_V1}
\end{aligned}
\end{equation}
\begin{equation}
\begin{aligned}
    -\eta \hat{V}_2 \sum_{S = \pm 1} \e^{ S i\hat{K}} \Psi_2(\Gamma_1,\Gamma_2;\mathbf{C}) & + \frac{\eta}{1-(1-\alpha)^2}\sum_{S = \pm 1} S \e^{ S i\hat{K}} \int_{-\infty}^{\Gamma_2} \dd Q_2 \frac{\e^{-\frac{1}{2} {\mathbf{Q}_2^\star}^\top \mathbf{C}^{-1} \mathbf{Q}_2^\star}}{2\pi \sqrt{\det \mathbf{C}}} \\
    & + \frac{\eta (1-\alpha)}{1-(1-\alpha)^2}\sum_{S = \pm 1} \e^{ S i\hat{K}} \int_{-\infty}^{\Gamma_1} \dd Q_1 \frac{\e^{-\frac{1}{2} {\mathbf{Q}_1^\star}^\top \mathbf{C}^{-1} \mathbf{Q}_1^\star}}{2\pi \sqrt{\det \mathbf{C}}} = 0
    \label{eq:saddle_V2}
\end{aligned}
\end{equation}
\begin{equation}
\begin{aligned}
    -\eta \hat{V}_{21} \sum_{S = \pm 1} \e^{ S i\hat{K}} \Psi_2(\Gamma_1,\Gamma_2;\mathbf{C}) & + \frac{\eta(1-\alpha)}{1-(1-\alpha)^2}\sum_{S = \pm 1} S \e^{ S i\hat{K}} \int_{-\infty}^{\Gamma_2} \dd Q_2 \frac{\e^{-\frac{1}{2} {\mathbf{Q}_2^\star}^\top \mathbf{C}^{-1} \mathbf{Q}_2^\star}}{2\pi \sqrt{\det \mathbf{C}}} \\
    & + \frac{\eta}{1-(1-\alpha)^2}\sum_{S = \pm 1} \e^{ S i\hat{K}} \int_{-\infty}^{\Gamma_1} \dd Q_1 \frac{\e^{-\frac{1}{2} {\mathbf{Q}_1^\star}^\top \mathbf{C}^{-1} \mathbf{Q}_1^\star}}{2\pi \sqrt{\det \mathbf{C}}} = 0
    \label{eq:saddle_V21}
\end{aligned}
\end{equation}
\begin{equation}
\begin{aligned}
    -\eta \hat{V}_{12} \sum_{S = \pm 1} \e^{ S i\hat{K}} \Psi_2(\Gamma_1,\Gamma_2;\mathbf{C}) & + \frac{\eta}{1-(1-\alpha)^2}\sum_{S = \pm 1} \e^{ S i\hat{K}} \int_{-\infty}^{\Gamma_2} \dd Q_2 \frac{\e^{-\frac{1}{2} {\mathbf{Q}_2^\star}^\top \mathbf{C}^{-1} \mathbf{Q}_2^\star}}{2\pi \sqrt{\det \mathbf{C}}} \\
    & + \frac{\eta (1-\alpha)}{1-(1-\alpha)^2}\sum_{S = \pm 1} S \e^{ S i\hat{K}} \int_{-\infty}^{\Gamma_1} \dd Q_1 \frac{\e^{-\frac{1}{2} {\mathbf{Q}_1^\star}^\top \mathbf{C}^{-1} \mathbf{Q}_1^\star}}{2\pi \sqrt{\det \mathbf{C}}} = 0,
    \label{eq:saddle_V12}
\end{aligned}
\end{equation}
with $\mathbf{Q}^*_1 = [Q_1 \quad \Gamma_2 ]^\top$ and similarly for $\mathbf{Q}^*_2$, and
\begin{equation}
    \frac{\partial \mathbf{C}^{-1}}{\partial i\hat{R}} = \frac{1}{(1-(i\hat{R})^2)^2} 
    \begin{bmatrix}
        2i\hat{R}(1 + (1-\alpha)^2) - 2(1-\alpha)((i\hat{R})^2 + 1) & S (-4i\hat{R}(1-\alpha) + (1+(1-\alpha)^2)((i\hat{R})^2+1))\\
        S (-4i\hat{R}(1-\alpha) + (1+(1-\alpha)^2)((i\hat{R})^2+1)) & 2i\hat{R}(1 + (1-\alpha)^2) - 2(1-\alpha)((i\hat{R})^2 + 1)
    \end{bmatrix}.
\end{equation}

We now take the particular case $\alpha = 1$. In this case, we expect two-cycles with a zero overlap in between consecutive steps, meaning $i\hat{R} = 0$. As a result, the matrix $\mathbf{C}$ is the identity matrix, and thus
\begin{equation}
    \Psi_2(\Gamma_1,\Gamma_2;\mathbf{C}) = \Phi\left( \eta(\hat{V}_{21} + S \hat{V}_2\right) \Phi\left( \eta(\hat{V}_{12} + S \hat{V}_1)\right).
\end{equation}
In order for $i\hat{R} = 0$ to satisfy Eq.~\eqref{eq:saddle_R}, we then require $i\hat{K} = 0$ and $\hat{V}_1 = \hat{V}_2 = 0$ such that $\sum_{S=\pm 1} S \e^{i\hat{K}} \Psi_2 (\Gamma_1,\Gamma_2;\mathbf{C}) = 0$. Given
\begin{equation}
    \frac{\partial \mathbf{C}^{-1}}{\partial i\hat{R}} \bigg\rvert_{i\hat{R} = 0} =
    \begin{bmatrix}
        0 & S\\
        S & 0
    \end{bmatrix}
\end{equation}
for $\alpha = 1$, the solution $i\hat{K} = 0$ satisfies the saddle point equation~\eqref{eq:saddle_K} while the independence of the integrands on $S$ ensures that $\hat{V}_1 = \hat{V}_2 = 0$ are compatible with Eqs.~\eqref{eq:saddle_V1}-\eqref{eq:saddle_V2}. Eqs.~\eqref{eq:saddle_V21}-\eqref{eq:saddle_V12} then suggest the ansatz $\hat{V}_{12} = \hat{V}_{21} = x$, in which case the problem is finally reduced to
\begin{equation}
\begin{aligned}
    \Sigma_2(\alpha = 1,\eta) &= \max_x \left\{ -\eta x^2 + 2\log 2 + 2\log \Phi(\eta x) \right\}\\
    &= 2 \Sigma_{\mathrm{FP}}(\eta),
\end{aligned}
\end{equation}
recovering the known solution of \cite{gutfreund1988nature,hwang2019number}.

\section{Derivation of the DMFT Equations}
\label{appendix:DMFT}
We start from the $N \gg 1$ discrete difference equations to which we have added an external field $h_i(t)$
\begin{equation}
    Q_i(t+1) = (1-\alpha) Q_i(t) + \alpha \sum_j J_{ij} m_j(t) + \alpha \eta_i(t) + h_i(t), \qquad \langle \eta_i(t) \eta_j^s \rangle \approx \upsilon(\varepsilon)(1-q(t)) \delta_{t,s} \delta_{i,j}
\end{equation}
and introduce a new agent at index $i = 0$. The influence of this newly introduced agent on the dynamic equation of agents $i \neq 0$ is then
\begin{equation}
    \sum_j J_{ij} m_j(t) + h_i(t) \to \sum_j J_{ij} m_j(t) + J_{i0} m_0(t) + h_i(t)
\end{equation}
Given $N$ is large, we consider the response at the linear order, meaning that the expected decision of all agents $i > 0$ becomes
\begin{equation}
    m_i(t) \to m_i(t) + \sum_{s < t} \sum_j \underbrace{\frac{\partial m_i(t)}{\partial h_j(s)}\bigg\rvert_{h = 0}}_{\chi_{ij}(t,s)} J_{j0} m_0(s),
\end{equation}
as $J_{j0} m_0(t)$ can be seen as the modification of the effective field ``felt'' by all agents $j$, and $\chi_{ij}(t,s)$ is simply the linear response function to a small external field. The dynamics that the newly introduced agent follows is then given by
\begin{equation}
    Q_0(t+1) = (1-\alpha) Q_0(t) + \alpha \sum_{i > 0} J_{0i} m_i(t) + \sum_{s < t} \Big( \sum_{ij} J_{0i} \chi_{ij}(t,s) J_{j0} \Big) m_0(s) + \alpha \eta_0(t) + h_0(t).
\end{equation}
The sum on $i,j$ can be split in its diagonal and off-diagonal parts. On the diagonal, we assume the central limit theorem holds, yielding
\begin{equation}
    \sum_{i} J_{0i} \chi_{ii}(t,s) J_{i0} = N \bigg[ \overline{\langle J_{0i} \chi_{ii}(t,s) J_{i0} \rangle }+ O\bigg(\frac{1}{\sqrt{N}}\bigg) \bigg] \approx (1-\varepsilon) \langle \chi_{ii}(t,s) \rangle,
\end{equation}
while the off-diagonal contribution will be sub-dominant as its mean will be zero given non-opposing entries in the interaction matrix are uncorrelated.  We can also assume the CLT is valid for the sum on indices $i$ at the leading order, in which case
\begin{equation}
    \sum_i J_{0i} m_i(t) \approx 0 + \upsilon(\varepsilon) \xi_0(t),
\end{equation}
where $\xi_0$ is a Gaussian of zero mean and correlated in time as $\langle \xi_0(t) \xi_0(s) \rangle = C_0(t,s) = \langle m_0(t) m_0(s) \rangle$. Bringing everything together, one realizes that there is no cross contributions between agents at the leading order and thus that we can drop the index $0$ and recover the equation in the main text
\begin{equation}
    Q(t+1) = (1-\alpha) Q(t) + \alpha^2 (1-\varepsilon) \sum_{s < t} G(t,s) m(s) + \alpha \phi(t) + \alpha h(t),
\end{equation}
with a new noise term combining the original thermal-like fluctuations and the effective contribution from the disorder averaging, with $\langle \phi(t) \rangle = 0$, and
\begin{equation}
    \langle \phi(t) \phi(s) \rangle = \upsilon(\varepsilon) [C(t,s) + (1-q(t)) \delta_{t,s}],
\end{equation}
and where the memory kernel and correlation function are to be determined self-consistently,
\begin{equation}
    G(t,s) = \langle \chi_{ii}(t,s) \rangle = \bigg\langle \frac{\partial m(t)}{\partial h(s)} \bigg\rvert_{h=0} \bigg\rangle, \qquad C(t,s) = \langle m(t) m(s) \rangle.
\end{equation}
Note that in order to eliminate the external field, we have expressed the susceptibility with the noise term, resulting in a rescaling $G(t,s) \to \alpha G(t,s)$. From this expression, the continuous limit can be taken, changing the sum in time weighted by $\alpha$ into an integral,
\begin{equation}
    \frac{\alpha}{2} \ddot{Q}(t) + \dot{Q}(t) = -Q(t) + (1-\varepsilon) \int_0^t \dd s\, G(t,s) m(s) + \phi(t) + h(t).
\end{equation}

\section{Adapting the Sompolinsky \& Crisanti result}
\label{appendix:Sompolinsky}
We start from the much simplified DMFT equation,
\begin{equation}
    \dot{Q}(t) = - Q(t) + \phi(t), \qquad \langle \phi(t) \phi(s) \rangle = \frac{1}{2} C(t,s)
\end{equation}
with $C(t,s) = \langle m(t) m(s) \rangle$, $m(t) = \sign (Q(t))$ in the $\beta \to \infty$ limit. As shown by Sompolinsky \& Crisanti \cite{sompolinsky1988chaos}, we can write a second order differential equation for the two-point autocorrelation of $Q(t)$,
\begin{equation}
    \ddot{\Delta}(\tau) = \Delta(\tau) - \frac{1}{2} C(\tau),
\end{equation}
with therefore $\Delta(\tau) = \langle Q(t+\tau) Q(t) \rangle$, and where thanks to the Gaussian nature of the fluctuations we have
\begin{equation}
\begin{aligned}
    C(\tau) &= \int_{-\infty}^{\infty} \frac{\dd z}{\sqrt{2\pi}} \, \e^{-\frac12 z^2} \bigg[ \int_{-\infty}^{\infty} \frac{\dd x}{\sqrt{2\pi}} \, \e^{-\frac12 x^2} \sign\Big(\sqrt{\Delta(0) - \lvert \Delta(\tau) \rvert} x + \sqrt{\lvert \Delta(\tau) \rvert} z\Big) \bigg]^2\\
    &= \int_{-\infty}^{\infty} \frac{\dd z}{\sqrt{2\pi}} \, \e^{-\frac12 z^2} \operatorname{erf}\bigg( \sqrt{\frac{\lvert \Delta(\tau) \rvert}{\Delta(0) - \lvert \Delta(\tau)}} \frac{z}{\sqrt{2}}\bigg)^2\\
    &= \frac{2}{\pi} \sin^{-1} \bigg( \frac{\Delta(\tau)}{\Delta(0)}\bigg),
\end{aligned}
\end{equation}
(see \cite{prudnikov1990more} for the last step above). Now, by identifying that the only difference with the original problem is a factor 2 in $\Delta$, we can directly use the result $\Delta(0) = 1-\frac2\pi$, found by enforcing the condition required for $\Delta(\tau)$ to decay monotonously \cite{crisanti2018path}. We may finally expand the inverse sine to recover
\begin{equation}
    \ddot{\Delta}(\tau) \underset{\tau \gg 1}{\sim} \bigg(1 - \frac{1}{\pi \Delta(0)}\bigg) \Delta(\tau)
\end{equation}
and therefore
\begin{equation}
    C(\tau) \underset{\tau \gg 1}{\sim} \frac{2}{\pi} \e^{-\frac{\tau}{\tau_1}}, \qquad \tau_1 = \sqrt{\frac{\pi-2}{\pi-3}} \approx 2.84.
\end{equation}
We notice that the value of the characteristic time is identical in the standard Sompolinsky \& Crisanti case (the only difference being a factor 2 in the magnitude of $\Delta(\tau)$), which is therefore \textit{not} equal to the value $\sqrt{\frac{\pi}{\pi - 2}} \approx 1.66$ originally given in \cite{sompolinsky1988chaos} and \cite{crisanti1988dynamics}.

\twocolumngrid

\bibliography{bibs}

\end{document}